\documentclass[a4paper,11pt]{article}
\pdfoutput=1
\usepackage{jheppub}

\usepackage[utf8x]{inputenc}

\usepackage{subfigure,graphicx,amsmath,amssymb,mathtools,hhline}
\usepackage{color,xcolor}
\usepackage{float,lscape}
\usepackage{rotating}
\usepackage{multirow}

\usepackage{soul}

\usepackage{pifont}
\newcommand{\cmark}{{\color{blue} \ding{51}}}%
\newcommand{\xmark}{{\color{red} \ding{55}}}%

\newcommand{\be}{\begin{eqnarray}}
\newcommand{\ee}{\end{eqnarray}}

\newcommand{\ba}{\begin{array}}
\newcommand{\ea}{\end{array}}
\newcommand{\bee}{\begin{equation}\ba{c}}
\newcommand{\eee}{\ea\end{equation}}

\newcommand{\bi}{\begin{itemize}}
\newcommand{\ei}{\end{itemize}}

\newcommand{\lsim}{{\;\raise0.3ex\hbox{$<$\kern-0.75em\raise-1.1ex\hbox{$\sim$}}\;}}
\newcommand{\gsim}{{\;\raise0.3ex\hbox{$>$\kern-0.75em\raise-1.1ex\hbox{$\sim$}}\;}}

\newcommand{\beq}{\begin{equation}}
\newcommand{\eeq}{\end{equation}}
\newcommand{\bea}{\begin{eqnarray}}
\newcommand{\eea}{\end{eqnarray}}
\mathchardef\minus="002D

\newcommand{\spp}{{p_p}}

\toccontinuoustrue

\title{Optimizing Energetic Light Dark Matter Searches in Dark Matter and Neutrino Experiments}

\author[a,b]{Doojin Kim,}
\author[c]{Pedro A. N. Machado,}
\author[d]{Jong-Chul Park,}
\author[e,f]{and Seodong Shin}
\affiliation[a]{Mitchell Institute for Fundamental Physics and Astronomy, Department of Physics and Astronomy, Texas A\&M University, College Station, TX 77843, USA}
\affiliation[b]{Department of Physics, University of Arizona, Tucson, AZ 85721, USA}
\affiliation[c]{Theory Department, Fermi National Accelerator Laboratory, P.O. Box 500, Batavia, IL 60510, USA}
\affiliation[d]{Department of Physics, Chungnam National University, Daejeon 34134, Republic of Korea}
\affiliation[e]{Enrico Fermi Institute, University of Chicago, Chicago, IL 60637, USA}
\affiliation[f]{Department of Physics, Jeonbuk National University, Jeonju, Jeonbuk 54896, Republic of Korea}
\emailAdd{doojin.kim@tamu.edu}
\emailAdd{pmachado@fnal.gov}
\emailAdd{jcpark@cnu.ac.kr}
\emailAdd{sshin@jbnu.ac.kr}

\preprint{
\begin{minipage}{5cm}
\begin{flushright}
EFI-18-20\\
FERMILAB-PUB-20-112-T \\
MI-TH-208
 \end{flushright}
\end{minipage}
}

\abstract{
Neutrino and dark matter experiments with large-volume ($\gtrsim 1$ ton) detectors can provide excellent sensitivity to signals induced by energetic light dark matter coming from the present universe.  
Taking boosted dark matter as a concrete example of energetic light dark matter, we scrutinize two representative search channels, electron scattering and proton scattering including deep inelastic scattering processes, in the context of elastic and inelastic boosted dark matter, in a completely detector-independent manner.
In this work, a dark gauge boson is adopted as the particle to mediate the interactions between the Standard Model particles and boosted dark matter. 
We find that the signal sensitivity of the two channels highly depends on the (mass-)parameter region to probe, so search strategies and channels should be designed sensibly especially at the earlier stage of experiments. 
In particular, the contribution from the boosted-dark-matter-initiated deep inelastic scattering can be subleading (important) compared to the quasi-elastic proton scattering, if the mass of the mediator is below (above) $\mathcal{O}({\rm GeV})$. 
We demonstrate how to practically perform searches and relevant analyses, employing example detectors such as DarkSide-20k, DUNE, Hyper-Kamiokande, and DeepCore, with their respective detector specifications taken into consideration. 
For other potential detectors we provide a summary table, collecting relevant information, from which similar studies can be fulfilled readily.
\newpage
}

\begin{document}

\maketitle

\section{Introduction}

The dark matter puzzle is a clear motivation for physics beyond the Standard Model (SM).
While the evidence for the existence of dark matter is all rooted in its gravitational interaction with ordinary matter, searches via its non-gravitational couplings are actively underway.
Of those trials, the strategy of dark matter direct detection has been playing a role of the major driver in searching for relevant signatures.
Most of direct detection experiments are designed to observe a recoil of target material (henceforth called primary signature) which is induced by the {\it elastic} scattering of {\it non}-relativistic dark matter (see e.g. Ref.~\cite{Undagoitia:2015gya} for a modern review on dark matter direct detection experiments). 

A variation in this search scheme is to look for inelastic scattering signals -- which was originally motivated by the DAMA annual modulation signature~\cite{Bernabei:2000qi} --, imagining the process that a dark matter particle scatters off to an excited state along with a target recoil whose energy spectrum differs from that in the elastic scattering mentioned above~\cite{TuckerSmith:2001hy}.
By construction, the excited state may be de-excited back to the dark matter state as previously expected in~\cite{Finkbeiner:2007kk}, potentially leaving visible signals at the detector (henceforth called secondary signature): for example, X-ray photon in neutrinoless double beta decay experiments~\cite{Pospelov:2013nea}. 
One may attempt to observe both primary and secondary signatures, but it is usually challenging for both of them to overcome the relevant detector threshold simultaneously due to inadequate dark matter kinetic energy. 
Indeed, the rich structure of inelastic dark matter models have endowed themselves with the potential to explain a diverse range of astrophysical phenomena~\cite{Finkbeiner:2007kk,ArkaniHamed:2008qn,Pospelov:2008jd,Finkbeiner:2009mi} and have inspired novel LHC search strategies~\cite{Essig:2007az,ArkaniHamed:2008qp,Bai:2011jg,Bell:2013wua,Berlin:2018jbm}.

A myriad of experimental efforts to observe dark matter signals have been devoted under the search schemes discussed above. 
However, none of them have recorded solid dark-matter-induced signatures yet, so they merely sets stringent bounds on parameter space of associated dark matter models.  
The null observation motivates alternative approaches.
One possible direction to pursue is to look for similar experimental signatures invoked by dark matter relics of different mass scales~\cite{Alexander:2016aln, Battaglieri:2017aum}.
Since most of the existing direct search experiments aim at weak-scale dark matter, hence the associated detectors are designed accordingly, new detector material and/or technology are often demanded in order to perform relevant experiments~\cite{Alexander:2016aln, Battaglieri:2017aum}.
Alternatively, one may search for any scattering signatures of {\it relativistically} incoming dark matter, usually having in mind the mass scale of standard thermal dark matter.

A straightforward production mechanism is to obtain relativistic dark matter at particle accelerators, where a certain fraction of initial-state beam energy is transferred to the dark matter.     
The elusive nature of typical dark matter often requires highly intensified particle beam essentially to increase signal statistics, e.g., fixed target experiments~\cite{LoSecco:1980nf, Alexander:2016aln, Battaglieri:2017aum}. 
The larger fraction of literature studied elastic scattering of relativistically produced dark matter, e.g., Refs.~\cite{Batell:2009di, deNiverville:2011it, DeRomeri:2019kic, Dutta:2019nbn, Batell:2019nwo}, but the energetic nature of such dark matter essentially allows decent cross section for its ``up''-scattering to an excited (or equivalently heavier unstable) state, under the framework of inelastic dark matter. 
Reference~\cite{Izaguirre:2014dua} pointed out the potential of detecting both the primary recoil induced by such relativistic dark matter and visible decay product(s) of the excited state and showed that it allows relevant signal searches to suffer from significantly less background contamination, hence inducing recent development in phenomenological investigations~\cite{Izaguirre:2017bqb, Berlin:2018pwi}. 

Another class of mechanisms invokes production of energetic (light) dark matter in the present universe.
We emphasize that our study here is straightforwardly applicable to several physics scenarios, models, or frameworks involving relativistically produced dark matter, but it is instructive to develop our argument in the context of a concrete example.
Such an example is the scenario of boosted dark matter (BDM)~\cite{Agashe:2014yua}.
In BDM models, it is assumed that a dark sector containing two dark matter species with a hierarchical mass spectrum~\cite{Belanger:2011ww}. 
Separate symmetries are usually employed to stabilize the two dark matter species, e.g., $Z_2 \times Z_2'$ or U(1)$'\times$ U(1)$''$.
The overall dark matter relic is set by the so-called ``assisted freeze-out'' mechanism~\cite{Belanger:2011ww}.
Suppose that the heavier and the lighter species are denoted by $\chi_0$ and $\chi_1$, respectively.
Typical models hypothesize that $\chi_0$ does not directly interact with SM particles but pair-annihilates into a $\chi_1$ pair which directly couples to SM particles, i.e., the $\chi_0$ relic abundance is determined by ``assistance'' of $\chi_1$. 
As a result, $\chi_0$ usually remains as the dominant relic component, whereas $\chi_1$ becomes subdominant. 
Under this setup, it is hard to detect relic $\chi_0$ at standard dark matter  
direct detection experiments due to its tiny coupling strength to SM particles, while it is again hard to detect relic $\chi_1$ due to its small statistics in the current universe.

However, the model setup allows for $\chi_1$ production via pair-annihilation of $\chi_0$ in the galactic halo. The produced light dark matter $\chi_1$ acquires a significant Lorentz boost factor due to the mass gap between $\chi_0$ and $\chi_1$, which opens a novel physics opportunity, the search for signatures of relativistic dark matter scattering.
The signal detection prospect is deeply related to the total $\chi_1$ flux $\mathcal{F}_1$~\cite{Agashe:2014yua}:
\bea
\mathcal{F}_1 = 1.6\times 10^{-8}\textrm{cm}^{-2}\textrm{s}^{-1} \left( \frac{\langle \sigma v \rangle_{0\to 1} }{5\times 10^{-26} \textrm{cm}^3\textrm{s}^{-1}}\right) \left(\frac{100\hbox{ GeV}}{m_0} \right)^2\,, \label{eq:baselineflux}
\eea
where $m_0$ denotes the mass of $\chi_0$ which consists of most of the dark matter relic in well-motivated regions of parameter space.
The chosen value for $\langle \sigma v \rangle_{0\to 1}$, the velocity-averaged annihilation cross section of $\chi_0$ to $\chi_1$, corresponds to a correct dark matter thermal relic density of $\chi_0$, under the assumption that an $s$-wave 
process dominates the annihilation.

The above relation implies that for $\chi_0$ of weak-scale mass (i.e., $\sim 100$ GeV), the incoming flux of lighter dark matter $\chi_1$ (near the earth) is as small as $\mathcal O(10^{-8} \,{\rm cm}^{-2} {\rm s}^{-1})$. 
Thus, large-volume neutrino detectors such as Super-Kamiokande (SK), Hyper-Kamiokande (HK), Deep Underground Neutrino Experiment (DUNE), and IceCube Neutrino Observatory are preferred in the search for experimental signatures induced by $\chi_1$ elastic scattering-off detector material~\cite{Agashe:2014yua,Berger:2014sqa,Kong:2014mia,Bhattacharya:2014yha,Kopp:2015bfa,Necib:2016aez,Alhazmi:2016qcs,Berger:2019ttc} or those by $\chi_1$ inelastic scattering (which can be viewed as a combination of the original boosted dark matter scenario and the inelastic dark matter model)~\cite{Kim:2016zjx}. 
We call the former and the latter scenarios $e$BDM and $i$BDM, respectively, as shorthand throughout the rest of this paper. 
The SK Collaboration has conducted the search for high-energy electron recoil ($ \gsim 0.1$ GeV) induced by $\chi_1$ elastic scattering and reported the first results~\cite{Kachulis:2017nci}.
On the other hand, it was shown that sub-GeV/GeV-range $m_0$ can increase the $\chi_1$ flux substantially, while keeping the resultant relic abundance consistent with the current measurement, so that ton-scale dark matter direct detection experiments such as Xenon1T and LUX-ZEPLIN can be sensitive enough to $e$BDM/$i$BDM signals~\cite{Cherry:2015oca, Giudice:2017zke,McKeen:2018pbb}.
Recently, the COSINE-100 Collaboration has searched for the electron-positron pair in coincidence with the primary electron signal by $\chi_1$ inelastic scattering as a signature of an $i$BDM interaction and reported the first results of direct search for $i$BDM~\cite{Ha:2018obm}.
See also a recent White Paper~\cite{Arguelles:2019xgp} surveying related physics opportunities in a wide range of large-volume neutrino experiments.

We remark that in terms of recoiling target particles, two channels are considered: an electron target and a proton target. 
The former is an elementary particle so that predicting a $\chi_1$ scattering cross section with electrons in the detector material is rather straightforward. 
On the other hand, the latter is a composite object and is typically bound in nuclei, and as a result, the corresponding scattering cross section involves form factors. Moreover, since incident light dark matter $\chi_1$ is relativistic, a deep inelastic scattering (DIS) process may arise.  
When it comes to $i$BDM, the proton target is often advantageous for $\chi_1$ to ``up''-scatter off to a heavier (unstable) dark-sector state~\cite{Kim:2016zjx, Giudice:2017zke}. 
Considering these factors altogether, therefore, it is of great importance to choose a better channel for a given parameter region to explore.  
This strategical approach is highly motivated, in particular, at the earlier stage of experiments.
For example, several sub-kiloton-scale neutrino detectors in Short Baseline Neutrino Program (SBN)~\cite{Antonello:2015lea, Acciarri:2016smi} and prototypical DUNE (ProtoDUNE)~\cite{Agostino:2014qoa, Abi:2017aow} are running or ready to take data within a few months, and physics opportunities in terms of $i$BDM~\cite{Chatterjee:2018mej} and $e$BDM~\cite{Kim:2018veo} have been proposed recently.

In light of this situation, we perform a dedicated study to provide useful guidance for boosted dark matter searches in this paper which can be taken as a reference search for energetic light dark matter coming from the universe. 
In more detail, we first show that the DIS contribution out of an entire proton scattering cross section is negligible, as long as the mass of the particle mediating the interaction between energetic light dark matter $\chi_1$ and SM particles is not much larger than the energy scale inducing a DIS process. 
Therefore, in many of the well-motivated scenarios, it is sufficient to consider only contributions by genuine proton scattering. 
We then compare the electron channel and the proton channel through their respective scattering cross sections for both $e$BDM and $i$BDM signals, including realistic factors such as detector energy threshold, cuts, and angular resolution at several benchmark detectors.
 
To deliver the main ideas efficiently, our paper is organized as follows. 
In Section~\ref{sec:model}, we briefly discuss the scenario of boosted dark matter and an example model to describe the interactions between lighter dark matter species $\chi_1$ and SM particles, followed by listing up several benchmark detectors and their key characteristics. 
We then look into scattering cross sections of an incoming $\chi_1$ with electron and proton targets in Section~\ref{sec:pvsDIS}, putting a particular emphasis on the proton DIS scattering.
Detailed comparison between electron and proton scattering channels for both $e$BDM and $i$BDM follows in Sections~\ref{sec:comparison} and~\ref{sec:comparisonwitheffects} at the theory level and the (semi-)detector level, respectively, while we scan over the mediator mass and the $\chi_1$ mass for a given $m_0$.
Example phenomenology will be discussed and demonstrated in Section~\ref{sec:pheno}, and our concluding remarks will appear in Section~\ref{sec:conclusion}.
For the sake of reference, we provide a couple of appendices. 
In Appendix~\ref{sec:derivation}, we provide our derivation for various scattering cross section formulas and some detailed description of our data analysis.
A summary of key specifications of detectors other than the benchmark ones is presented in Appendix~\ref{sec:detectors}.

\section{Benchmark Models and Detectors}
\label{sec:model}

We begin with setting up benchmark models and detectors with which our detailed analysis will be demonstrated.
The method can be straightforwardly extended to other models and detectors, but the case studies that we are performing in this paper will provide a baseline for other applications.   

\subsection{Dark matter models and experimental signatures}

The simplified model under consideration is divided into two parts, the one delineating the production mechanism of boosted dark matter at the universe today and the one describing the interaction between SM particles and the boosted dark matter. 
Expected experimental signatures follow once the latter part is defined. 

We note that there are several ways to create relativistic (or at least fast-moving) dark matter particles in the universe: for example, two-component dark matter scenario~\cite{Agashe:2014yua, Belanger:2011ww, Kim:2017qaw, Aoki:2018gjf}, models with a $Z_3$ symmetry which may induce semi-annihilation processes~\cite{DEramo:2010keq}, models involving anti-baryon-numbered dark matter-induced nucleon decays inside the sun~\cite{Huang:2013xfa}, scenarios with decaying super-heavy particles~\cite{Bhattacharya:2014yha, Kopp:2015bfa, Heurtier:2019rkz}, or energetic cosmic-ray induced (semi-)relativistic dark matter scenarios~\cite{Yin:2018yjn, Bringmann:2018cvk, Ema:2018bih}.
One can also think of various places from which boosted dark matter dominantly comes. 
Examples include the galactic center (GC)~\cite{Agashe:2014yua}, the sun~\cite{Berger:2014sqa, Kong:2014mia}, and dwarf galaxies~\cite{Necib:2016aez}.  
Among those possibilities, we simply choose the two-component dark matter scenario in which the dominant flux of boosted dark matter comes from the GC, as our benchmark model. 
In this scenario, the boost factor of BDM is a free parameter, determined by the mass gap between the two dark matter species. 
Thus, we can change the boost factor freely and study/show the resulting phenomenological effect without any model restriction. 

As briefly explained in the Introduction, one of the two dark matter species (usually the heavier $\chi_0$) indirectly couples to SM-sector particles via the other dark matter species (usually the lighter $\chi_1$) which directly communicates with SM particles.  
Assuming that dark matter relic abundance is determined thermally, we see that the $\chi_0$ relic is set with the aid of $\chi_1$~\cite{Belanger:2011ww}. 
In more detail, $\chi_0$ is in (indirect) contact with the thermal bath via $\chi_0\chi_0 \to \chi_1\chi_1$ followed by sufficiently large $\chi_1 \chi_1 \to \hbox{SM\,SM}$, in annihilation models described by an effective operator, e.g.,
\bea
\mathcal{L} \supset \frac{1}{\Lambda^2} \bar{\chi}_0 \chi_0 \bar{\chi}_1 \chi_1\,,
\eea
where $\Lambda$ parameterizes some high-scale physics. 
The thermally averaged annihilation cross section of $\chi_0 \chi_0 \to \chi_1 \chi_1$ at the present universe is fixed to be $\sim 10^{-26}\,{\rm cm}^3\,{\rm s}^{-1}$ yielding the observed value of dark matter relic abundance, with the assumption that an $s$-wave annihilation dominates $\chi_0 \chi_0 \to \chi_1 \chi_1$.
The calculation procedure to find the $\chi_1$ flux is similar to that for the photon flux via dark matter pair-annihilation, so we write it as follows:
\begin{align}
\mathcal F_1 &= \frac{1}{2} \cdot \frac{1}{4\pi} \int d\Omega \int_{\rm l.o.s.}ds \langle \sigma v \rangle_{0 \to 1}\left(\frac{\rho(s,\theta)}{m_0} \right)^2  \label{eq:fluxformula} \\
&\approx 1.6 \times 10^{-8}\,{\rm cm}^{-2} {\rm s}^{-1}  
\, \times \, \left( \frac{\langle \sigma v \rangle_{0 \to 1}}{5 \times 10^{-26}\,{\rm cm}^3 {\rm s}^{-1}} \right) \times \left( \frac{100\,{\rm GeV}}{m_0}\, \right)^2 \nonumber \,,
\end{align}
where $\rho$ describes the $\chi_0$ density distribution in terms of the line of sight (l.o.s.) $s$ and solid angle $\Omega$ with $\theta$ being the angle between the direction of l.o.s. and the axis connecting the GC and the earth.
We essentially reproduce the result in Eq.~\eqref{eq:baselineflux} in the second line of the above formula, assuming that the $\chi_0$ is distinguishable from its anti-particle, say $\bar \chi_0$, and that the dark matter halo is distributed according to the Navarro-Frenk-White profile~\cite{Navarro:1995iw,Navarro:1996gj}.
In the other case in which $\chi_0$ and $\bar \chi_0$ are indistinguishable, one can simply drop the prefactor 1/2 in the formula.
It is noteworthy that the flux formula \eqref{eq:fluxformula} is independent of the type of interactions between the dark sector and SM sector as we have not assumed any particular type.

As for the interactions between the dark sector and the SM sector, we adopt a dark gauge boson scenario as our reference model.
For a general analysis, we take the framework of inelastic boosted dark matter~\cite{Kim:2016zjx,Giudice:2017zke,Chatterjee:2018mej} which includes an additional (unstable) dark sector state $\chi_2$ heavier than $\chi_1$.
We allow both diagonal interaction of $\chi_1$ with target (i.e., $e$BDM) and off-diagonal interaction (i.e., $i$BDM).
Minimal ingredients forming the relevant sector of our reference model are $\chi_1$, $\chi_2$, and a hidden massive gauge boson $X^\mu$.
The interaction Lagrangian includes the following operators:
\bea
\mathcal{L} \supset &-&\frac{\epsilon}{2}F_{\mu\nu}X^{\mu\nu} + g_{11}\bar{\chi}_1\gamma^{\mu}\chi_1 X_\mu +g_{12}\bar{\chi}_2\gamma^{\mu}\chi_1 X_\mu + {\rm h.c.}\,, \label{eq:lagrangian}
\eea
where the first term describes the kinetic mixing between U(1)$_{\textrm{EM}}$ and U(1)$_X$~\cite{Okun:1982xi, Galison:1983pa, Holdom:1985ag, Huh:2007zw, Pospelov:2007mp, Chun:2010ve, Park:2012xq, Belanger:2013tla, PARK:2016wip}, that is, field strength tensors for the ordinary photon and a hidden gauge boson $F_{\mu\nu}$ and $X_{\mu\nu}$ are mixed by parameter $\epsilon$.
The diagonal and off-diagonal gauge interactions are parameterized by the couplings $g_{11}$ and $g_{12}$.
The example realization of such interactions (in particular, flavor-changing currents) in a model construction was discussed in Refs.~\cite{TuckerSmith:2001hy,Giudice:2017zke}.\footnote{One can also apply the idea that $\chi_1$ and $\chi_2$ transform under different representations of a dark gauge group with a mass mixing, inspired from Refs.~\cite{Kim:2010gx,Dermisek:2014qca,Dermisek:2015oja,Dermisek:2015hue,Dermisek:2019vkc}}
Of course, one can alternatively consider Higgs portal type~\cite{McDonald:1993ex,Kim:2008pp,Kim:2009ke,Kim:2016csm,Kim:2018uov} or dipole type~\cite{Giudice:2017zke} interactions in replacement of $X^\mu$ and other types of dark sector particles such as scalars and vectors. 
However, the analysis method here still goes through even in those alternative scenarios.

\begin{figure}[t]
\begin{center}
\includegraphics[width=15.cm]{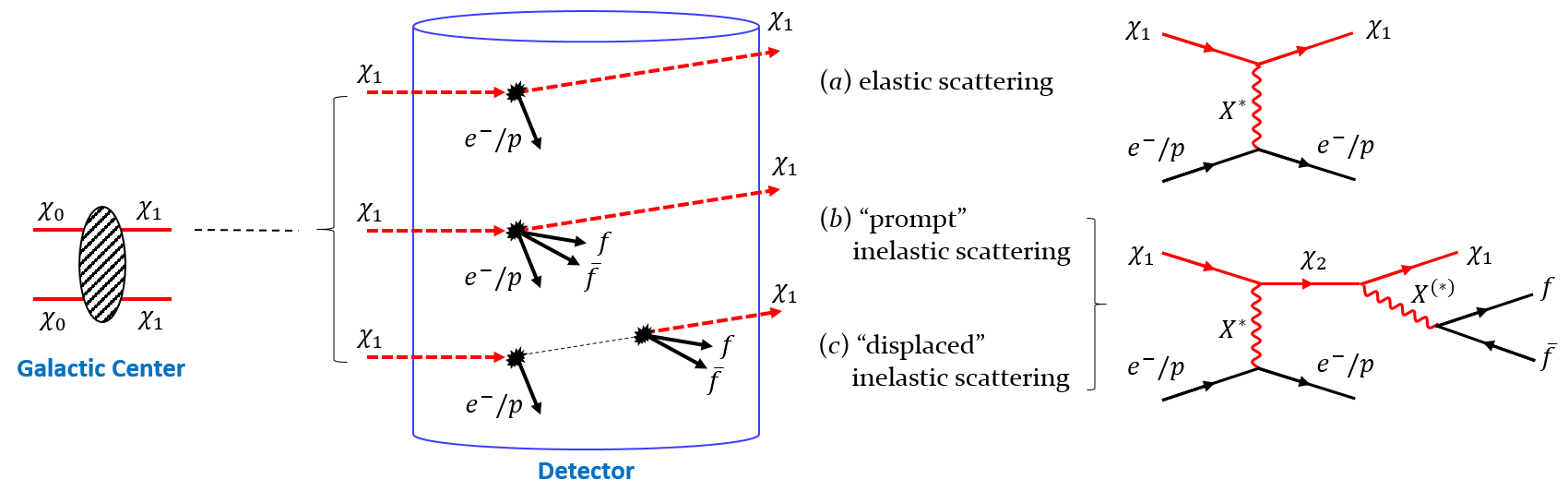}
\caption{A two-component dark matter scenario, where the boosted $\chi_1$ is dominantly produced at the galactic center, and possible experimental signatures based on the interactions in~\eqref{eq:lagrangian} together with relevant Feynman diagrams. 
}
\label{fig:scenario}
\end{center}
\end{figure}

Given the interactions in~\eqref{eq:lagrangian}, three classes of experimental signatures can arise as shown in Figure~\ref{fig:scenario}.
\begin{itemize}
\item Scenario ($a$) depicts nothing but an ordinary elastic scattering process of $\chi_1$ involving an electron or proton recoil. See also the Feynman diagram next to it.
\item Scenario ($b$) sketches a scattering signature where the secondary process (i.e., the decay of $\chi_2$ to $\chi_1$ and $f\bar{f}$) happens ``promptly'' so that a fermion pair $f\bar{f}$ from the decay of the dark gauge boson comes out of the primary vertex, within detector vertex resolution. See also the Feynman diagram next to it.
\item On the other hand, in scenario ($c$), either $X$ or $\chi_2$ is relatively long-lived so that a pair of time-correlated primary and secondary vertices appear displaced, while the same Feynman diagram as for scenario ($b$) is relevant. 
\end{itemize}
Obviously, the first one is relevant to $e$BDM, whereas the other two are $i$BDM-induced. 
Here we denote the decay products of $X$ by generic fermion $f$, but in the detection level some clarification is needed. In particular, if $f$ is either $\tau$ lepton or quark, its appearance can be traced back by its decay products or hadronized objects. We enumerate several well-motivated truth-level final states and briefly discuss their experimental features.
\begin{itemize}
\item $e^+e^-$: Electron (positron) is one of the easily reconstructible objects in various types of detectors including Cherenkov and Liquid Argon Time Projection
Chamber (LArTPC) ones. 
It actively emits electromagnetic radiation while traveling through a detector medium. 
Such radiation manifests as a set of Cherenkov light in Cherenkov detectors and a series of showering in LArTPC detectors.  

\item $\mu^+\mu^-$: Muon is the cleanest object in typical detectors. 
Due to its massiveness relative to electron, electromagnetic radiation is less vigorous. 
So, muons have Cherenkov radiation rings with sharp outer edges unlike electrons having rings with blurred outer edges.
In LArTPC detectors, it does not leave much showering, but ionizes atoms in the vicinity of its trajectory, creating a clean track. 
Furthermore, if it stops inside a detector, it decays and emits a Michel electron resulting in a ``kink'' at the end of the track.\footnote{Since muon is rather long-lived, high-energetic ones do not usually decay inside a detector but leave a kink{\it less} track until escaping from the detector.}

\item $\pi^+\pi^-$: This channel opens up once the mass difference between $\chi_2$ and $\chi_1$ is greater than twice the mass of charged pion. 
A pion behaves similar to a muon as both have the same electric charge and similar mass values. 
Since a pion undergoes nuclear interactions, it typically travels a shorter distance than a muon for a given energy. 
Nevertheless, distinguishing a pion from a muon (or vice versa) is highly nontrivial, so that some non-conventional techniques may be desired.
For example, Ref.~\cite{Acciarri:2016ryt} takes convolutional neural networks to differentiate charged pions from the others, in particular, muons, in a LArTPC detector, and reports $70-75$\% pion tagging efficiency (modulo $\sim20$\% muon contamination). 
\end{itemize}
As the mass gap between $\chi_2$ and $\chi_1$ increases, more variety of modes (e.g., $\pi^+\pi^0\pi^-$, $\pi^+\pi^-\pi^+\pi^-$, etc.) become available. 
In other words, they often accompany multiples of the above bulleted particles, increasing the complexity of particle identification (PID).  

Before closing this subsection, we shall make a few comments for our study with $i$BDM-initiated processes.
First, while the analysis method is completely applicable to any SM charged leptons and quark-induced final states, we take electron and positron for simplicity. 
Kinematically, they just require a small mass difference between $\chi_1$ and $\chi_2$ ($\gsim 1$ MeV); however, the minimum mass gap to open a dimuon channel is 211 MeV. 
A large boost factor of $\chi_1$, $\gamma_1$ is usually demanded to access a $\chi_2$ with such a mass gap~\cite{Kim:2016zjx}. Therefore, lighter $\chi_2$ (if exist) are preferred for a given $\gamma_1$, or a too heavy $\chi_2$ is inaccessible if $\gamma_1$ is not large enough. 
In this sense, the $e^+e^-$ pair is highly motivated over the other signatures. 
Second, all three final state particles may be collimated, if an $i$BDM process is initiated by a $\chi_1$ with a large boost factor. 
Angular resolution of the detector becomes crucial to separate the three particles. 
Depending on detectors, available are some features to identify or tag merged objects; for example, $dE/dx$ in LArTPC detectors~\cite{DeRoeck:2020ntj}.
In our analysis, we consider the issue of angular separation when reporting the results for $i$BDM signals. 
Third, given the experimental signatures shown in Figure~\ref{fig:scenario}, potential backgrounds should be carefully identified and assessed especially for the sensitivity calculation. While $i$BDM suffers far less from background contamination, $e$BDM actually does because of its simple signature. 
Our main focus is comparisons among different signal channels. So, we for the moment pretend to be safe from background issues while referring to e.g., \cite{Kachulis:2017nci,Chatterjee:2018mej,DeRoeck:2020ntj} for more systematic discussions. Finally, even $e$BDM models would give rise to signatures similar to the $i$BDM ones if a secondary process is accompanied via dark gauge gauge boson radiation-off of initial/final-state $\chi_1$ (i.e., dark-strahlung)~\cite{Kim:2019had}. 
$i$BDM search strategies are essentially relevant to the dark-strahlung channel, but the interpretations of experimental results need to be conducted under a proper model hypothesis. 

\subsection{Benchmark detectors \label{sec:benchmarkdetectors}}

\begin{figure}[!ht]
\noindent\rotatebox{-90}{\noindent\begin{minipage}[b][\paperheight][b]{\paperwidth}
    \centering
    \includegraphics[width=\textwidth]{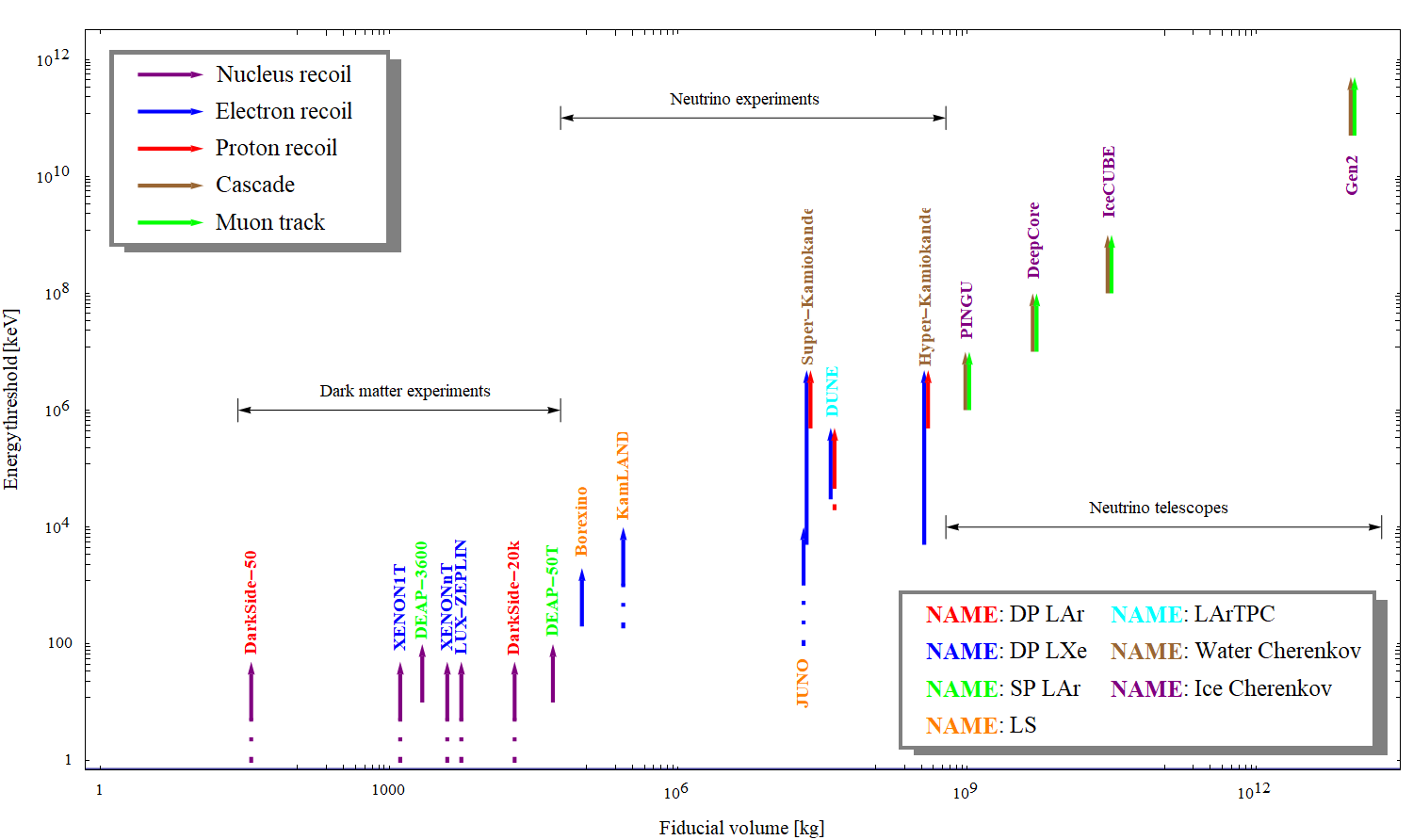}
    \caption{\label{fig:detpic}
    Detectors in various dark matter and neutrino experiments in terms of their fiducial volume and energy threshold.
    The starting points of arrows mark the values of energy threshold, while the dotted lines indicate that the energy thresholds are lying within the dotted-line segments but not specified.
    The lengths of arrows have no meaning.
    [LAr/LXe: liquid argon/xenon, TPC: time projection chamber, DP/SP: dual/single phase, LS: liquid scintillator]
    }     
\end{minipage}}
\end{figure}

Numerous large-volume ($\gsim 1$ ton) experiments are currently in operation or planned, aiming at dark matter and neutrino physics.
In this work, we will choose only several benchmark experiments with various target volumes ranging $\mathcal{O}(10)$ ton -- $\mathcal{O}(10)$ Mton such as DarkSide-20k~\cite{Aalseth:2017fik}, DUNE~\cite{Abi:2018alz, Abi:2018rgm, Abi:2020wmh, Abi:2020evt, Abi:2020loh}, HK~\cite{Abe:2011ts, Abe:2016ero, Abe:2018uyc}, and DeepCore~\cite{Collaboration:2011ym, Aartsen:2016nxy}.
For later use, we present some of the detector specifications of the benchmark experiments.
\begin{itemize}
\item DarkSide-20k~\cite{Aalseth:2017fik}:
 As a unified dark matter direct detection program of the four LAr-based projects (ArDM, DarkSide-50, DEAP-3600, and MiniCLEAN), the DarkSide-20k experiment has been approved due to the successful experience in operating the DarkSide-50 detector.
In the framework of the DarkSide-20k experiment, a dual-phase LArTPC with an active (fiducial) mass of 23 t (20 t) will be deployed at Laboratori Nazionali del Gran Sasso [a depth of $\sim 3800$ meter water equivalent (m.w.e.)] in Italy, where the rate of cosmic rays is reduced to $\sim 1.1\,{\rm m}^{-2}{\rm hr}^{-1}$, in 2021.
The LArTPC, i.e., the active (fiducial) volume, will be an octagonal shape with a height of 2.39 m (2.27 m) and a distance between parallel walls of 2.9 m (2.78 m).\footnote{We assume that each border of the fiducial volume is defined as the same distance inward from the corresponding border of the active volume.}

\item DUNE~\cite{Abi:2018alz, Abi:2018rgm, Abi:2020wmh, Abi:2020evt, Abi:2020loh}:
The far detector of DUNE will consist of four LArTPC modules and be located about 1500 m ($\approx$ 4300 m.w.e.) underground at the Sanford Underground Research Facility in South Dakota, USA, where the expected rate of cosmic rays is $\sim 0.6\,{\rm m}^{-2}{\rm hr}^{-1}$. 
The first (second) module is planned to be ready for operations in 2026 (2027).
Two LArTPC technologies, single-phase (SP) and dual-phase (DP), are planned.
Each module will consist of a cryostat with internal dimensions 15.1 m (width) $\times$ 14.0 m (height) $\times$ 62.0 m (length) which will contain a total (fiducial) LAr mass of about 17.5 kt (at least 10 kt).
These LArTPC detectors will have excellent angular resolution ($\theta_{\rm res}$), good PID capability, and relatively low energy threshold ($E_{\rm th}$), e.g., $\theta_{\rm res} \sim 1^\circ$ and $E_{\rm th} \sim 30$ MeV for an electron.

\item HK~\cite{Abe:2011ts, Abe:2016ero, Abe:2018uyc}:
HK is a next generation underground water Cherenkov experiment based on the very successful operation of SK and will serve as a far detector of T2HK/T2HKK, a long baseline neutrino experiment for the upgraded J-PARC beam.
It will consist of two detectors, each of which will have a cylindrical water tank with 60 m (51.8 m) in height and 74 m (67.8 m) in diameter and holds a total (fiducial) water mass of 258 kt (187 kt).
The first detector will be hosted at Tochibora mine (a depth of 650 m $\approx 1750$ m.w.e.) near the current SK site in Japan with the expected cosmic-ray rate of $\sim 27\,{\rm m}^{-2}{\rm hr}^{-1}$.
The second one is currently considered to be constructed under the Mt. Bisul or Mt. Bohyun with about 1000 m overburden ($\approx$ 2700 m.w.e.) in Korea with the reduced cosmic-ray rate of $\sim 5.7\,{\rm m}^{-2}{\rm hr}^{-1}$.
The operation of the first detector in Japan will be ready in 2027.
For electrons, HK will be able to have quite low energy threshold of $\sim 5$ MeV, but much higher energy threshold is required to achieve good angular resolution, e.g., $E_{\rm th}=100$ MeV for $\theta_{\rm res} \sim 3^\circ$.

\item DeepCore~\cite{Collaboration:2011ym, Aartsen:2016nxy}:
DeepCore is a subarray of IceCube, an ice Cherenkov experiment, that has an approximately five times higher detector module density than that of the original IceCube array.
It has been fully installed between 2100 and 2450 m below the surface of the icecap at the South Pole, where the expected cosmic-ray rate is $\sim 21\,{\rm m}^{-2}{\rm hr}^{-1}$, and taking physics data since May 2010.  
Due to a denser module array, DeepCore can lower the energy threshold by over an order of magnitude ($E_{\rm th}\approx 10$ GeV) than that of IceCube.
The effective target mass of DeepCore varies from $\sim 5$ Mt to $\sim$ 30 Mt as the signal energy does from $\sim 10$ GeV to $\sim$ 100 GeV since it has no clear boundary of target material.
In this study, we take a cylindrical lump of ice  with 350 m in height and 70 m in radius holding a total mass of 5 Mt as a conservative effective target mass with $E_{\rm th}\approx 10$ GeV.
PID is only good for the muon, and $\theta_{\rm res} \sim 1^\circ$ for a muon-track event while $\theta_{\rm res} \gtrsim 10^\circ$ for a shower event.
\end{itemize}
A summary of key characteristics of various relevant detectors including benchmark ones are tabulated in Appendix~\ref{sec:detectors} for convenience of reference.
Here we show their fiducial volumes and energy thresholds in a two-dimensional plane in Figure~\ref{fig:detpic}, as the two are closely related to appropriate detector selection for a given set of the signal flux and the typical energy carried.  
The starting points of arrows mark the values of energy threshold, while the dotted lines indicate that the energy thresholds lie within the dotted-line segments but the exact values are not specified. 
No particular meanings are associated with the lengths of arrows, and target materials are categorized by color-coding the name of experiments. 

\section{Signal Cross Sections \label{sec:xs}}

The cross section formulas of the primary scattering of $\chi_1$ with a target $T$ -- which are categorized as the quasi-elastic $e$-scattering, $p$-scattering, and DIS by the types of the primary signature -- for the benchmark model are summarized in Appendix~\ref{sec:derivation}.
In this section, we compare these primary cross sections in a wide range of parameter space.
Then, we take the detector effects into account and expect the preferred primary signatures in the reference experiments for any given parameter set within the range of our consideration.

\subsection{$p$-scattering vs. DIS}
\label{sec:pvsDIS}

We first contrast the (quasi-elastic) $p$-scattering with DIS in this scenario.
In our terminology (as in much of BDM literature), the $p$-scattering means that the incoming light dark matter $\chi_1$ scatters off proton elastically,  $\chi_1 p \to \chi_1 p$ ($e$BDM), or inelastically, $\chi_1 p \to \chi_2 p$ ({\it i}BDM), in which the proton does not break apart, hence not accompany additional hadronized objects.
In reality, it is probabilistic, competing with other scattering channels including DIS, and can be given by a function of the modulus of the spatial momentum of the recoiling proton, $p_p \equiv |\vec p_p|$.
In this paper, we assume that it is a step function around $\spp = 2$ GeV for simplicity, i.e., the events with $\spp < 2$ GeV is categorized as $p$-scattering. 
Here the value $p_p = 2$ GeV is the boundary value for which the probability of producing a pion or a charged secondary in the water is 50\% based on the Monte Carlo study by the SK Collaboration~\cite{Fechner:2009aa}.\footnote{Note that the energy threshold values are written in $p_p$ in SK and HK, while they are kinetic energies, i.e., $\sqrt{m_p^2 + p_p^2} - m_p$ in the other reference experiments.} 
As we will argue shortly, the differential cross section in $\spp$ peaks around $\spp \ll 1$ GeV in the parameter space of interest (see also for example, FIG.~2 of Ref.~\cite{Kim:2016zjx}), and thus the precise $\spp$ value differentiating the DIS and $p$-scattering regimes does not significantly affect our analysis results. 
In addition, we require energy threshold $E_{\rm th}$ for any recoil proton, whereas no corresponding cuts are imposed on visible particles in the DIS process as it may involve unnecessary complication. 
Indeed, as we shall show, the $p$-scattering is more important than the DIS in most of parameter space of interest. 
Therefore, this seemingly ``unfair'' treatment on the $p$-scattering nevertheless does not affect our final conclusion.

\begin{figure}
\begin{center}
\includegraphics[width=.325\textwidth]{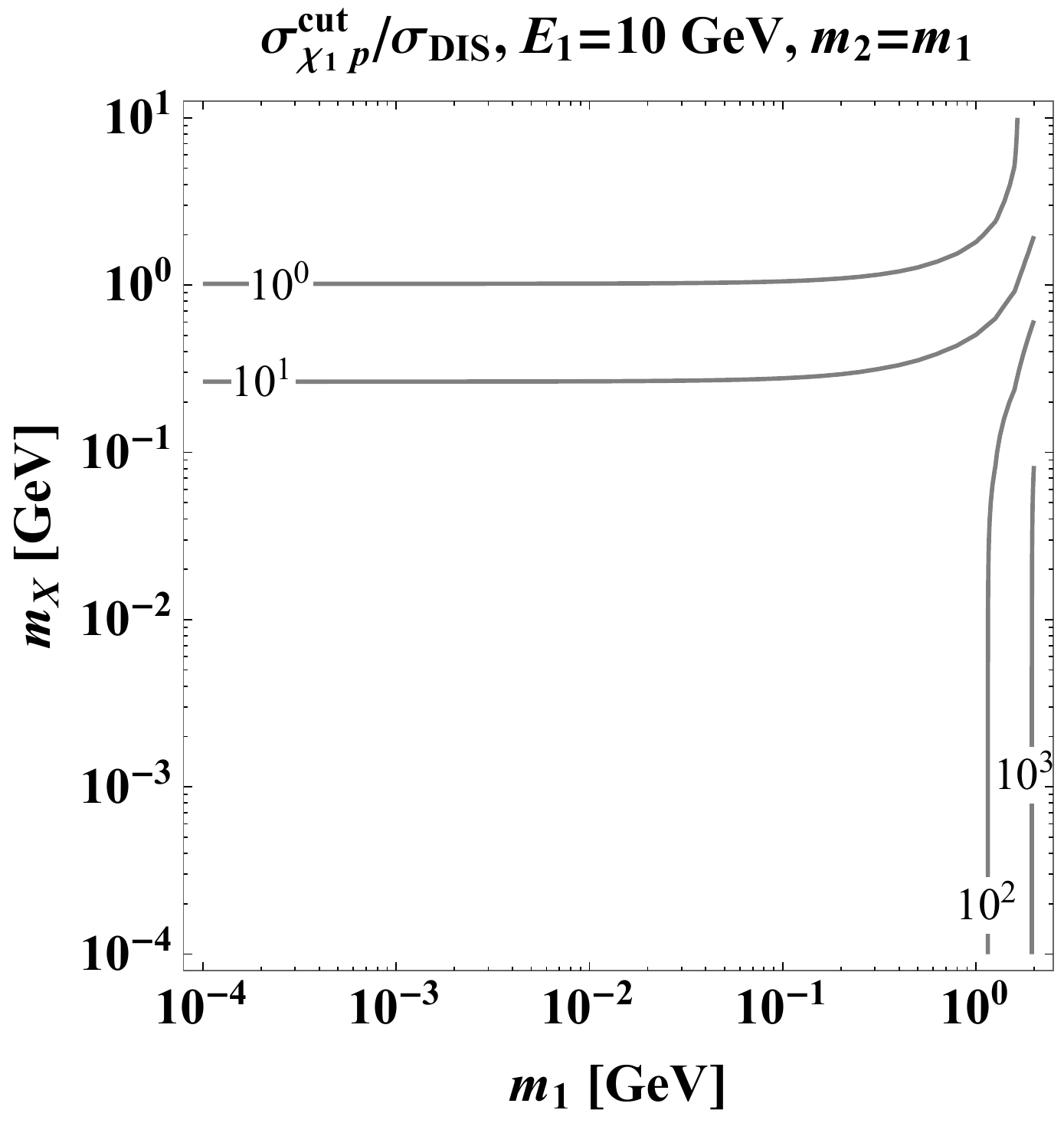}
\includegraphics[width=.325\textwidth]{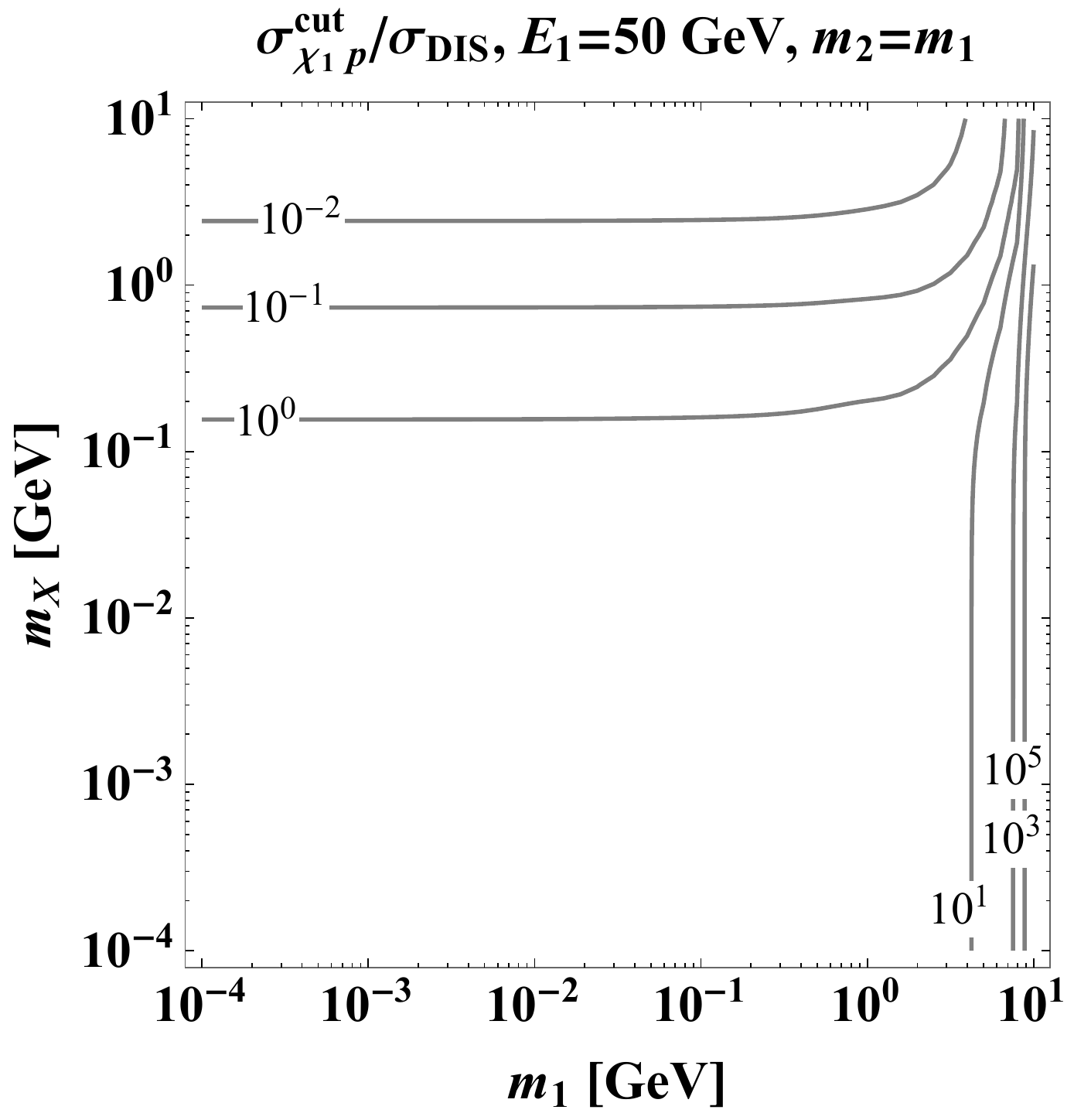}
\includegraphics[width=.325\textwidth]{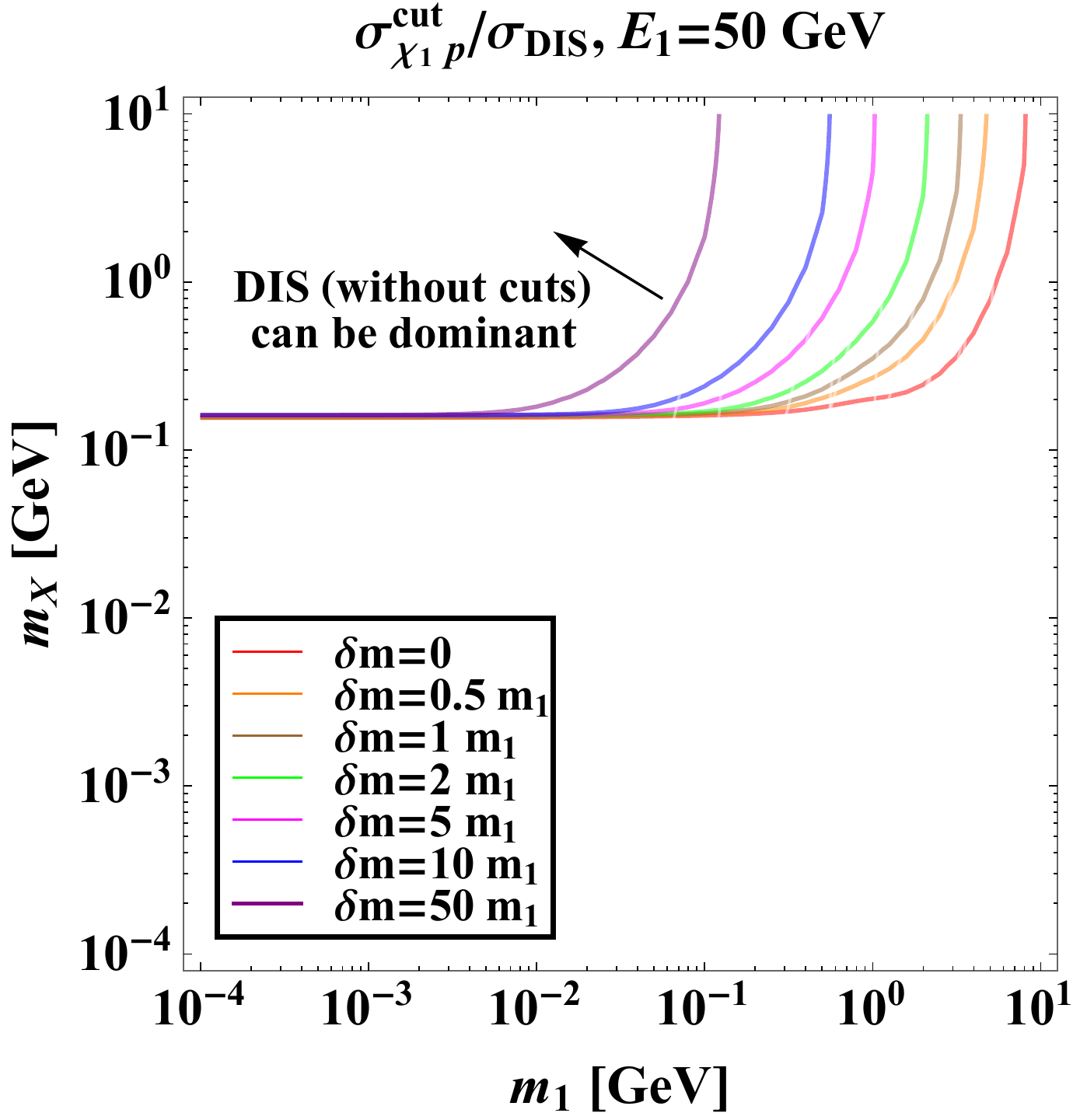}
\includegraphics[width=.325\textwidth]{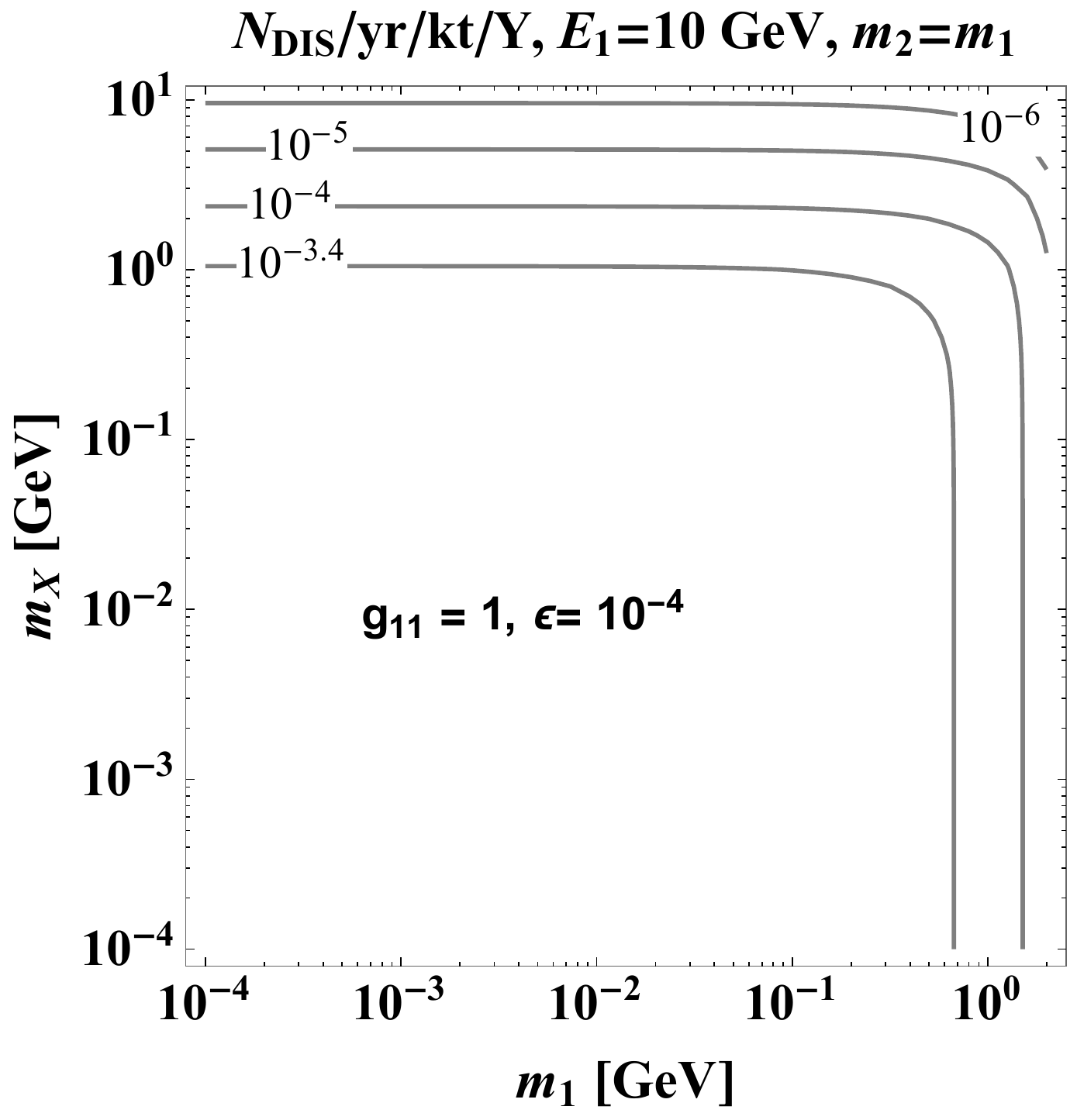}
\includegraphics[width=.325\textwidth]{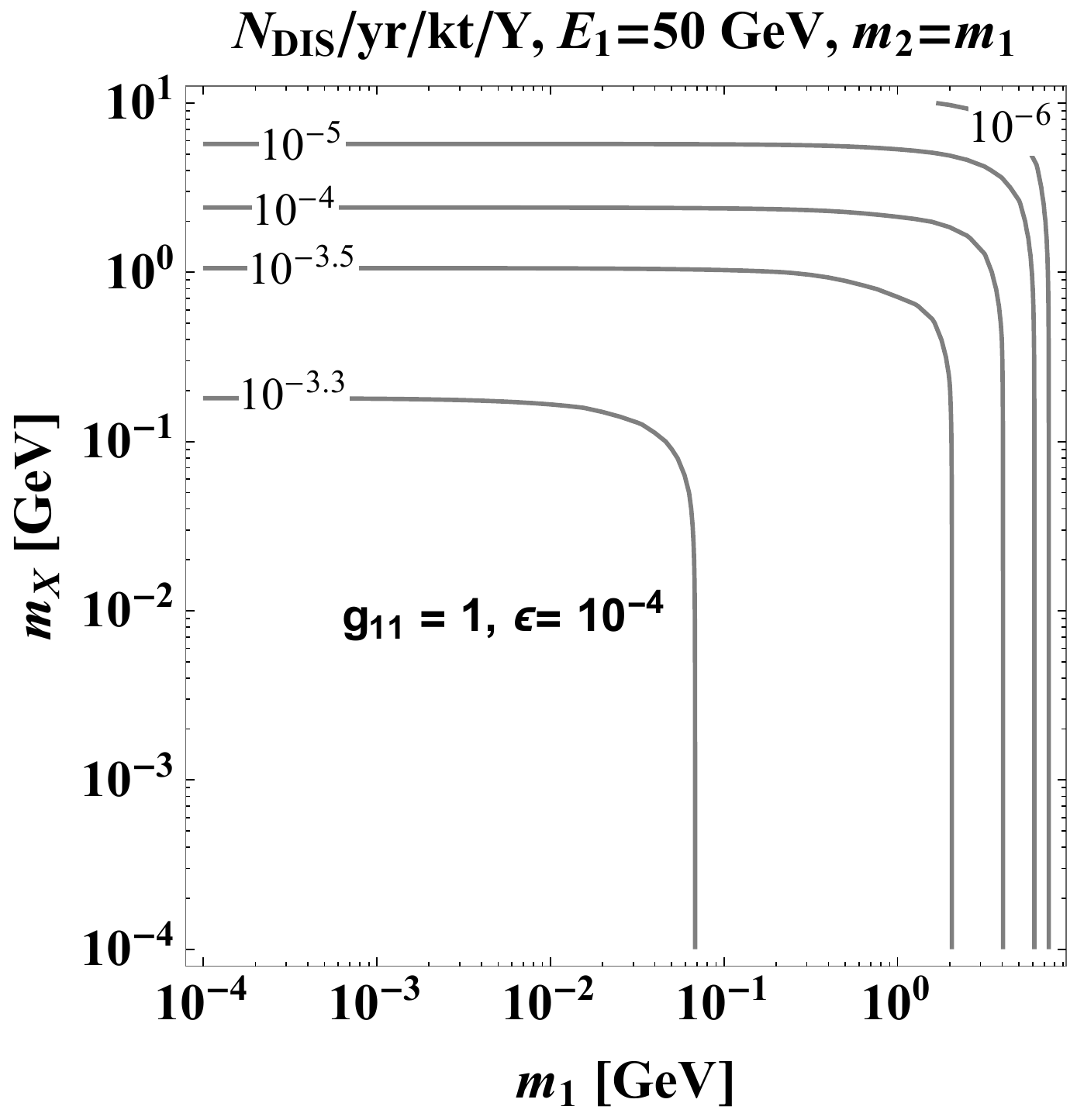}
\includegraphics[width=.325\textwidth]{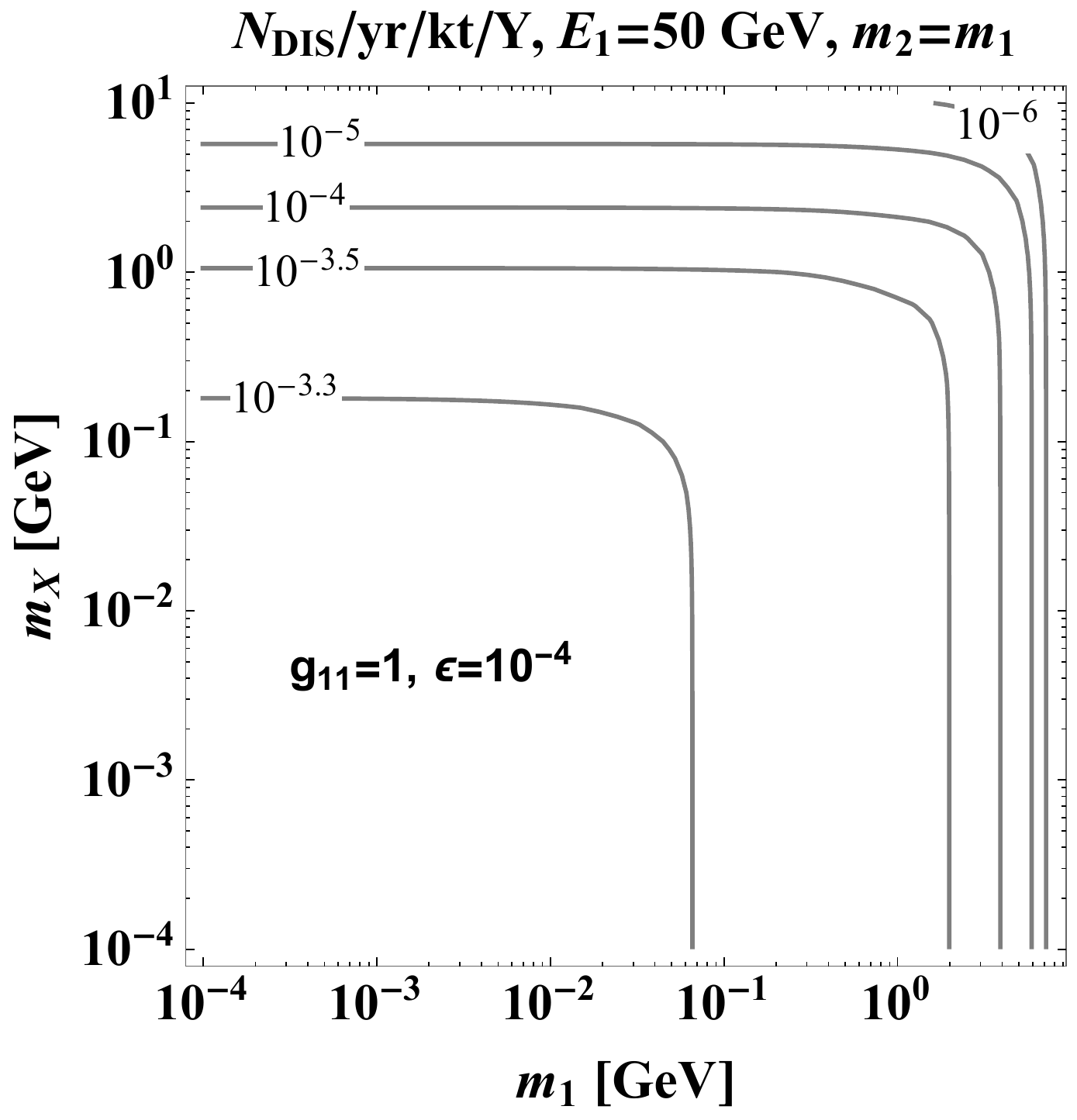}
\caption{Upper panels: Contours of $\sigma_{\chi_1 p}^{\rm cut}/\sigma_{\rm DIS}$ expected at a ``DUNE-like'' detector. The first two plots show the results for $e$BDM (i.e., $m_2=m_1$) with $E_1$ being 10 GeV (leftmost) and 50 GeV (middle). The rightmost panel displays the $\sigma_{\chi_1 p}^{\rm cut}/\sigma_{\rm DIS}=1$ lines corresponding to various $\delta m$ values, with $E_1$ being 50 GeV.
Lower panels: The expected number of DIS events. The first two plots are for $e$BDM (i.e., $m_2=m_1$) with $E_1$ being 10 GeV (leftmost) and 50 GeV (middle), while the last one is for $i$BDM with $m_2=1.5m_1$ and $E_1=50$ GeV.
}
\label{fig:dis}
\end{center}
\end{figure}

In the first two panels of the upper row of Figure~\ref{fig:dis}, we show the contours of $\sigma_{\chi_1 p}^{\rm cut} / \sigma_{\rm DIS}$ in the $m_1 - m_X$ plane for $E_1 (= m_0) = 10$ GeV (upper-left) and 50 GeV (upper-middle) fixing $m_2 = m_1$ (i.e., $e$BDM).
Here $\sigma_{\chi_1 p}^{\rm cut}$ denotes the primary $p$-scattering cross section with the energy threshold cut imposed.
We specifically require the recoil kinetic energy of the proton to be above 21 MeV~\cite{Necib:2016aez,Acciarri:2014gev} which would be adopted in a ``DUNE-like'' detector in the next section. 
In calculating the DIS scattering cross section, $\sigma_{\rm DIS}$, we take the MSTW2008NNLO parton distribution functions~\cite{Martin:2009iq} and require the energy transfer $Q > 1.5$ GeV.
We confirm that our $\sigma_{\rm DIS}$ reproduces the shape of the differential cross section of neutrino-induced DIS in Ref.~\cite{Formaggio:2013kya} after replacing the dark gauge boson by a $W$ boson.
For the neutrino scattering, it is well known that the DIS dominates over the quasi-elastic scattering (and $\Sigma$ resonance) once the energy of incoming neutrino is large enough, $E_\nu > \mathcal{O}(10)$ GeV~\cite{Formaggio:2013kya}. 
On the other hand, we find that the $p$-scattering cross section in BDM scenarios dominates over the DIS cross section even when the energy of incident $\chi_1$ is as large as 50 GeV (and up to 100 GeV which is not shown here), as long as the mediating particle (i.e., the dark gauge boson in our study) is lighter than $\mathcal O (1)$ GeV.
Regarding the fact that we imposed a ``penalty'' to $p$-scattering cross section by an $E_{\rm th}$ cut, our statement is rather robust.
In the upper-right panel of Figure~\ref{fig:dis}, we display the results corresponding to non-zero $\delta m$, i.e., {\it i}BDM processes.
Contours are defined by $\sigma_{\chi_1 p}^{\rm cut} / \sigma_{\rm DIS} = 1$ 
with $E_1 = 50$ GeV.
The overall behavior is similar to the cases with $\delta m = 0$. However, the regime where the $p$-scattering is in favor widens as $\delta m/m_1$ increases, because an up-scattering process is easier with proton itself than with a parton inside the proton.

We can semi-analytically understand these behaviors of $p$ scattering and DIS in this regime in the following manner.
First of all, the differential $p$-scattering cross section in proton recoil momentum is peaking toward small $\spp (\ll m_p)$ due to the $t$-channel exchange of $X$ (see Appendix~\ref{sec:derivation} for details).
From Eqs.~\eqref{eq:matrixX} through~\eqref{eq:ETspec}, we obtain
\begin{align}
\frac{d\sigma_{\chi_1 p}}{d\spp} \propto \frac{1}{\{2 m_p (E_2 - E_1) - m_X^2\}^2} \simeq \frac{1}{(\spp^2 + m_X^2)^2}\,,
\label{eq:psigma}
\end{align}
in the limit of $\spp \ll m_p$.
Here we omit the overall factor from the phase space integral and terms in the numerator of the relevant matrix element. 
This relation implies that the $p$-scattering cross section rises in decreasing $m_X (\ll m_p \approx 1 \hbox{ GeV})$ where the quartic dependence is kept as long as $\spp \lesssim m_X$.
By contrast, the energy transfer $Q$ in the DIS should be larger than $\sim 1.5$ GeV, and in turn, much larger than $m_X$ under consideration. 
The DIS differential cross section is given by
\bea 
\frac{d^2\sigma_{\rm DIS}}{dx dy} \propto \frac{1}{(Q^2+m_X^2)^2}\approx \frac{1}{Q^4}
\eea
for $m_X \ll 1$ GeV (see also Appendix~\ref{sec:derivationDIS}). 
This implies that the DIS cross section does not vary much for $m_X \ll 1$ GeV. 
Our numerical study further suggests that $\sigma_{\rm DIS}$ be less than $\sigma_{\chi_1 p}^{\rm cut}$ for $m_X \approx 0.1$ GeV and $E_1 \lesssim 50$ GeV, and therefore, the ratio of $\sigma_{\chi_1 p}^{\rm cut}$ to $\sigma_{\rm DIS}$ should be greater than 1 for $m_X \ll 1$ GeV. 

Moving onto the lower panels of Figure~\ref{fig:dis}, we now show the contours of the theory-level number of DIS-induced events $N_{\rm DIS}$ per year$\cdot$kt$\cdot$Y with Y defined by (atomic number)/(atomic weight).
We calculate $N_{\rm DIS}$ as follows:
\bea 
N_{\rm DIS} = \sum_{i=p,n}N_i\cdot \sigma_{\rm DIS}^i\cdot \mathcal{F}_1 \cdot t_{\rm exp},
\eea
where $\mathcal{F}_1$ is the $\chi_1$ flux coming from the GC, given in Eq.~\eqref{eq:fluxformula}, $N_i$ is the number of protons or neutrons inside a fiducial volume of a detector, and $t_{\rm exp}$ is the amount of time exposure. 
The leftmost and middle panels show results for {\it e}BDM case with $E_1 = 10$ GeV and 50 GeV, respectively.
The dark-sector coupling $g_{11}$ is assumed unity, and the kinetic mixing parameter $\epsilon$ is set to be $10^{-4}$ for illustration.
By contrast, in the rightmost panel, we set $m_2 = 1.5\, m_1$ together with $g_{12} = 1$ and $\epsilon = 10^{-4}$ for $E_1 = 50$ GeV.
For simplicity, the contours are shown only in the parameter region that $\sigma_{\chi_1 p}^{\rm cut}/\sigma_{\rm DIS} \lesssim 1$, i.e., where $\sigma_{\rm DIS}$ can be at least comparable or larger.
Nevertheless, we observe that $N_{\rm DIS}$ does not change much when $m_X \lesssim 1$ GeV and decreases in increasing $m_X$ and $m_1$ in the entire parameter space.
The former two results are consistent with our argument that the differential cross section $d \sigma_{\rm DIS} / dx dy \propto 1/(Q^2 + m_X^2)^2$.
The last result (dependence on $m_1$) is due to the fact that the incoming $\chi_1$ actually scatters off a parton instead of a proton for the DIS case.

In case of HK which uses water as the target material, i.e., $Y = 10/18$,  with the fiducial volume of 380~kt, the contour $N_{\rm DIS}\cdot{\rm yr}^{-1}{\rm kt}^{-1}{\rm Y}^{-1} = 10^{-3.3}$ corresponds to $0.1 \,{\rm yr}^{-1}$.
Since this is the expected number without considering any detector effects such as energy threshold, angular and position resolutions, the actual $N_{\rm DIS}$ would be negligible even after running the HK for more than 10 years.
Consequently, it is generically hard to observe DIS-induced signal events.\footnote{For some parameter space where $m_X > 100$ MeV, $\epsilon \sim 10^{-3}$ is still allowed. 
This can result in increasing $N_{\rm DIS}$ by a factor of tens, so there might be a narrow window that $N_{\rm DIS} =$ a few per year in HK. 
However, we expect that the actual $N_{\rm DIS}$ would be still negligible, applying the detector resolutions.}
Reminding that the {\it i}BDM cross section is further suppressed for DIS, as stated previously, the aforementioned conclusion applies both for {\it e}BDM and {\it i}BDM signals.
We remark that this can be avoided if a much larger detector is simply introduced or the boosted $\chi_1$ flux is enhanced by the source close to the earth and/or dark matter (self-)interactions, e.g., solar-capture BDM scenarios~\cite{Berger:2014sqa,Kong:2014mia,Alhazmi:2016qcs,Berger:2019ttc}.

In summary, we find that the $p$-scattering is larger than the DIS as far as the mediator $X$ is lighter than $\sim 1$ GeV even in the case of $E_1\approx 50$ GeV.
Furthermore, we anticipate that it is unlikely to observe DIS-initiated signal events in the near future belonging to the parameter regime where the DIS cross section dominates over the $p$-scattering cross section, modulo the $\chi_1$ flux given by Eq.~\eqref{eq:baselineflux}.
These series of observations are truly contrasted to neutrino-initiated events. 
The main difference stems from the fact that the weak gauge bosons $W$ and $Z$ involved in neutrino scattering processes are much heavier than typical dark gauge boson considered in this work. 
In this context, one may wonder whether or not examining the DIS channel is motivated essentially to probe the model space where $m_X$ is sizable (say, a few tens of GeV).
In the vector portal scenario, an $\epsilon^2$ suppression is unavoidable in any scattering channels, on top of the suppression by $\sim 1/m_X^4$. 
Therefore, the signal sensitivity relevant to the DIS channel, in general, quickly drops as $\epsilon$ reduces. 
Again this is clearly supported by our exercise shown in the lower panels of Figure~\ref{fig:dis}.
We henceforth focus on the (quasi-elastic) $p$-scattering and $e$-scattering processes throughout the rest of this paper unless specified otherwise.

\subsection{$e$-scattering vs. $p$-scattering: basic considerations}
\label{sec:comparison}

Given our observation in the previous section that the contribution from the boosted $\chi_1$-induced DIS is negligible as far as a light mediator (much lighter than say, $W$ boson mass) is concerned, we can now compare the $e$-scattering (i.e., involving an electron recoil) with $p$-scattering (i.e., involving a proton recoil). 
The beginning discussion is devoted to pure theoretical results which would have been obtained with ``perfect'' detectors; that is, we include neither characteristics of a detector such as resolutions and energy threshold nor nuclear effects. 
Although this setup is unrealistic, the relevant exercise allows us to develop intuitions and paves a road to truth-level understanding.
Once done with the theoretical investigation, we discuss, in the next section, how the preferred channels are affected by the inclusion of detector effects such as energy threshold, angular resolution, and the acceptance of the secondary signatures. 
Elastic BDM processes require the first effect as they accompany a target recoil only, whereas inelastic BDM ones are affected by all three.
To demonstrate the difference according to detector types, we consider two representative ones, a DUNE-like LArTPC detector and a HK-like Cherenkov light detector.

\begin{figure}[t]
\centering 
\includegraphics[width=7.2cm]{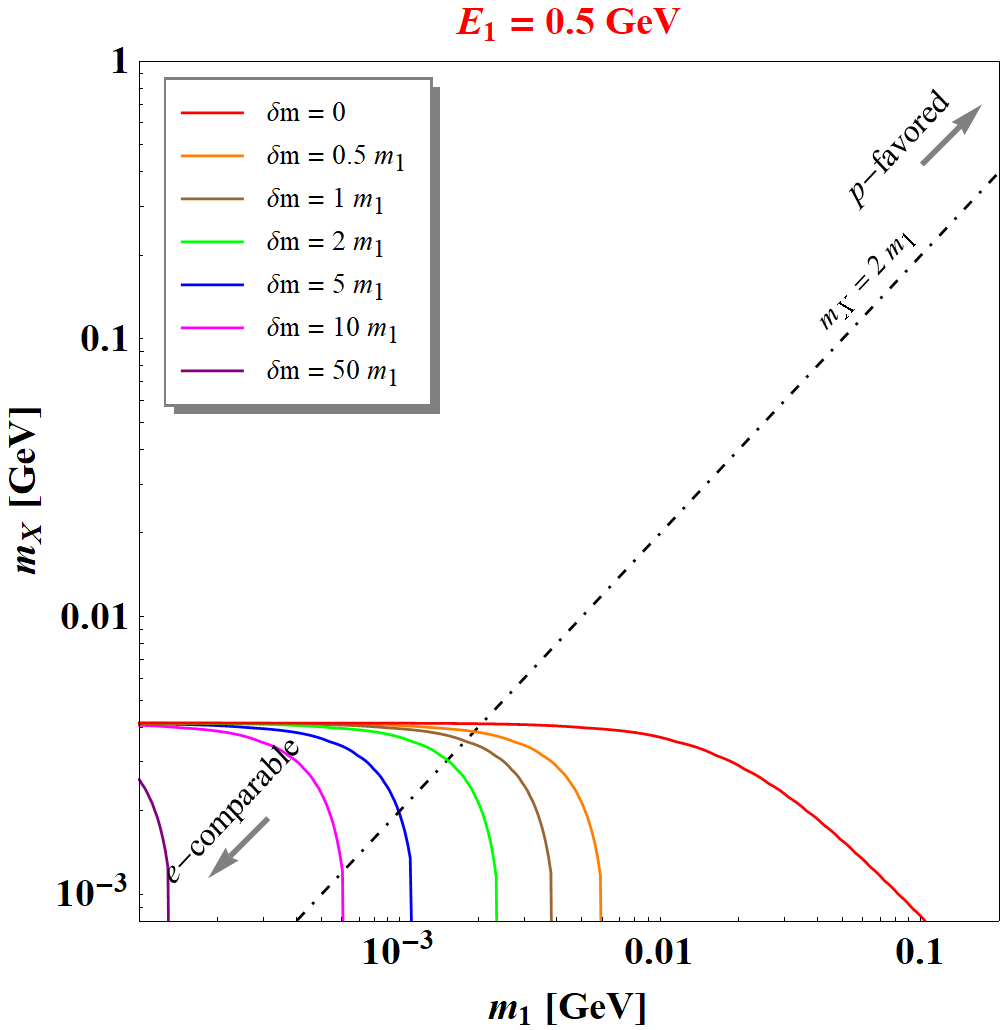} \hspace{0.2cm}
\includegraphics[width=7.2cm]{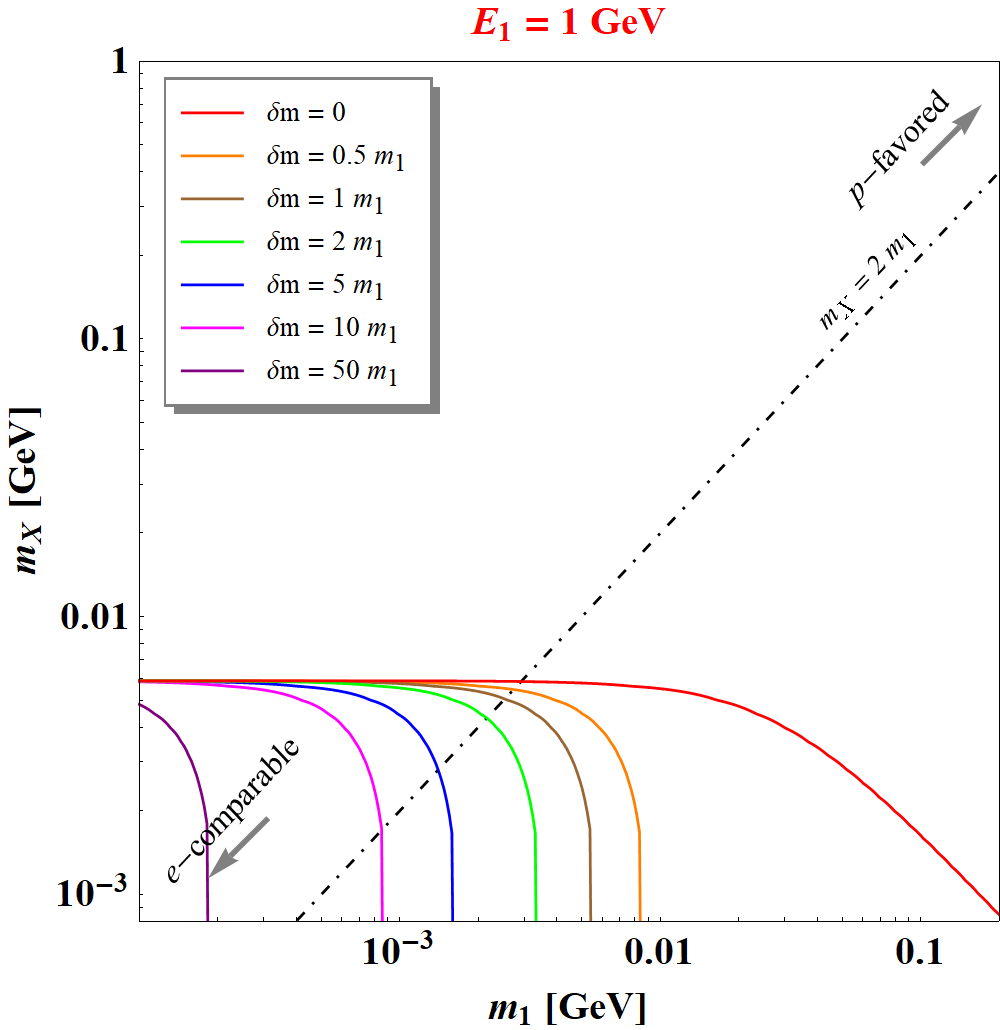} \\
\vspace{0.2cm}
\includegraphics[width=7.2cm]{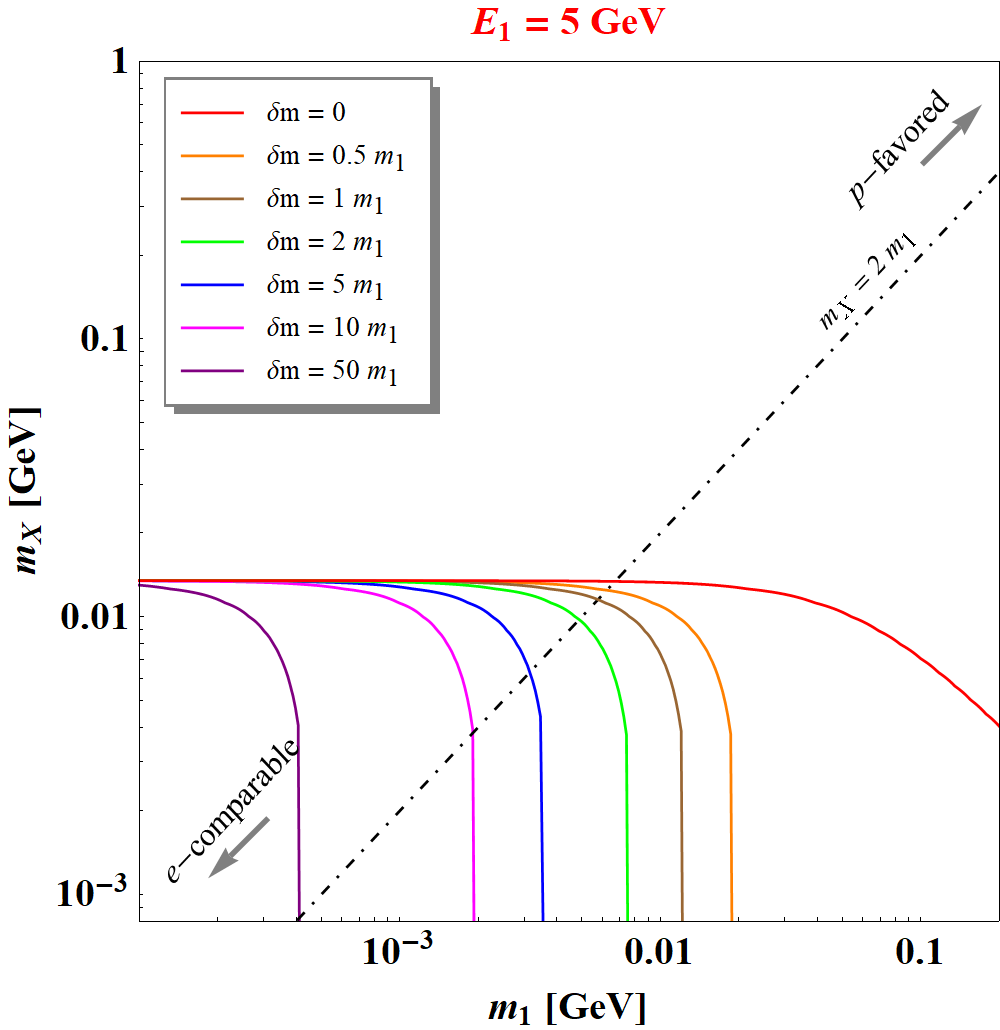}\hspace{0.2cm}
\includegraphics[width=7.2cm]{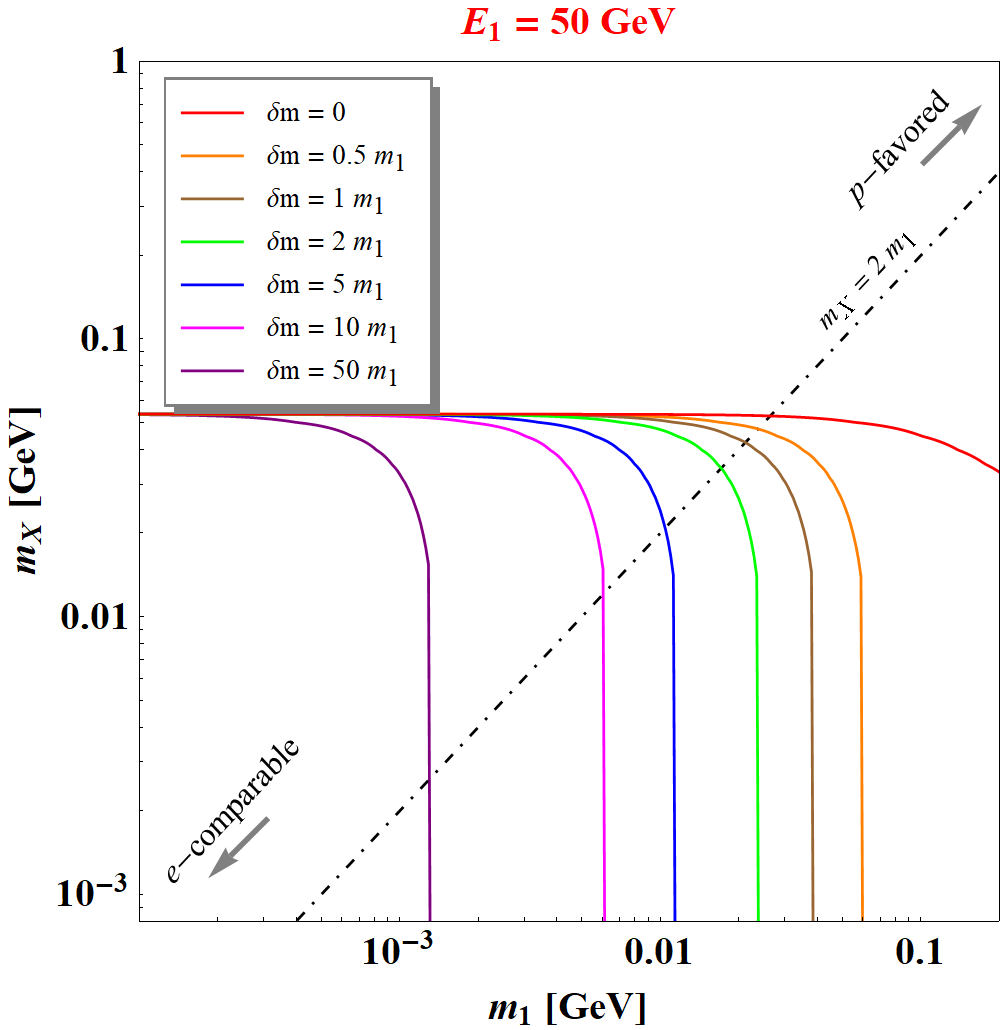}
\caption{\label{fig:compNoCuts} Comparison of $e$-scattering and $p$-scattering cross sections for $E_1=0.5$ GeV (upper-left), $E_1=1$ GeV (upper-right), $E_1=5$ GeV (lower-left), and $E_1=50$ GeV (lower-right) with a ``perfect'' detector where all visible final state particles are correctly tagged and reconstructed. 
For the proton channel, the phase space within $p_p=2$ GeV is calculated, if phase space is allowed beyond it. 
In most of parameter space, the proton scattering is showing a greater cross section than the electron one is, so the contours are drawn along $\sigma_{\chi_1 e}=0.9\sigma_{\chi_1 p}$. 
The proton (electron) channel becomes advantageous (comparable) as $m_X$ and $m_1$ increase (decrease). 
The black dot-dashed lines are defined by $m_X = 2m_1$. 
See the text for more detailed discussions.
}
\label{fig:xscompth}
\end{figure}

We show our findings with respect to $E_1$, the incoming energy of $\chi_1$; again, results will be relevant for the same $E_1$ value, irrespective of the production mechanism of boosted $\chi_1$. 
Figure~\ref{fig:compNoCuts} demonstrates comparisons between $e$-scattering cross section and $p$-scattering cross section for four example $E_1$ (or equivalently the mass of $\chi_0$ in the annihilating two-component dark matter scenario) values: $E_1 =0.5$, 1, 5, and 50 GeV in the upper-left, upper-right, lower-left, and lower-right panels, respectively. 
Results are shown in the plane of the $\chi_1$ mass, $m_1$, vs. the dark gauge boson mass, $m_X$, with seven different mass gaps from $\delta m = 0$ (red) to $\delta m = 50 m_1$ (purple). Note that the case with a vanishing mass gap corresponds to the elastic BDM scenario. 
We essentially scan over the two-dimensional $m_1-m_X$ parameter space, and at each scan point we compute the total scattering cross section within allowed phase space.\footnote{We collect relevant formulas in Appendix~\ref{sec:derivation}.}
For the proton channel, we consider it up to recoil proton spatial momentum being 2 GeV beyond which the proton may mainly break apart and give rise to a DIS, as stated previously.
While performing the scan, we observe that the proton scattering cross section is always (at least, slightly) greater than the one corresponding to electron in the scanning ranges of interest, unless $m_X \sim$ a few keV.
So, we choose to divide the region to the $p$-preferred and the $e$-comparable along the boundary defined by $\sigma_{\chi_1 e} = 0.9\, \sigma_{\chi_1 p}$.
Above (below) the boundary, the $p$-scattering ($e$-scattering) channel comes with more (comparable) number of signal events.  

Together with the boundary curves, we add black dot-dashed diagonal lines to mark the border defined by $m_X=2m_1$.
Above the line an {\it on}-shell $X$ decays invisibly into dark matter pair $\chi_1 \bar{\chi_1}$, whereas below the line $X$ is allowed to decay to visible SM particles 
assuming that no lighter dark-sector particle exists. We henceforth call the former and the latter scenarios I and II.
In both scenarios, the mass spectrum of the dark sector will dictate how and under which circumstances the visible decays of $\chi_2$ can take place.
We list all possibilities that give rise to three visible particles below and provide a summary in Table~\ref{tab:scenarios}.
\begin{itemize}
\item Scenario I-i: The mass spectrum satisfies $2 m_1 < m_X$, $\delta m < m_X$,  and $m_2 \le 3m_1$.
Hence, $\chi_2$ can decay visibly via an {\it off}-shell $X$ exchange.
\item Scenario I-ii: The mass spectrum satisfies $2 m_1 < m_X$ and $\delta m < m_X$, but $3m_1 < m_2$. 
If $g_{11}$ is suppressed or vanishes (i.e., a model maximizes the off-diagonal gauge interaction $\bar \chi_2 \gamma^\mu \chi_1 X_\mu$ as in the examples in Ref.~\cite{Giudice:2017zke}), the channel of $\chi_2 \to 3\chi_1$ is not allowed, hence a visible decay is possible. 
\item Scenario II-i: The mass spectrum satisfies $m_X \le 2 m_1$ and $m_X<\delta m $  (or $\le m_1 + m_2$ if $g_{11} = 0$). 
$\chi_2$ emits an on-shell $X$ as a decay product, and the $X$ decays visibly.
\item Scenario II-ii: The mass spectrum satisfies $\delta m < m_X\le 2 m_1$ (or $\le m_1 + m_2$ if $g_{11} = 0$). 
$\chi_2$ decays visibly via a three-body process just like Scenarios I-i and I-ii.
\end{itemize}

\begin{table}[t]
    \centering
    \begin{tabular}{c c c c c}
         &
         \begin{minipage}{.175\textwidth}
            \includegraphics[width=\linewidth]{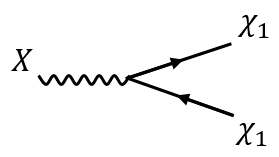}
         \end{minipage}
         &  
         \begin{minipage}{.175\textwidth}
            \includegraphics[width=\linewidth]{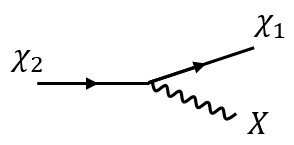}
         \end{minipage}
         &  
         \begin{minipage}{.175\textwidth}
            \includegraphics[width=\linewidth]{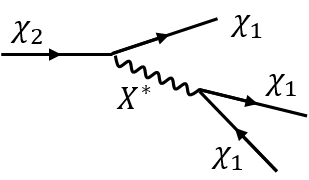}
         \end{minipage}
         &
         \\
         \hline \hline
    Scenario & \multicolumn{3}{ |c| }{Allowed?} & Comment \\
    \hline 
    \multicolumn{1}{c|}{I-i} & \cmark & \xmark & \xmark & \multicolumn{1}{|l}{visible 3-body decay}  \\
    \multicolumn{1}{c|}{I-ii} & \cmark & \xmark & \cmark & \multicolumn{1}{|l}{needs suppressed $g_{11}$} \\
    \multicolumn{1}{c|}{II-i} & \xmark & \cmark &  {\color{gray} {\Huge $\boldsymbol{\minus}$}} & \multicolumn{1}{|l}{visible 2-body decay} \\
    \multicolumn{1}{c|}{II-ii} & \xmark & \xmark &  {\color{gray} {\Huge $\boldsymbol{\minus}$}} & \multicolumn{1}{|l}{visible 3-body decay} \\
    \hline \hline
    \end{tabular}
    \caption{
    Different scenarios depending on the dark gauge boson decay being visible or not.
    We show which processes (Feynman diagrams) are allowed (check mark) or not allowed (cross mark) due to  the dark sector mass spectrum in each scenario. Gray horizontal lines mean that the phenomenology is relatively independent of the respective process. The column on the right provides a comment on  possible ways of having a visible signature in each scenario.
    }
    \label{tab:scenarios}
\end{table}

Looking into plots in Figure~\ref{fig:compNoCuts}, first of all, we see the trend that the region, where the $e$-scattering channel stays competitive, expands as $\delta m$ becomes smaller (i.e., less inelastic) and incident $\chi_1$ comes with more energy. 
The revealed dependence on the former is not surprising because smashing a (lighter) $\chi_1$ on a (heavier) proton target is much more advantageous in transiting to a much heavier $\chi_2$ state, considering the maximally allowed $m_2$ for a given pair of $E_1$ and $m_T$ ($T=e, p$). 
The relation is given by 
\bea 
m_2 \leq \sqrt{m_T^2+2E_1 m_T+m_1^2}-m_T\,, \label{eq:m2limit}
\eea
which can be approximated to
\begin{eqnarray}
    m_2 &\lsim & m_1 +\frac{m_T}{m_1} E_1  \hspace{0.5cm} \hbox{ for } m_1 \gg m_T\,, \label{eq:etarget} \\
    m_2 &\lsim & E_1 \hspace{2.2cm} \hbox{ for } m_1 \ll m_T\,. \label{eq:ptarget}
\end{eqnarray}
The relation in \eqref{eq:etarget} corresponds to the electron target case (i.e., $m_T = m_e$), whereas that in \eqref{eq:ptarget} corresponds to the proton target case (i.e., $m_T=m_p$). 
These formulas for the two limiting cases clearly support the fact that for a given $E_1$ the proton target is better to access much larger values of $m_2$ than the electron target. 

Now defining $\delta m = r\, m_1$, taking the electron target, and solving the above inequality \eqref{eq:m2limit}, we obtain the below inequality for $m_1$ in term of $E_1$ and $r$:
\bea 
m_1 \leq \frac{-(r+1)m_e+\sqrt{(r+1)^2m_e^2+2r(r+2)E_1m_e} }{r(r+2)}\, \stackrel{E_1 \gg m_e}{\longrightarrow}\, \sqrt{\frac{2E_1m_e}{r(r+2)}}\,. \label{eq:bar}
\eea
This implies that vertical drops take place below the saturation point defined in the right-hand side (henceforth called kinematic ``barrier'').
In reality, as $E_1$ increases, the actual drop happens close to this kinematic ``barrier''.
When it comes to the case of larger $E_1$, the value in~\eqref{eq:bar} increases so that the $e$-comparable area extends toward higher $m_1$. 
There is another barrier in the $m_X$ direction mainly due to the faster drop of $\sigma_{\chi_1 e}$ in increasing $m_X$, compared to $\sigma_{\chi_1 p}$.
Indeed, our numerical study finds that the $p$-scattering cross section is roughly constant or independent of $E_1$ within the parameter space of interest, whereas the $e$-scattering one is enhanced in rising $E_1$. Therefore, the ``$e$-comparable'' area is widened along the direction of $m_X$ in growing $E_1$.

We now close this section, summarizing the main observations as follows (see Table~\ref{tab:summary-channels}).
\begin{itemize}
\item If a BDM search hypothesizes a heavy dark gauge boson (but still not much above the GeV scale),
the proton scattering channel expedites discovery.
\item If a model conceiving $i$BDM signals allow for large mass gaps between $\chi_1$ and $\chi_2$, the proton channel is more advantageous.
\item On the other hand, the electron channel becomes comparable/complementary in probing the parameter regions with smaller $m_1$ and $m_X$. 
\item As the boosted $\chi_1$ comes with more energy, more parameter space where the $e$-scattering is comparable opens up.
\end{itemize}
\begin{table}[t]
     \begin{center}
     \begin{tabular}{l|l}
     \hline \hline
     Channel & Advantageous scenarios  \\ 
     \hline
     $p$-scattering~~ & Heavier dark matter \& mediator masses ~\\
     & Large dark-matter particles mass gap\\ 
     \hline
     $e$-scattering & Smaller dark matter \& mediator masses \\
     & Small dark-matter particles mass gap \\ & Larger boosts of dark matter\\
     \hline \hline
    \end{tabular}
    \end{center}
    \caption{Summary of the  advantages of proton and electron scattering scenarios in probing inelastic boosted dark matter models.}
    \label{tab:summary-channels}
\end{table}
Again we emphasize that even if these observations are predicated upon simplified calculations, they are not too far away from the reality.
As we will find out in the next section, many of generic trends are retained even in the presence of various realistic effects. Furthermore, depending on detector designs, some effects may be negligible; for example, energy threshold could be extremely small, highly collimated objects could be resolved, and so on. 
We will come back to this issue and make a brief discussion in Section~\ref{sec:pheno}.

\subsection{$e$-scattering vs. $p$-scattering with realistic effects}
\label{sec:comparisonwitheffects}

In order to compare the $e$-scattering and the $p$-scattering channels with detector effects taken into account, a detector type is first specified because different ones are characterized by different dimensions, energy threshold, resolution, and so on. 
Our study in this section begins with a LArTPC detector. 
More specifically, we employ a ``DUNE-like'' experiment, meaning that it has four modules each of which comes with a cubic-shaped 17.5 kt total volume formed by \{width $\times$ height $\times$ length\} $=$ \{15.1 m $\times$ 14.0 m $\times$ 62.0 m\}.
In our analysis, we assume a fiducial volume of 10 kt for each module in a ``DUNE-like'' experiment. 
More complete DUNE detector specifications and properties are collected in Table~\ref{table:exp} in Appendix~\ref{sec:detectors}.

We essentially scan over a set of mass points in the $m_1-m_X$ plane for a given pair of $E_1$ and $m_2/m_1$. 
The values of $m_1$ are varied from $10^{-4}$ GeV to $10^{-0.5}$ GeV with 70 points equally spaced in the logarithmic scale, while $m_X$ ones are varied from $10^{-4}$ GeV to $10^0$ GeV with 80 points equally spaced in the logarithmic scale. 
Namely, we perform simulation for 5,600 points. 
In each point, we generate 5 million events using the TGenPhaseSpace class implemented in the \texttt{ROOT} package and reweight them with respective matrix element values. 
In this study, we consider the case that the secondary vertex accompanies only an $e^-e^+$ pair for simplicity. 
However, a sequence of procedures described later on are straightforwardly applicable to other channels. For the final state particles, we require the following set of criteria:
\begin{itemize}
\item[i)] $p_e>30$ MeV, $200\, {\rm MeV}< p_p < 2\, {\rm GeV}$,
\item[ii)] $\Delta\theta_{e-i}> 1^\circ$, $\Delta\theta_{p-i} > 5^\circ$ with $i$ denoting the other visible final state particles, and
\item[iii)] both primary and secondary vertices should appear in the detector fiducial volume.\footnote{We do not require that particle tracks are fully contained, which is beyond the scope of this work.}
\end{itemize}
We warn the reader that the minimum spatial momentum for proton (i.e., $p_p^{\rm th} = 200$ MeV) corresponds to $E_{\rm th}=21$ MeV, predicated upon the assumption that the level of $E_{\rm th}$ in ArgoNeuT~\cite{Acciarri:2014gev} can be achievable in DUNE.
The last criterion is applied if any intermediary particle (either $\chi_2$ or $X$) is long-lived. However, we do not calculate the laboratory-frame decay length event-by-event because it would take an enormous amount of time. 
To save the time and computing resource, we instead adopt a ``shortcut'' but rather conservative strategy. A detailed description can be found in Appendix~\ref{app:shortcut}. 

\begin{figure}[t]
\centering
\includegraphics[width=4.9cm]{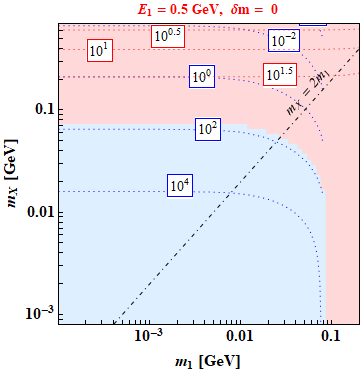}
\includegraphics[width=4.9cm]{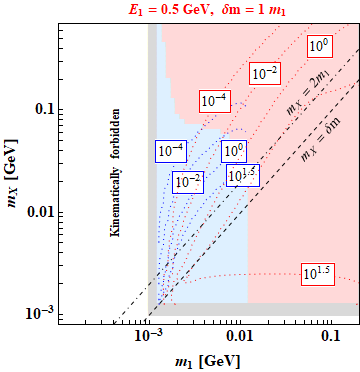}
\includegraphics[width=4.9cm]{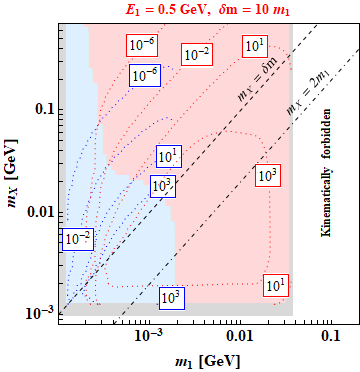} \\
\vspace{0.2cm}
\includegraphics[width=4.9cm]{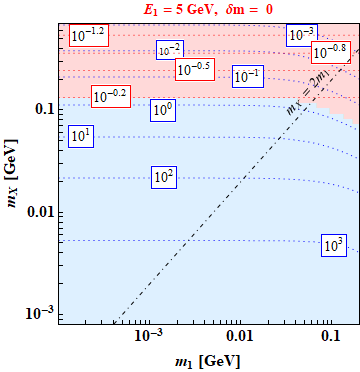}
\includegraphics[width=4.9cm]{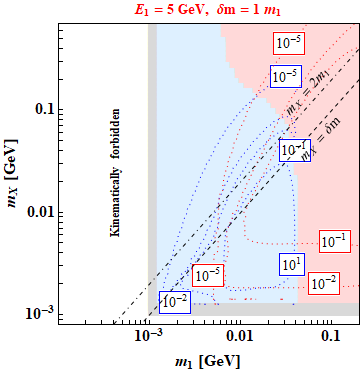}
\includegraphics[width=4.9cm]{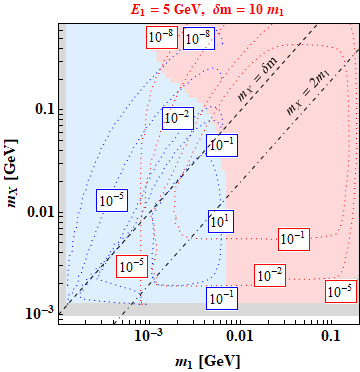}
\caption{
Comparisons between the $e$-scattering and the $p$-scattering channels at a DUNE-like detector, including various realistic effects such as energy threshold, angular separation, acceptance with respect to the displaced vertex.
Contours are the numbers of signal events calculated by Eq.~\eqref{eq:Nsig} with statistics of 40~kt$\cdot$yr.
The upper row is for $E_1=0.5$ GeV while the lower row is for $E_1=5$ GeV. 
From the left to the right panels, $\delta m$ are set to be 0, $m_1$, and $10m_1$, correspondingly.  Red and blue dotted contours show the expected number of signal events in the proton and electron channels with 40 kt$\cdot$yr statistics. 
Red-shaded and blue-shaded regions denote $p$-preferred and $e$-preferred ones, respectively, and the boundaries between them are given at $\sigma_{\chi_1 e}=\sigma_{\chi_1 p}$. 
The white regions are where decay process $\chi_2\to \chi_1 e^+e^-$ is not kinematically allowed, while the gray regions are where some or all of the final-state particles are not isolated or detected. See the text for more detailed explanations.
}
\label{fig:xscompcutsDUNE}
\end{figure}

The results with $E_1=0.5$ GeV are shown in the upper row of Figure~\ref{fig:xscompcutsDUNE}. Let us first take a look at the leftmost one corresponding to the $e$BDM case. Due to the fact that the proton differential cross section $d\sigma/d\spp$ is quickly rises toward smaller $\spp$ and the cuts harder on the proton, the fiducial cross section of the $e$-scattering now can be the same as or larger than that of the $p$-scattering unlike the comparisons in the previous section. 
So, we divide the entire space of interest into the $e$-preferred (blue shaded) and the $p$-preferred (red shaded) regions by $\sigma_{\chi_1 e}=\sigma_{\chi_1 p}$.
In fact, we find that the fiducial cross section of the $p$-scattering varies very mildly in the $e$-preferred region.  
The reason is as follows. Since $m_X \ll m_p$, events densely populate toward the lower energy regime (see also Eq.~\eqref{eq:psigma} and the nearby discussion) so that the fiducial cross section is essentially given by its tail part where the $m_X$ dependence is limited.
Moreover, since $E_1 \gg m_1$ in the $e$-preferred region, the fiducial $p$-scattering cross section is roughly constant in the variation of $m_1$.
Just for developing the intuition on the expected number of signal events, we add contours by red and blue dotted curves corresponding to $p$-scattering and $e$-scattering events, respectively.
The number of events, $N_{\rm sig}$, can be calculated by
\bea 
N_{\rm sig} = \sigma_{\chi_1 p(e)} \cdot \mathcal{F}_1 \cdot A\cdot t_{\rm exp} \cdot N_{p(e)}\,, \label{eq:Nsig}
\eea
where $\mathcal{F}_1$ is the signal flux shown in Eq.~\eqref{eq:fluxformula}, $t_{\rm exp}$ is the amount of time exposure, and $N_{p(e)}$ is the number of total target protons (electrons) inside a detector fiducial volume. We assume statistics of 40 kt$\cdot$yr. 
Here $A$ denotes the final signal efficiency including $A_\ell$ (see Appendix~\ref{app:shortcut}).
As in the previous section, the diagonal black dot-dashed line describes the boundary beyond (below) which an {\it on}-shell dark gauge boson $X$ predominantly decays invisibly (visibly). 
Therefore, any limit or interpretation with respect to the mass points above and below the line should be associated with corresponding exclusion limits in the dark gauge boson mass vs. kinetic mixing parameter. 
We will come back this point in Section~\ref{sec:pheno}.

Basically, the $e$-preferred region expands, compared to the corresponding one displayed in the upper-left panel of Figure~\ref{fig:compNoCuts}.
As argued in Appendix~\ref{sec:derivation}, this is because the distribution of the differential cross section $d\sigma / d \spp$ of the $p$-scattering is steeper than that corresponding to the $e$-scattering (see also Ref.~\cite{Kim:2016zjx} for generic shapes of proton and electron recoil energy spectra). 
This implies that the application of the $E_{\rm th}$ cut could reject the region around which $d\sigma / d \spp$ is peaking, in particular when $m_X$ gets much smaller than $m_p$ [see Eq.~\eqref{eq:maxpp}] 
and $m_1$ is too small to boost up the recoiling proton to overcome the energy threshold.
Possible search strategies inferred from this exercise are not much different from what is summarized in the previous section; the electron channel is better in probing smaller $m_1$ and $m_X$, and the other way around. 
However, care should be taken here.
Although many points in the above parameter space allow for quite a few signal events, $e$BDM suffers from large backgrounds such as atmospheric neutral current neutrino scattering events since it has only target recoil in the final state. 
Therefore, search strategies targeted at (seemingly) well-motivated parameter points should be designed with the potential of background contamination taken explicitly into account.\footnote{Note also that the $e$BDM signal could be degenerate with that of WIMP pair-annihilation to neutrinos~\cite{Arguelles:2019ouk}.} 

When it comes to the $i$BDM scenario, relevant phenomenology becomes even richer. 
Now all selection criteria i) through iii) are imposed.
The third one regarding the displaced vertex requires explicit values for kinetic mixing parameter $\epsilon$ and dark-sector gauge coupling $g_{12}$. 
Unfortunately, these factors are not completely canceled out in the ratio of $\sigma_{\chi_1 e}$ to $\sigma_{\chi_1 p}$ for a given $\{E_1, m_1, m_2\}$ because the decay length also depends on the boost factor of the slowly decaying particle which is in turn a function of target mass as well. 
So, we fix $\epsilon$ and $g_{12}$ to be $10^{-4}$ and 1 throughout this section for illustration, but the expected results with other $\epsilon$ and $g_{12}$ choices differ not much (i.e., the same intuition should go through).

We start by looking at the white regions. 
Since we focus on the case of an electron-positron pair coming out of the secondary vertex, the immediate parent particle for them should be massive enough to create them. 
This is translated as follows. 
If $\chi_2$ decays to $\chi_1$ and an $e^-e^+$ pair via an off-shell $X$ (i.e., $m_2 < m_1+m_X$), the mass gap between $\chi_2$ and $\chi_1$, $\delta m$ should be greater than $2m_e$. 
The relevant region of $\delta m < 2m_e$ is shown by the wider one sweeping through the space vertically.\footnote{Of course, in this case, $\chi_2$ may decay to $\chi_1$ and a photon through for example, dipole-type interaction~\cite{Giudice:2017zke}, but it requires an introduction of additional particles to induce the relevant operator. We do not examine this possibility for the sake of minimality.}
By contrast, if $\chi_2$ decays to an on-shell $X$ along with $\chi_1$ (i.e., $m_2 > m_1+m_X$), the dark gauge boson mass should be at least twice the electron mass to disintegrate to the $e^-e^+$.\footnote{We do not consider the lighter on-shell $X$ dominantly decaying to three photons, due to the Landau-Yang theorem~\cite{Yang:1950rg}, by the 8-dimensional operators with four field strength tensors. 
In this case, the mean decay time of $X$ is too long to provide a secondary signature in a terrestrial detector and various cosmological and astrophysical bounds can apply~\cite{Pospelov:2008jk}.} 
This case of $m_X < 2m_e$ is mapped to the horizontally-lying white region at the bottom of the plot. 
Another type of white region arises in the case of $\delta m = 10 m_1$ (see the rightmost panel). 
As discussed before, there is the kinematically allowed maximum $m_2$ value for a given set of $E_1$, $m_1$, and target mass $m_T$. 
Although the proton scattering generously allows for a wide range of $\delta m$, it also encounters the associated kinematic ``barrier'' which is actually realized in the right-hand side of the plot.  

The gray-colored regions are somewhat different in the sense that $i$BDM events can be produced but some or all of final state particles are not isolated or detected. 
For the vertically-stretching gray region, the associated mass spectra marginally allows for the creation of the $e^-e^+$ pair in the $\chi_2$ decay, but they are too soft to get over the threshold for electron.
However, for the horizontally-extended gray region, the (too) light $X$ is so significantly boosted that its decay products $e^+$ and $e^-$ are too close to each other, hence not isolated by criteria ii). 
The remaining narrow gray band shown in the upper rightmost panel of Figure~\ref{fig:xscompcutsDUNE} emerges because of the threshold for protons. 
Remember that nearby space is already close to the kinematic ``barrier'' for the $p$-scattering process. 
As a result, most of incoming $E_1$ is spent for forming a massive $\chi_2$, but only a tiny fraction of energy is transferred to a target, inducing a soft recoiling proton.

Speaking of diagonal lines, there arises a different type denoted by the black dashed. 
One can easily see that with respect to the line $\chi_2$ decays to $\chi_1 e^+e^-$ via from on-shell $X$ to off-shell $X$ as moving bottom to top, or vice versa. Therefore, the signal acceptance may be affected substantially due to this kinematical transition, along the line.

Just like the $e$BDM case, the $e$-preferred region becomes wider than the corresponding theoretical prediction again because cuts on energy and angle diminish the signal acceptance in the $p$-scattering channel. 
The lower vertical drop arises because the up-scatter of $\chi_1$ becomes kinematically challenging beyond it. Indeed, the associated $N_{\rm sig}$ contours show that the $e$-scattering cross section falls rapidly near the border. 
By contrast, the vertical boundary toward bigger $m_X$ was absent in the corresponding theory prediction, but now emerges. The reason for this lies in the angular separation between the $e^+$ and $e^-$ from the $\chi_2$ decay. 
Since the proton is usually heavier than the other particles involved in the process, the proton recoil is not likely to carry out a large amount of energy, but $\chi_2$ is. 
Thus, $\chi_2$ is significantly boosted so that its decay products are inclined to be too collimated to satisfy the angle cut. 
Of course, a similar behavior is anticipated in the $e$-scattering channel. 
However, the recoiling electron and the $e^-e^+$ pair are more likely to share the incoming energy $E_1$ (together with the outgoing $\chi_1$). 
Three electron tracks are often moving in the same direction, but some fraction of events can pass the angle cut. 
We see that this sort of $e$-preferred region becomes even wider as $\delta m$ increases (see the upper rightmost panel of Figure~\ref{fig:xscompcutsDUNE}). 
Basically, a wider range of low-mass $m_1$ is accessible so that the resultant low-mass $\chi_2$ will give a merged $e^-e^+$ pair in the proton channel. 

We remark that care should be taken in interpreting the results in the rightmost panels where $m_2 = 11 m_1 > 3m_1$. 
In this case, the result above $m_X=2m_1$ line would lose its meaning, if $g_{11}$ in Eq.~\eqref{eq:lagrangian} were sizable. $\chi_2$ would decay invisibly to three $\chi_1$ particles.
Therefore, the comparison in that regime is relevant to models with vanishing or suppressed $g_{11}$ (see Scenarios I-ii and II-ii described in Section~\ref{sec:comparison}).

We next show the results corresponding to $E_1 = 5$ GeV in the lower panels of Figure~\ref{fig:xscompcutsDUNE}.
Similar behaviors and trends are essentially retained as in the case of $E_1=0.5$ GeV. 
A larger value of $E_1$ simply opens more accessible phase space in the $\chi_1$ mass, so one can understand the results as some sort of ``stretching'' of those in the $E_1=0.5$ GeV case. 
In summary, the $e$ channel is complementary to or even favored over the $p$ channel toward smaller $m_1$ or smaller $m_X$. 
On the other hand, the $p$ scattering will be a discovery channel when a search is targeting at the regime where $\chi_1$ and $X$ are heavier.\footnote{The expected number of events in the $p$-scattering channel looks small because a small value of $\epsilon$ and one-year data collection are assumed. 
One can reach the discovery-level events by increasing the hypothesized value of $\epsilon$ and/or data collection.} 
Finally, we emphasize that background-free searches for $i$BDM are possible as relevant signal events come up with several unique features. 
Hence, hunting for $i$BDM signals will provide complementary information in investigating dark-sector physics.

\begin{figure}[t]
\centering
\includegraphics[width=4.9cm]{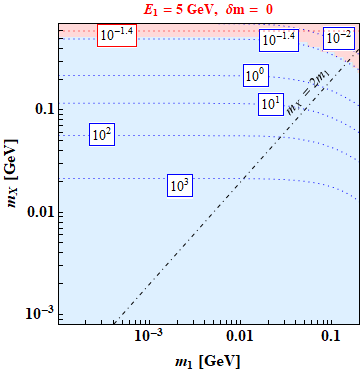}
\includegraphics[width=4.9cm]{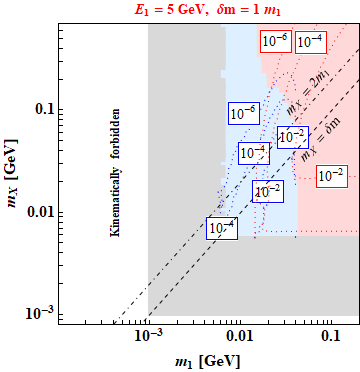}
\includegraphics[width=4.9cm]{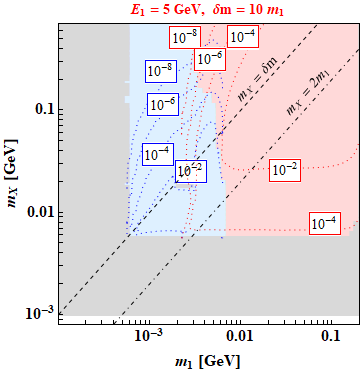}
\caption{
Comparisons between the $e$-scattering and the $p$-scattering channels at a HK-like detector, including various realistic effects such as energy threshold, angular separation, acceptance with respect to the displaced vertex. Contours are the numbers of signal events calculated by Eq.~\eqref{eq:Nsig} with statistics of 380~kt$\cdot$yr.
See the caption in Figure~\ref{fig:xscompcutsDUNE} for more details.
}
\label{fig:xscompcutsHK}
\end{figure}

Lastly, we discuss how the phenomena that we have observed at a DUNE-like LArTPC detector are affected by different detector specifications. As advertised before, we take a ``HK-like'' Cherenkov detector with a fiducial mass of 380 kt, as an example. Again we refer to Table~\ref{table:exp} in Appendix~\ref{sec:detectors} for detailed specifications and properties.
Similarly to the DUNE-like detector, we demand the final state particles to satisfy the following conditions:
\begin{itemize}
\item[i)] $p_e > 100$ MeV, 1.07 GeV $<p_p<$ 2 GeV,
\item[ii)] $\Delta \theta_{e-i}>3^\circ$ ($\Delta \theta_{e-i}>1.2^\circ$) for $p_e<1.33$ GeV ($p_e>1.33$ GeV) and $\Delta \theta_{p-i}>3^\circ$ for all $p_p$ with $i$ running over the other visible final state particles, and
\item[iii)] both primary and secondary vertices should appear in the detector fiducial volume. 
\end{itemize}
Note that the minimum required energy to observe an electron in SK/HK is expected to be $\sim 5$ MeV, but the associated angular resolution is too poor. 
So, we instead adopt 100 MeV to keep a decent level of angular resolution as per the search for elastic BDM interacting with an electron performed by the SK Collaboration~\cite{Kachulis:2017nci}.
One should also notice that a momentum threshold of $p_p\simeq1.07$ GeV, which corresponds to $E_{\rm th}\simeq485$ MeV, is required for the relatively heavy proton to emit Cherenkov light.
Finally, the acceptance $A_\ell$ associated with criterion iii) is evaluated in the same way as explained earlier.

The comparisons between the $e$-scattering and $p$-scattering channels are shown in Figure~\ref{fig:xscompcutsHK}. 
We report the results only with $E_1=5$ GeV, because $E_1=0.5$ GeV is so small that entire events in the proton channel would not be selected due to the threshold for protons. Hence, the associated comparison is meaningless.
Nevertheless, the results with $E_1=5$ GeV will suffice to discuss phenomenological differences.
Looking at the $e$BDM case first (the leftmost panel), we see that the $e$-preferred region is significantly extended in comparison with the corresponding theory prediction in Section~\ref{sec:comparison}. 
This is because a proton should have enough kinetic energy to create Cherenkov radiation while the proton scattering cross section is typically peaking toward small energy/momentum transfer especially for $m_X \lsim \mathcal{O}(1)$ GeV.
Therefore, the electron channel may be considered more important in the search for BDM signals. 
When it comes to $i$BDM cases (see the middle and the rightmost panels of Figure~\ref{fig:xscompcutsHK}), similar understanding goes through. 
A crucial difference is the fact that the gray regions become much wider for a given $E_1$. This is purely due to the larger thresholds and angular resolution. Therefore, HK-like Cherenkov detectors are not ideal for probing parameter space  with small $m_X$ and/or small $m_1$ through the $i$BDM channel. 
However, at the same time, this implies that devising a search strategy getting around the threshold and resolution issues would enable to explore the above-mentioned regime. 

\section{Example Data Analysis}
\label{sec:pheno}

Furnished with various observations and guidelines discussed in the previous section, in this section we demonstrate how actual analyses would be conducted, focusing on $i$BDM signals. 
Since our benchmark model described in Eq.~\eqref{eq:lagrangian} involves interactions with a dark gauge boson $X$, it is natural to study expected experimental sensitivities in dark gauge boson parameter space with respect to the four benchmark experiments listed in Section~\ref{sec:benchmarkdetectors}.
We essentially contrast parameter coverages for several reference model points. 
By doing so, we develop the intuition, what experiment and channel would be better motivated or more advantageous for a given reference point. 
Conversely, each experiment would design optimized search strategies and choose best-motivated parameter space, performing similar exercises demonstrated here.

A parameter scan is done in the following way. 
We first fix values of $E_1$, $m_1$, and $m_2$, and then divide the expected $m_X$ coverage into 50 segments equal-sized in the logarithmic scale. 
For each parameter point defined by $(E_1, m_1, m_2, m_X)$, one million events reweighted by the matrix element are generated and energy and angle cuts are applied in order to calculate a cut-related acceptance. 
We next scan $\epsilon$ space from $10^{-6}$ to $3\times 10^{-3}$ after chopping the range into 350 intervals equal-spaced again in the logarithmic scale. 
For each $\epsilon$, the maximum laboratory-frame mean decay length of the long-lived particle (either $\chi_2$ or $X$ depending on the mass spectrum) $\bar{\ell}_{\rm lab}^{\max}$ is calculated with $g_{12}$ set to be unity for simplicity. 
Plugging $\bar{\ell}_{\rm lab}^{\max}$ to the relevant fitted function, we extract an $A_\ell$ value. 
The final acceptance $A$ can be obtained by multiplying the cut-related acceptance by $A_\ell$. 
One can calculate the number of expected signal events, utilizing Eq.~\eqref{eq:Nsig}. 
Here $N_{p(e)}$ and $t_{\rm exp}$ can be easily determined, once one chooses a detector and the amount of time exposure. 
Cross sections (before imposing any selection criteria) are calculable according to the formulation detailed in Appendix~\ref{sec:derivation}, and the flux value for a given $E_1 (=m_0)$ comes from Eq.~\eqref{eq:fluxformula} in the annihilating two-component dark matter scenario. Labeling the 90\% confidence level (C.L.) exclusion limit by $N^{90}$ which depends on the number of expected background events, we calculate it with a modified frequentist construction~\cite{Read:2000ru, ATLAS:2011tau}. 
An experiment is said to be sensitive to a given signal if $N_{\rm sig} \geq N^{90}$, i.e., parameter points whose $N_{\rm sig}$ is greater than $N^{90}$ will be excluded with 90\% C.L. under the associated background assumption.
We hereafter take a zero-background assumption for simplicity, as $i$BDM-induced events come with distinctive features that known SM processes (e.g., atmospheric neutrino-initiated ones) are hard to mimic (see also discussions in Section~\ref{sec:model}).\footnote{Nevertheless, one may argue with a few plausible possibilities such as neutrino-induced resonance and DIS events involving a handful of mesons which subsequently decay to charged particles. 
A more systematic discussion will be elaborated in Ref.~\cite{DeRoeck:2020ntj}.}

\subsection{DarkSide-20k vs. DUNE} 
Our first exercise is to compare DarkSide-20k and DUNE experiments. Before showing relevant results, we need to explain our event selection scheme for DarkSide-20k. 
It is basically designed to be best sensitive to the signatures induced by the scattering process between slowly-moving WIMP or WIMP-like dark matter and a nucleon.
The scale of typical recoil kinetic energy is expected to be order of $1-100$ keV, so that detectors are most sensitive to this energy range. 
In particular, the ionization signal (called S2) is sort of ``amplified'' in the gas-phase Argon to measure such a small energy deposit. 
However, in many well-motivated cases, $E_1$ (or equivalently $m_0$) is of sub-GeV range, and as a consequence, the typical energy scale of final state visible particles in our signal processes will be several tens of MeV or greater. 
A potential problem with a large energy deposit is that photomultiplier tubes -- which are the agents to measure the amount of deposited energy -- may be saturated due to too much amplification beyond their measurement capacity, and thus the associated data analysis would face severe challenges. 
To avoid this potential issue, we here take a rather tricky strategy to look at only scintillation signal (called S1).\footnote{We expect that $i$BDM-induced S1 are rather strong unlike WIMP-induced ones.} 
The $i$BDM process under consideration often involves a displaced secondary vertex as discussed in Section~\ref{sec:model}. 
Therefore, it would be possible to identify two separated vertices by some dedicated S1-pattern analysis.\footnote{Here we do not claim or justify at all that this way of analysis is viable, which is certainly up to experimentalists' effort and not within the scope of this paper. 
However, we bring readers' attention to e.g., DEAP-3600~\cite{Amaudruz:2014nsa, Amaudruz:2017ekt, Amaudruz:2017ibl} in which an interaction vertex is identified only with a set of S1 signals.}
For illustration, we assume that a $\geq 30$ cm separation is discernible at the DarkSide-20k detector.\footnote{For example, DEAP-3600~\cite{Amaudruz:2014nsa, Amaudruz:2017ekt, Amaudruz:2017ibl} claims a resolution of $\sim 10$ cm for a single vertex. 
As there are two vertices in a signal event, we believe that our assumption here is reasonable.} 
So, the corresponding acceptance $A_\ell$ at the DarkSide-20k detector is calculated by checking if the secondary vertex is not only within the associated detector fiducial volume but separated from the primary vertex by more than 30 cm. 
We find that the resultant $A_\ell$ is well accommodated by a different, empirical fit template\footnote{It is inspired by the fit template introduced in Ref.~\cite{Agashe:2012bn}.}:
\bea 
A_\ell = c_1' \exp\left[-c_2'\left(\frac{\ell}{c_3'}+\frac{c_3'}{\ell} \right)^{c_4'} \right] \hbox{ for }\ell \geq 0\,,
\eea
where $c_i'$ ($i=1,2,3,4$) are fit parameters. 
No other cuts associated with energy and angle are imposed, so that $A_\ell$ can be directly translated to final signal acceptance $A$. 

\begin{figure}[t]
\centering
\includegraphics[width=7.2cm]{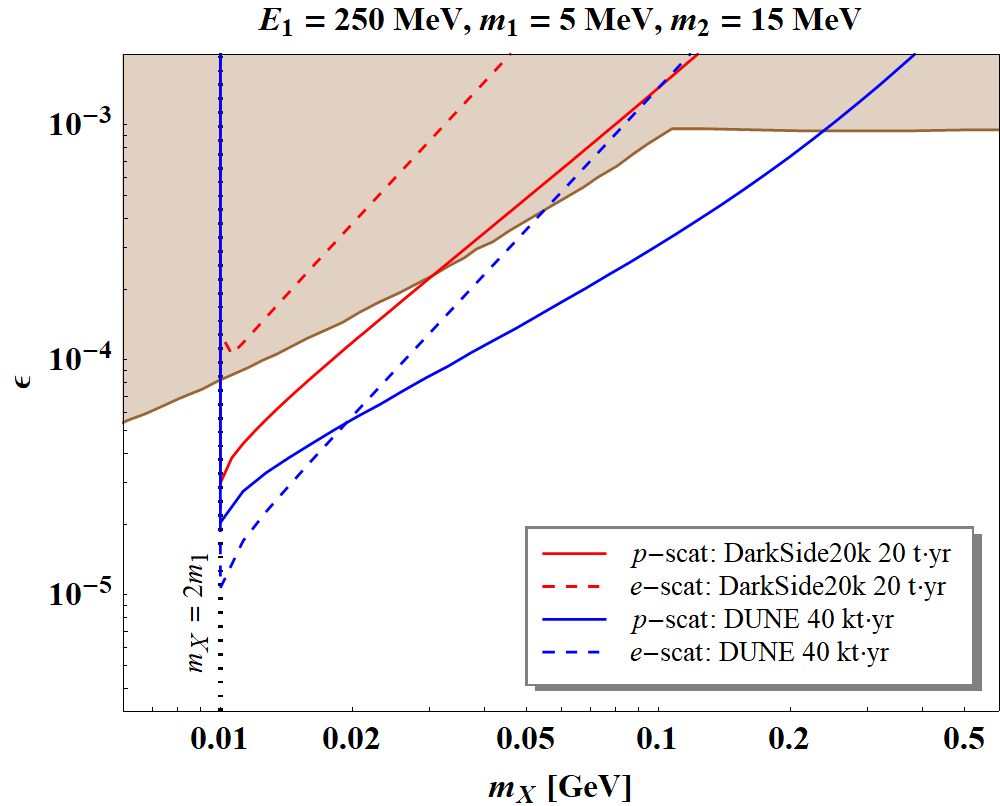} \hspace{0.2cm} 
\includegraphics[width=7.2cm]{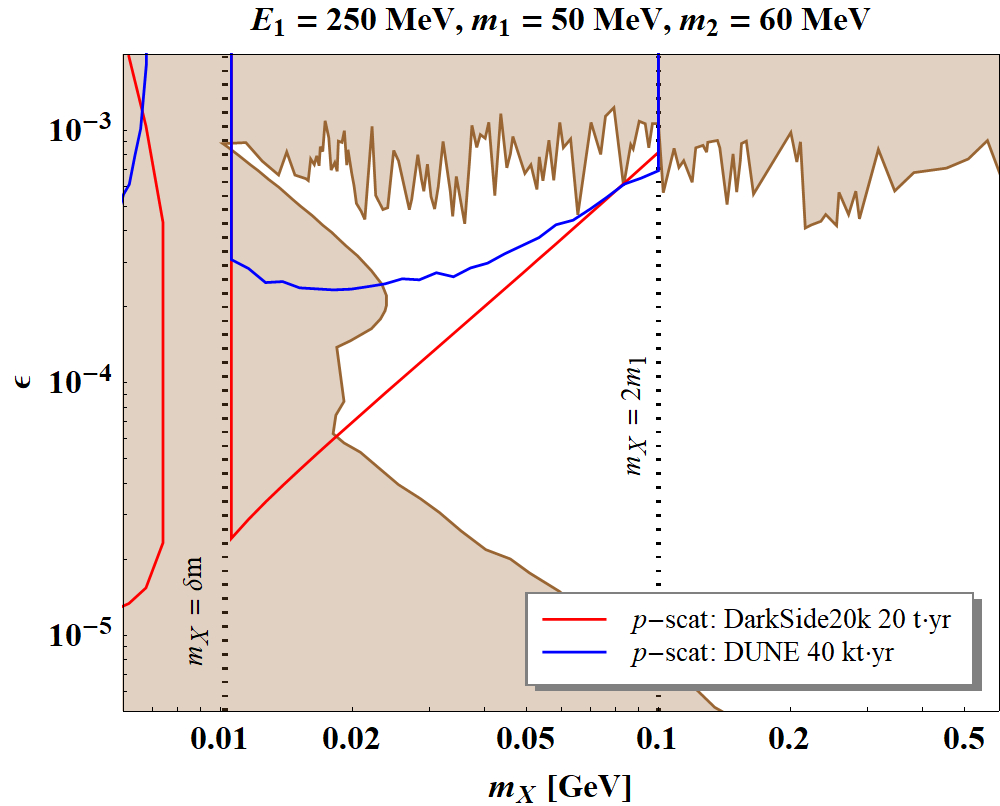} \\
\vspace{0.2cm}
\includegraphics[width=7.2cm]{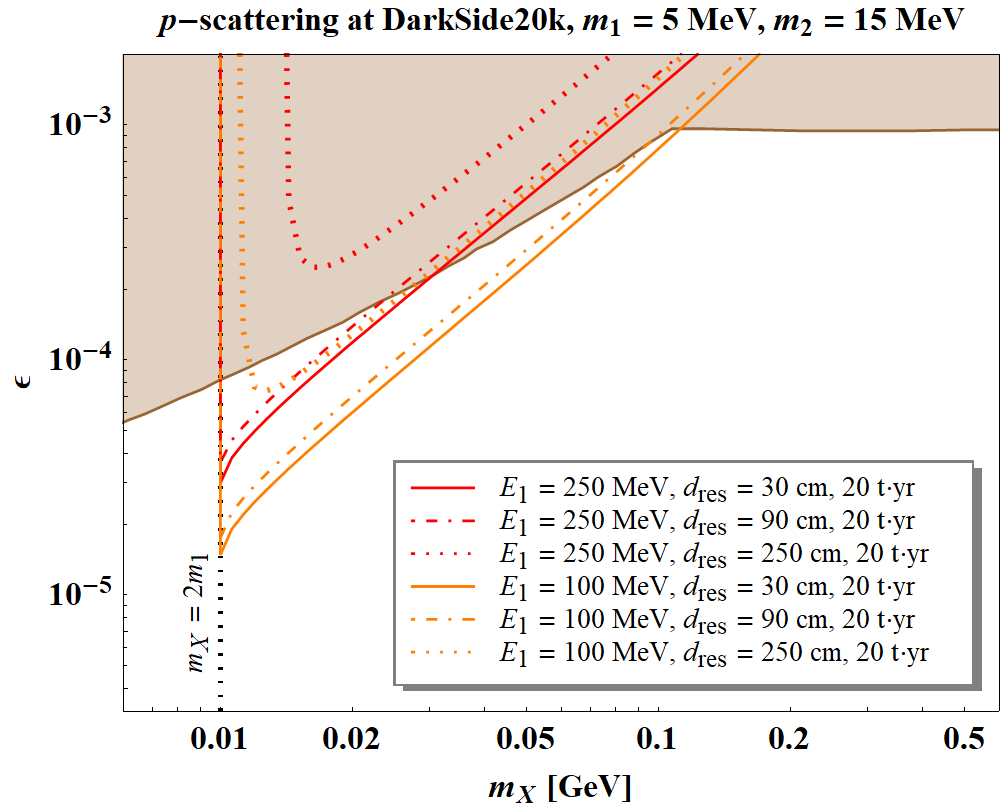} \hspace{0.2cm}
\includegraphics[width=7.2cm]{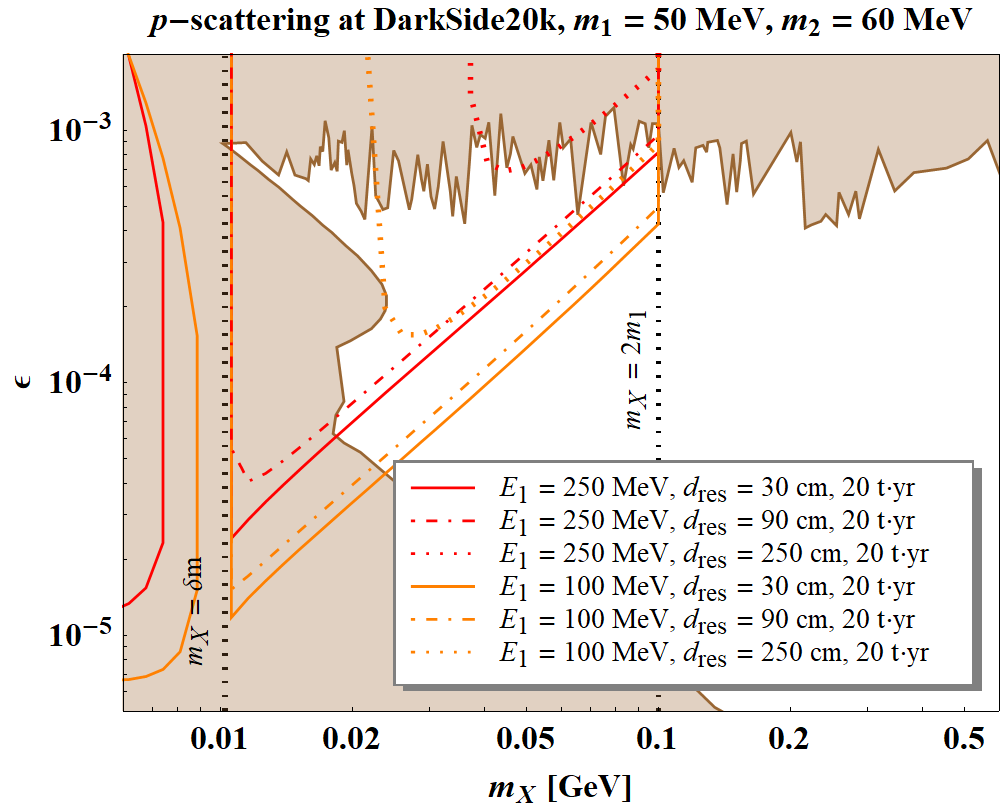}
\caption{ \label{fig:DSvsDUNE}
Experimental sensitivities for Scenario I (left panels) and Scenario II (right panels) in the $m_X-\epsilon$ plane. 
The brown-shaded regions are the current exclusion limits which are extracted from Refs.~\cite{Banerjee:2017hhz} and~\cite{Banerjee:2018vgk} for Scenario I and Scenario II, respectively.
The upper-left panel compares DarkSide-20k--20t$\cdot$yr (red lines) and DUNE--40kt$\cdot$yr (blue lines) in the proton (solid lines) and electron (dashed lines) channels for $(E_1, m_1, m_2)=(250, 5, 15)$ MeV, while the upper-right panel compares them only in the proton channel for $(E_1, m_1, m_2)=(250, 50, 60)$ MeV. 
The lower panels compare proton-channel coverages between two $E_1$ values with three different $d_{\rm res}$ values only for DarkSide-20k--20t$\cdot$yr. 
See the text for a more detailed discussion.
}
\end{figure}

The results are displayed in Figure~\ref{fig:DSvsDUNE}, where we compare experimental sensitivities which would be achieved in the DarkSide-20k and DUNE experiments in the plane of $m_X$ vs. $\epsilon$. 
The brown-shaded covers the current excluded regions whose boundary values are extracted from Refs.~\cite{Banerjee:2017hhz} and~\cite{Banerjee:2018vgk} for Scenario I and Scenario II, respectively. 
We assume 1-year exposure for concreteness and apply the same selection criteria listed in the previous section, for DUNE.
The upper-left panel is for Scenario I, contrasting DarkSide-20k--20t$\cdot$yr (red lines) and DUNE--40kt$\cdot$yr (blue lines) in the proton (solid lines) and electron (dashed lines) channels for a reference point defined by $(E_1, m_1, m_2)=(250, 5, 15)$ MeV. 
We see that DarkSide-20k can cover a decent range of parameter space in spite of 2,000 times smaller fiducial mass than DUNE. 
Being aware that the scattering cross section is quadratic in $\epsilon$ and $A_\ell$ decreases as $\epsilon^2$, we estimate that a 2,000 times smaller target mass would lead about 6.7 ($\approx 2000^{1/4}$) times degradation in probing $\epsilon$ values.
This sort of rough estimation goes through the electron channel, as DUNE can explore down to $\sim10$ times smaller $\epsilon$ for a given $m_X$. 
$E_{\rm th}$ for electrons at DUNE is sufficiently low relative to the chosen $E_1$ and effectively no minimum required length for displaced vertices (due to its mm-scale position resolution) unlike DarkSide-20k together with a bigger detector length scale.
However, when it comes to the proton channel, the $\epsilon$ reach at DarkSide-20k is degraded only by a factor of $\sim 2$ at $m_X=0.01$ GeV.
The reason for this is that no energy and angular cuts are imposed for DarkSide-20k whereas an application of sizable energy threshold and angular separation cuts away a certain fraction of events at DUNE.
On the other hand, recoil proton gets harder in increasing $m_X$ so that threshold affects less and the difference in the $\epsilon$ coverage gets gradually larger.
Another phenomenon to find here is that the proton channel is preferred over the electron one in both DarkSide-20k and DUNE experiments. 
In the previous section we observe that the $p$-scattering is more advantageous as $E_1$ decreases. Especially, in order for DarkSide-20k to attain some signal sensitivity, smaller $E_1$ values are necessary because the relatively small mass of DarkSide-20k can be compensated by a large flux induced by a small $m_0 (=E_1)$ [see Eq.~\eqref{eq:fluxformula}]. Therefore, the search for BDM signals in the proton channel is better motivated at DarkSide-20k.

In the lower-left panel, we examine the expected experimental sensitivity for $E_1=100$ MeV. 
Because of energy threshold, DUNE has no sensitivity to this reference point. 
Since a smaller $E_1$ implies a larger $\chi_1$ flux, the resultant parameter coverage becomes wider. 
We further investigate the dependence of experimental sensitivities on the vertex resolution (denoted by $d_{\rm res}$), employing two more different values, $d_{\rm res}=90$ cm and $d_{\rm res}=250$ cm.
We see that $d_{\rm res}=90$ cm still allows to explore a similar range of parameter space. 
By contrast, $d_{\rm res}=250$ cm rapidly reduces the regions to probe because $d_{\rm res}$ becomes comparable to the detector length scale of DarkSide-20k. 

Moving onto the experimental sensitivities for Scenario II, we demonstrate our results for $(E_1, m_1, m_2)=(250, 50, 60)$ MeV, in the upper-right panel of Figure~\ref{fig:DSvsDUNE}.
Since the chosen parameter values forbid $\chi_1$ from up-scattering to a $\chi_2$ of 60 MeV in the electron channel, we report only results in the proton channel. 
Unlike the previous case, DarkSide-20k can explore a wider range of parameter space than DUNE can, mainly due to the threshold for proton. As we argued in Section~\ref{sec:xs}, recoil proton is likely to be softer in decreasing $m_X$ so that many events at DUNE do not pass the $E_{\rm th}$ criterion for proton.
Again this clearly shows how the energy threshold affects the experimental sensitivity.

An interesting phenomenon worth mentioning is that there exists a finite range of $m_X$ near $m_X=\delta m$ for which no signal sensitivity is essentially available in both experiments.
But the underlying reasons for the individual ones are quite different. 
Note that $\chi_2$ decays to an on-shell $X$ below the $m_X=\delta m$ line so that the potentially long-lived particle is $X$ ($\chi_2$) below (above) the line. 
For a given $\epsilon$ value, $\chi_2$ (undergoing a three-body decay) is much more long-lived than the on-shell $X$. 
For DarkSide-20k, the on-shell $X$ usually decays rather ``promptly'' with respect to the given $d_{\rm res}$, and as a result, two vertices are not resolvable. 
However, in decreasing $m_X$, the on-shell $X$ becomes significantly boosted, so that the laboratory-frame life time can be significantly dilated. 
Hence, DarkSide-20k restores the signal sensitivity for lighter $m_X$. 
When it comes to DUNE, the electron and positron coming from the on-shell $X$ are not energetic enough to overcome the energy threshold. 
For the chosen set of parameter values, the scattered $\chi_2$ is not much boosted. 
Just below $m_X=\delta m$, the produced $X$ is (roughly) as boosted as $\chi_2$ so that its energy is no more than $m_X\times E_1/m_2\approx 40$ MeV with $m_X =10$ MeV and the maximal boost of $\chi_2$ assumed (i.e., $\gamma_2 <E_1/m_2$). 
Therefore, each of the positron and electron carries away $\sim 20$ MeV which is below the threshold. 
However, as $m_X$ gets smaller, the electron (positron) emission direction relative to its own boost direction and/or the boost direction of $X$ comes into play. 
Therefore, a certain fraction of events can pass the selection criteria and the signal sensitivity starts to grow again. 

We next show the expected sensitivity for $E_1=100$ MeV in the lower-right panel of Figure~\ref{fig:DSvsDUNE} just as in the case of Scenario I and see the potential of probing a wider range of parameter space. 
The dependence of experimental sensitivities on the vertex resolution is also investigated for completeness. 
We again find that experimental sensitivities mildly depend on $d_{\rm res}$ unless it is comparable to the detector length scale of DarkSide-20k.

One should be mindful that in this analysis we did not include a potential contribution from the coherent scattering of $\chi_1$ off target nuclei.
The reason is that the energy of incoming $\chi_1$ is 100 MeV or more so that we expect that the majority of events happen via an incoherent scattering process. 
Nevertheless, it is fair to say that the chance that the momentum transfer (to the target) falls below $\mathcal O (10\,{\rm MeV})$  may not be negligible even for $E_1 = 250$ MeV.\footnote{A more quantitative statement will be available, provided that a dedicated study is preceded. 
However, this is beyond the scope of our work here, so we leave it for future. See also a very recent study in Ref.~\cite{McKeen:2018pbb}.}
In this context, our reference $E_1$ values may be in a regime far from concluding either a full coherent or full incoherent scattering. 
However, we believe that our approach using the incoherent scattering cross section $\sigma_{\chi_1 p}$ is rather conservative since the coherent nuclear scattering would be enhanced by the atomic number of a target nucleus for dark gauge boson scenarios.

\subsection{DUNE vs. HK} 
\begin{figure}[t]
\centering
\includegraphics[width=7.2cm]{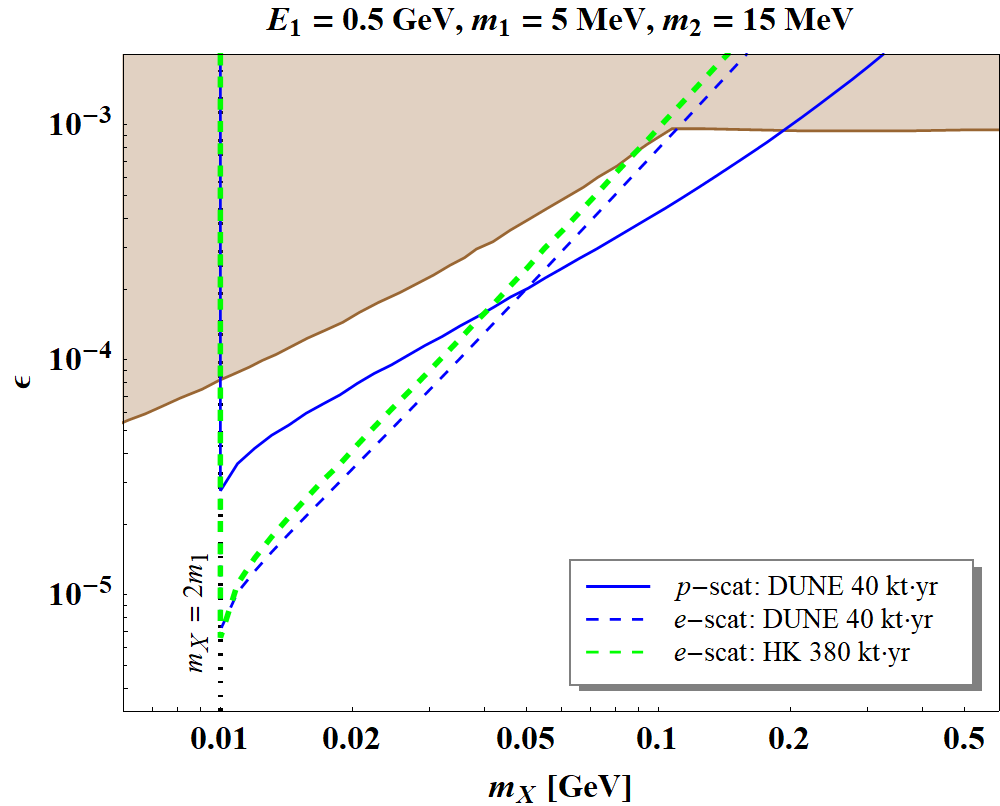} \hspace{0.2cm}
\includegraphics[width=7.2cm]{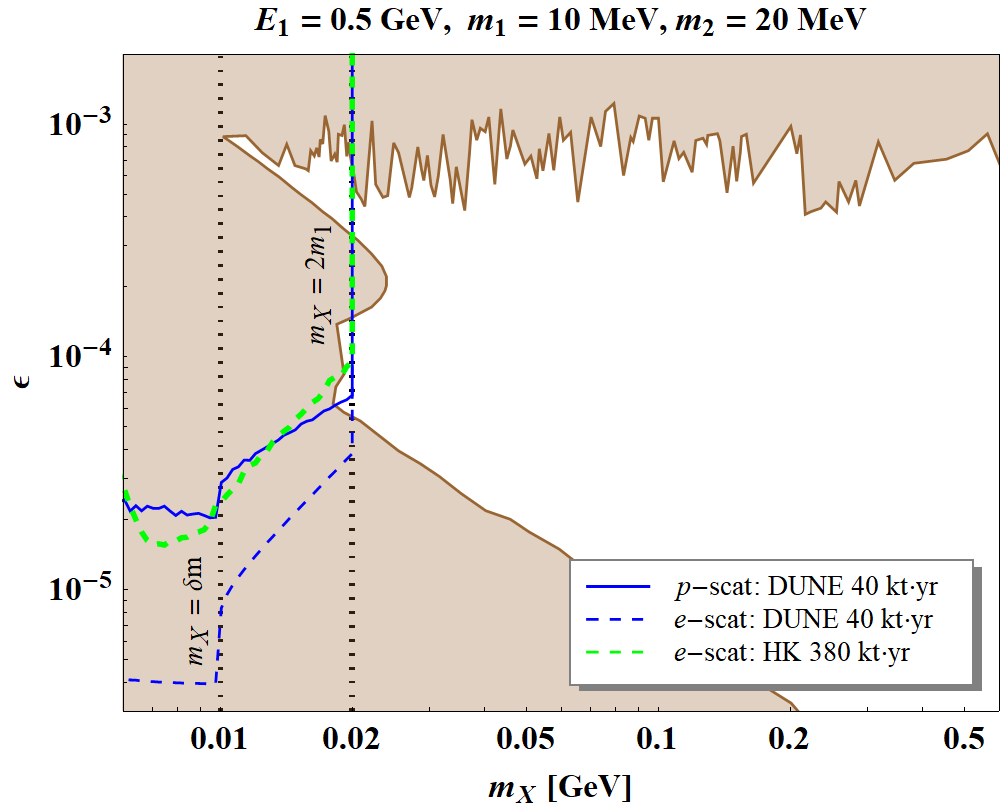} \\
\vspace{0.2cm}
\includegraphics[width=7.2cm]{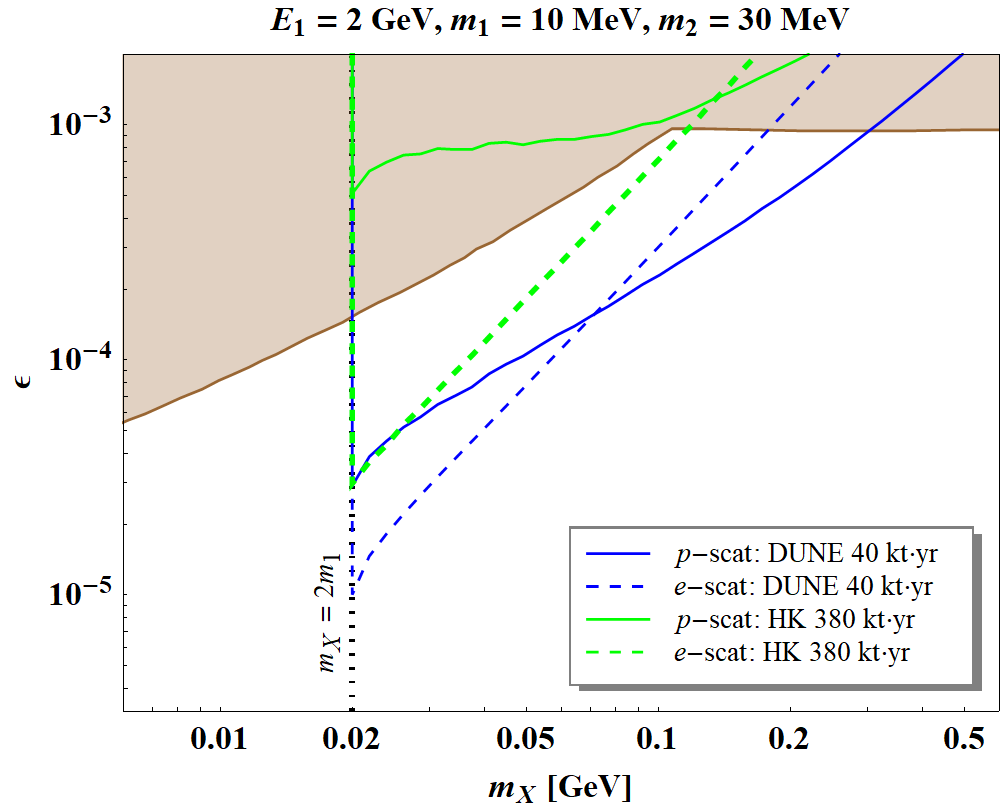} \hspace{0.2cm}
\includegraphics[width=7.2cm]{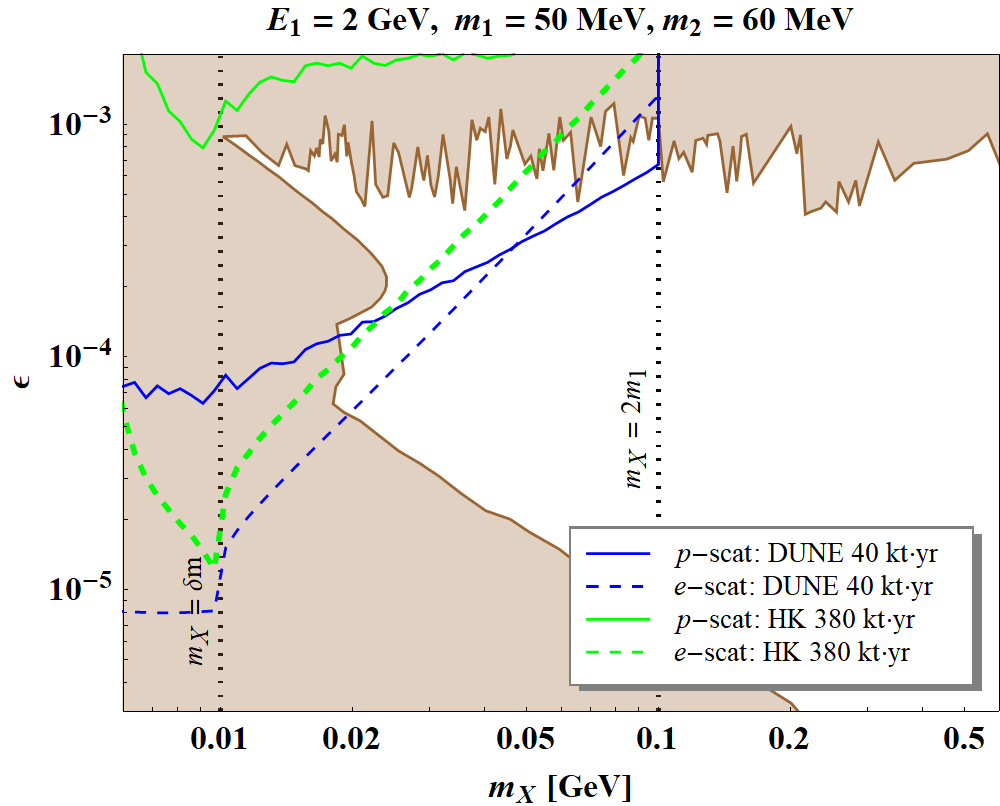}
\caption{\label{fig:DUNEvsHK} Experimental sensitivities achievable in the electron (dashed lines) and proton (solid lines) channels, in the $m_X-\epsilon$ plane. 
The left (right) panels compare DUNE--40kt$\cdot$yr (blue lines) and HK--380kt$\cdot$yr (green lines) for Scenario I (Scenario II) with two different reference points. }
\end{figure}

The next exercise is to contrast DUNE and HK experiments. 
The total mass of HK detectors is about 10 times larger than that of DUNE far detectors. On the other hand, the DUNE detectors are expected to fulfill more precise measurements thanks to the LArTPC technology. Therefore, it is expected that they both would show a similar level of performance in the search for $i$BDM signals for a given amount of time exposure. Our main results are displayed in Figure~\ref{fig:DUNEvsHK}.

First of all, the upper-left panel shows sensitivities with respect to Scenario I, which would be achievable in the electron (dashed lines) and proton (solid line) channels in the DUNE (blue lines) and HK (green line) experiments, for a reference point defined by $(E_1, m_1, m_2)=(500, 5, 15)$ MeV. 
Event selections for DUNE and HK are done according to selection criteria described in Section~\ref{sec:comparisonwitheffects}, and again 1-year data collection is assumed.
Note that here the proton channel is not available in HK because $E_1$ is too small for any single recoil proton to overcome the relevant threshold.
Comparing the electron channel, we see that both experiments show similar coverage. 
The chosen $m_1$ is close to $m_e$ so that the four final state particles (i.e., $e^-, e^-, e^+$, and $\chi_1$,) tend to equally share the incoming energy. 
Therefore, in a large fraction of events, all three electrons have an energy greater than 0.1 GeV. 
This implies that other choices of $m_1$ would result in smaller coverage of HK because of harder selection cuts. 
Now comparing the $e$-scattering and $p$-scattering in DUNE, one should notice that the latter wins over the former as $m_X$ increases whereas the trend is reverse toward smaller $m_X$. 
This agrees with the observation made in Sections~\ref{sec:comparison} and~\ref{sec:comparisonwitheffects} that the $e$-scattering channel becomes competitive as $m_1$ and $m_X$ decrease. 

We next raise $E_1$, $m_1$ and $m_2$, reporting the associated result in the lower-left panel of Figure~\ref{fig:DUNEvsHK}. 
Increasing $E_1$ essentially means a reduction of the $\chi_1$ flux, so one would naively expect that HK shows wider coverage due to its 10 times bigger mass. 
We find, however, that selection cuts indeed matter over the detector fiducial mass. 
Relatively larger energy threshold often forces events to fall in the collimated regime where the angular separation plays a crucial role, in turn. 
On the other hand, keeping a sufficient angular distance often forces (some) final state particles to be soft relative to threshold. 
Therefore, a right balance between the two factors, in general, would be desired like the previous case in order to accomplish better experimental sensitivities. 
Now the proton channel becomes available because $E_1$ is large enough to allow for $p_p>1.07$ GeV. 
However, the same, above-described argument applies to the proton scattering more stringently, and as a result, it enables us to probe only excluded regions. 

We perform similar exercises for Scenario II and present the results in the right panels of Figure~\ref{fig:DUNEvsHK}. 
As before, the upper and lower ones are for $E_1=0.5$ GeV and $E_1=2$ GeV, respectively. 
Basically, similar interpretations are relevant here. 
Unlike the study in DarkSide-20k vs. DUNE, no discontinuity around the line of $m_X=\delta m$ arises because both $E_1$ values are sufficiently large. 
Nevertheless, we observe some sort of transition about the line. 
Interestingly, sensitivities in the proton channel get worse in decreasing $m_X$. 
For the $p$ scattering, $\chi_2$ typically takes away more energy than the proton because $m_p \gg m_1, m_2$. 
This implies that the on-shell $X$, which is now long-lived, becomes more boosted in decreasing $m_X$, so that not only resultant $A_\ell$ drops rapidly but the $e^-e^+$ pair from the $X$ decay becomes more merged. 
In particular, HK is affected by the latter effect more than DUNE. 
On the contrary, $\chi_2$ in the electron channel is not as much boosted as that in the proton channel so that the above-mentioned effects are not substantial. 

\subsection{HK vs. DeepCore}
As the final exercise, the DeepCore experiment is compared to the HK experiment. 
Unlike DUNE and HK, DeepCore does not see charged particle-induced tracks (except muons) but measures cascade-type signatures. 
Therefore, if all three final state particles come out of a single point (within the detector vertex resolution), it is (almost) impossible to distinguish it from a true recoil-only event. 
To get around this issue, we take a rather simplified strategy; any event showing double (displaced) cascades will be tagged as signal. 
(See Ref.~\cite{Coloma:2017ppo} for a double-cascade signal in the context of heavy neutrino-accompanying processes.) 
A $\tau$ neutrino would give rise to a double-cascade event, but it would have to carry (at least) sub-PeV energy which is much away from the energy scale of interest here. 
From a comparison between IceCube and DeepCore detector arrays, we estimate that $\sim 5$ meters distance is resolvable in DeepCore. 
The following selection criteria are applied: 
\begin{itemize}
\item[i)] $p_e^{\rm recoil}>10$ GeV, $p_{e^+e^-}^{\rm secondary}>10$ GeV, and
\item[ii)] the secondary vertex should appear in the detector fiducial volume and be at least 5 meters away from the primary vertex.
\end{itemize}
The second requirement in i) means that the energy deposit by the $e^-e^+$ pair should be collectively greater than 10 GeV threshold. 
The main spirit behind criterion ii) is similar to that for DarkSide-20k, so we adopt the same strategy to calculate the corresponding $A_\ell$.

In this analysis, we slightly modify the selection criteria for HK as well; for a fairer comparison we do not demand an angular separation of $3^\circ$ for the electron-position pair emitted from the secondary vertex. 
As we will show shortly, the $E_1$ value under consideration is as large as order several tens of GeV. 
In such an energy regime, potential backgrounds are expected scarce so that some selection cuts may be relaxed appropriately. 
However, a more detailed discussion is not a focus of this work. We simply assume that the aforementioned selections and modifications will allow for a zero-background analysis.\footnote{Even in the case of non-zero background, the main messages that we will deliver still hold.} 

Unlike the previous two exercises, we report the results only in the electron channel. 
In particular, any (isolated) recoil proton cannot exceed the energy threshold of the DeepCore experiment.
Furthermore, considering the required energy scales relevant to DeepCore, the proton-involved (incoherent) scattering processes are not dominant. 
The $\chi_1$-induced DIS would come into play. 
According to our discussion in Section~\ref{sec:pvsDIS}, however, we may have to collect the data for $\gsim 10$ years in much of the parameter space of interest in order to reach the expectation of having a single event, before multiplying the acceptance.
Nevertheless, it is worth to look into DIS-induced events with large values of $E_1$ as the effective mass of DeepCore is at least $\sim5$ Mt. 
We leave the relevant dedicated analysis to the future~\cite{futurework}.

\begin{figure}[t]
\centering
\includegraphics[width=7.2cm]{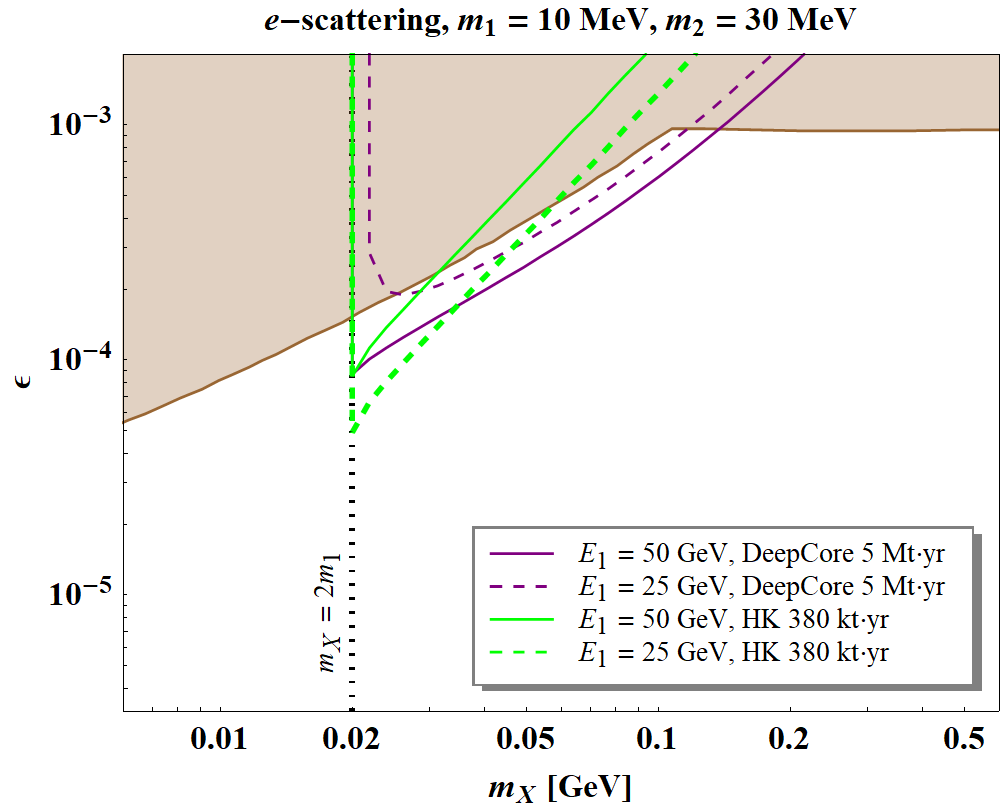} \hspace{0.2cm} 
\includegraphics[width=7.2cm]{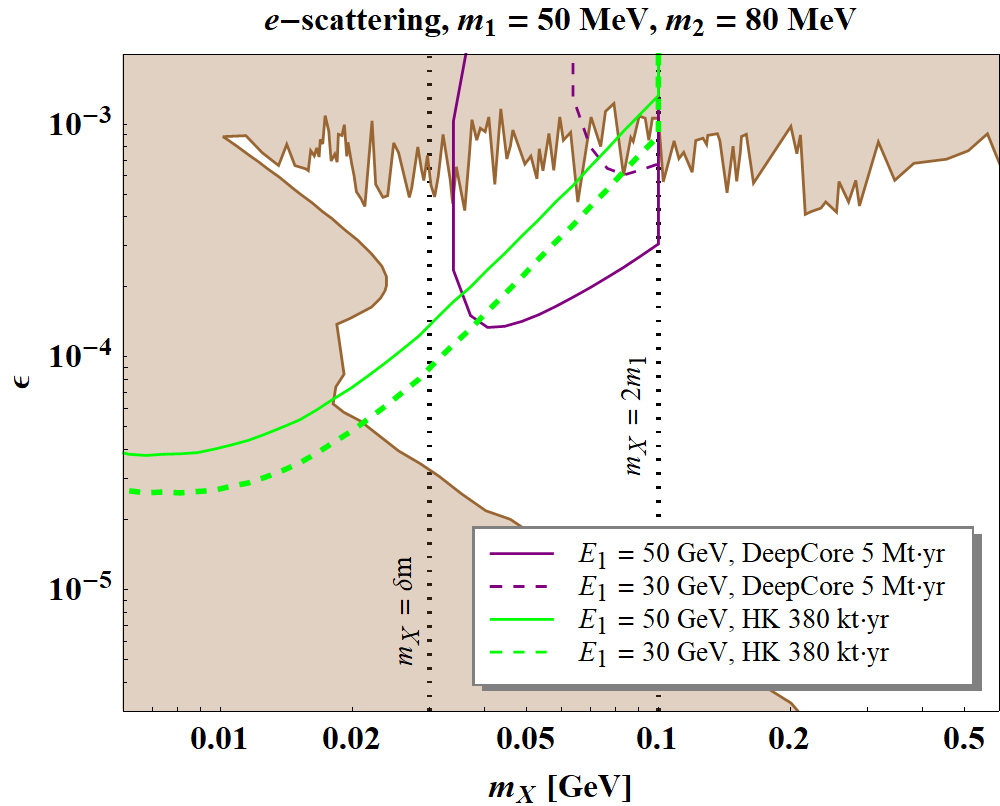}
\caption{\label{fig:DCvsHK} 
Experimental sensitivities achievable in the electron channel, in the $m_X-\epsilon$ plane. 
The left panel (right panel) compares DeepCore--5Mt$\cdot$yr (purple lines) and HK--380kt$\cdot$yr (green lines) for Scenario I (Scenario II) with two different reference points.
}
\end{figure}

We finally demonstrate experimental sensitivities for Scenarios I and II in the left and right panels of Figure~\ref{fig:DCvsHK}, respectively. 
DeepCore--5Mt$\cdot$yr and HK--380kt$\cdot$yr are color-coded by purple and green. 
In the left panel, we first contrast the sensitivities with $(E_1, m_1, m_2)=(50, 0.01, 0.03)$ GeV. 
The incoming energy is enough to overcome the threshold in DeepCore. 
On the other hand, the associated $\chi_1$ flux is not large, so DeepCore is better motivated due to its $\sim 13$ times larger effective mass. 
However, we see that as smaller $E_1$ (here $E_1=25$ GeV) is assumed, DeepCore quickly loses signal sensitivity because of the energy threshold, while HK achieves improved sensitivity due to a higher $\chi_1$ flux. 

Similar phenomenon can be observed for Scenario II (see the right panel). 
For larger $E_1$, DeepCore would be better motivated, but for smaller $E_1$ hypotheses, the sensitivity of DeepCore falls rapidly, motivating detectors which are sensitive to smaller energy deposit. 
Here we see that DeepCore has no signal sensitivity below the $m_X=\delta m$ line. 
The reason is that in this regime an on-shell $X$ is created and decays quickly within 5 meters, i.e., no double-cascade signature arises.
The chosen mass spectrum indeed does not let $\chi_2$ decay late even when $m_X>\delta m$. 
So, the acceptance $A_\ell$ at HK does not change drastically across the $m_X=\delta m$ line and both of the HK curves do not show an appreciable kink structure around the line. 

Before moving on to the concluding section, we make a couple of noteworthy comments. 
Since DeepCore comes with an effective mass of 5 Mt, a few years of data taking could make it sensitive to parameter space with $E_1\gsim 1$ TeV, which is far beyond the reach of other experiments. 
More interestingly, such a large $E_1$ allows $\chi_1$ to up-scatter to a much heavier state so that visible particles other than an electron-positron pair can be emitted from the secondary interaction vertex (see Section~\ref{sec:model}). 
For example, the $e$-scattering channel with $E_1 > 100$ GeV may accompany $\mu^+ \mu^-$ in the final state.
This is certainly an interesting enough possibility, as it may induce a signature of a cascade plus a track (from different vertices). 
We reserve the subject for future work~\cite{futurework}.

\section{Conclusions}
\label{sec:conclusion}

The search for boosted dark matter has received rising interest as an alternative approach to probe dark sector physics including cosmological dark matter. 
Several dark matter model frameworks have been proposed in order to give rise to boosted dark matter in the universe today: for example, two-component dark matter scenario~\cite{Belanger:2011ww, Agashe:2014yua, Kim:2017qaw, Aoki:2018gjf}, $Z_3$-stabilized dark matter models carrying semi-annihilation processes~\cite{DEramo:2010keq}, models involving dark matter-induced nucleon decays~\cite{Huang:2013xfa}, models with decaying super-heavy particles~\cite{Bhattacharya:2014yha, Kopp:2015bfa, Heurtier:2019rkz}, or cosmic-ray induced energetic dark matter scenarios~\cite{Yin:2018yjn, Bringmann:2018cvk, Ema:2018bih}.
Many ongoing and future dark matter direct detection and neutrino experiments can observe signals induced by boosted dark matter, and a host of phenomenological studies have proposed search strategies, channels, and sources of boosted dark matter~\cite{Agashe:2014yua, Berger:2014sqa, Kong:2014mia, Bhattacharya:2014yha, Kopp:2015bfa, Necib:2016aez, Alhazmi:2016qcs, Kim:2016zjx, Cherry:2015oca, Giudice:2017zke, McKeen:2018pbb, Chatterjee:2018mej, Kim:2018veo, Kim:2017qaw, Aoki:2018gjf, Huang:2013xfa, Yin:2018yjn, Bringmann:2018cvk, Ema:2018bih, Kim:2019had, Heurtier:2019rkz, Berger:2019ttc, DeRoeck:2020ntj}, with regard to those experiments. 
The relevant field of research is rather new and systematic approaches toward devising clever search schemes are absent. 
A more efficient design to dig out elusive boosted dark matter signals is of great importance especially at the earlier stage of experiments, improving sensitivity to BDM models. 

In light of this situation, we performed a dedicated study to address the issue above, taking a benchmark model, namely, a two-component dark matter scenario with boosted (lighter) dark matter scattering off either electron or proton. 
We allowed the boosted dark matter $\chi_1$ to scatter off not only elastically (i.e., $e$BDM) but inelastically (i.e., $i$BDM) to a heavier unstable dark-sector state $\chi_2$ which further 
decays back into the lighter dark matter and additional secondary (visible) particles.
For each of the two possible scenarios, the scattering cross sections of boosted dark matter via electron and proton targets were carefully investigated.
Deep inelastic scattering cross sections were included in association with the proton scattering.\footnote{Recently, a DIS module for $e$BDM is implemented in \texttt{GENIE}~\cite{Berger:2018urf}.}
We have found that the DIS contribution is negligible in the total proton scattering cross section, as far as the mass of the particle mediating the interaction between boosted dark matter and SM particles is not significantly larger than the typical energy scale to induce a DIS. 
A parameter choice for which the DIS dominates over the proton scattering cross section typically results in a tiny total cross section so that the proton channel becomes irrelevant unless the detector of interest comes with a huge volume to compensate for the small signal rate. 
Therefore, treating the proton itself 
as a target suffices when one considers the cross section of boosted dark matter scattering-off a proton.
We emphasize that the generic conclusions here readily hold for other scenarios involving boosted/relativistic dark matter (e.g., scenarios with cosmological dark matter decaying to lighter dark matter, fixed target experiments), as our study was done with respect to the energy of incoming boosted dark matter. 

We then compared electron scattering and proton scattering channels in both of elastic and inelastic boosted dark matter signals. 
We first performed simplified comparisons assuming a perfect detector (i.e., perfect energy/angular resolutions, no energy thresholds, etc.) simply to develop our intuition. 
We found that the proton scattering channel is better motivated as hypothesized $m_1$ and $m_X$ increase. 
The proton channel is also advantageous if the underlying model assumption is an $i$BDM process with a large mass difference between $\chi_1$ and $\chi_2$. 
We then took some realistic effects into account such as energy threshold, angular resolution, and tagging efficiency for the displaced vertex. 
To show the dependence on detector types, we performed separate studies on a DUNE-like LArTPC detector and a HK-like Cherenkov detector. 
For the former type, most of the observations made in the simplified study are relevant. 
On the other hand, for the latter type, the electron channel becomes competitive in a broader range of kinematically allowed parameter space. 
We, however, observed that (relatively) hard cuts associated to the HK-like detector significantly reduce not only the range of accessible parameter space but the expected number of signal events in the search for $i$BDM. 
Therefore, it seems that ameliorating experimental performance relevant to the event selection is an inevitable task toward improved signal sensitivities.

We finally conducted example analyses in our benchmark detectors: DarkSide-20k, DUNE, Hyper-K, and DeepCore. 
Experimental data, which would be collected at those detectors, was interpreted in terms of coverage in the plane of dark gauge boson mass vs. kinetic mixing parameter, as our benchmark model to describe interactions between $\chi_1$ and SM particles involves a dark gauge boson. 
We found that relevant experimental sensitivities highly depend on various detector specifications such as geometry, energy threshold, resolutions, and so on. 
Thus, an optimized (BDM or BDM-like signal) search scheme can be established up on multilateral considerations about those factors.
Detectors with a small volume and small threshold (e.g., DarkSide-20k) are better for searching for signatures induced by less energetic boosted $\chi_1$, whereas those with a large volume and large threshold (e.g., DeepCore) can be sensitive to signal events coming from high energetic boosted $\chi_1$ although the associated $\chi_1$ flux is often small. 
A more precise measurement/reconstruction of final state visible particles can help smaller-volume detectors achieve experimental sensitivities comparable to larger-volume detectors.

\section*{Acknowledgments}We thank Joshua Berger, Joshua Isaacson, Anne Schukraft, Yun-Tse Tsai, and Jaehoon Yu for insightful discussions. 
Fermilab is operated by Fermi Research Alliance, LLC, under Contract No. DE-AC02-07CH11359 with the US Department of Energy.
DK, JCP, and SS appreciate the hospitality of Fermi National Accelerator Laboratory.
The work of DK was supported in part by the Department of Energy under Grant No. DE-FG02-13ER41976/DE-SC0009913 and is supported in part by the Department of Energy under Grant de-sc0010813.
The work of JCP is supported by the National Research Foundation of Korea (NRF-2019R1C1C1005073 and NRF-2018R1A4A1025334). 
The work of SS was supported by the National Research Foundation of Korea (NRF-2017R1D1A1B03032076 and NRF-2020R1I1A3072747).
This work was performed at the Aspen Center for Physics, which is supported by National Science Foundation grant PHY-1607611.
SS would like to express a special thanks to the Mainz Institute for Theoretical Physics (MITP) of the Cluster of Excellence PRISMA+ (Project ID 39083149) for its hospitality and support.

\appendix
\section{Cross Section Formulas and Analysis Details}
\label{sec:derivation}
\numberwithin{equation}{section}

\subsection{Incoherent scattering}
We first recall the matrix element squared for the process $\chi_1 T \rightarrow \chi_2 T$ with $T$ being the associated target (i.e., $T=e$ or $p$)~\cite{Kim:2016zjx}:
\begin{align}
\overline{\left|\mathcal{M} \right|}^2
&= \frac{8(\epsilon e g_{12})^2 m_T}{\{2m_T(E_{2}-E_{1})-m_X^2\}^2} \nonumber \\
&\hspace{0.2cm} \times \left[\mathcal{M}_0  (F_1 + \kappa F_2)^2   +  \mathcal{M}_1 \left\{ - (F_1 + \kappa F_2)\kappa F_2  + \frac{E_{1} - E_{2} + 2 m_T}{4 m_T}(\kappa F_2)^2 \right\}\right], \label{eq:matrixX}
\end{align}
where $E_{1(2)}$ is the $\chi_{1(2)}$ energy measured in the laboratory frame.
Here $\mathcal{M}_0$ and $\mathcal{M}_1$ are defined as 
\bea
\mathcal{M}_0 &=& \left[m_T( E_{1}^2+E_{2}^2)-\frac{\delta m^2}{2}(E_{2} -E_{1} +m_T) + m_T^2(E_{2}-E_{1})+m_{1}^2E_{2} - m_{2}^2 E_{1} \right],~~~ \label{eq:expm0} \\
\mathcal{M}_1 &=& m_T \left[ \left(E_{1} + E_{2} -\frac{m_{2}^2 - m_{1}^2}{2m_T} \right)^2 + (E_{1} - E_{2} + 2 m_T) \left(E_{2} - E_{1} - \frac{\delta m^2}{2m_T} \right) \right], \label{eq:expm1}
\eea
where $\delta m \equiv m_{2}-m_{1}$. 
In the laboratory frame, the differential scattering cross section is
\begin{align}
\frac{d\sigma}{dE_T}=\frac{m_T }{8\pi\lambda(s,m_T^2,m_{1}^2)}\overline{\left|\mathcal{M} \right|}^2 \,, \label{eq:ETspec}
\end{align}
where $E_T =m_T+E_1-E_2$ is the energy of the recoiling target and $\lambda(x,y,z) = (x - y - z)^2 - 4yz$.
The kinematically allowed maximum and minimum of $E_T$, defined as $E_T^+$ and $E_T^-$, are
\begin{align}
E_T^\pm = \frac{s + m_T^2 - m_2^2}{2 \sqrt{s}} \frac{E_1 + m_T}{\sqrt{s}} \pm \frac{\lambda^{1/2}(s,m_T^2,m_2^2)}{2\sqrt{s}}\frac{p_{\chi_1}}{\sqrt{s}}\,,
\end{align}
where $p_{\chi_1} = \sqrt{E_1^2 - m_1^2}$.

Let us explain how to calculate the terms in $\mathcal M_1$ $\times$ (form factors), following Ref.~\cite{Borie:2012tu}.
In the matrix element, the vertex $\bar u (p_p) \gamma^\mu u(p_0)$ should be replaced by $\bar u (p_p) \Gamma^\mu u(p_0)$, where $p_p$ is the 4-momentum of the recoiling proton while $p_0$ is the 4-momentum of the proton at rest before the scattering. 
We then express $\Gamma^\mu$ in terms of $F_1$ and $F_2$ as follows:
\begin{align}
\Gamma^\mu = F_1 (q^2) \gamma^\mu + \kappa F_2 (q^2) \frac{i \sigma^{\mu \nu} q_\nu}{2 m_p}~,
\end{align}
where $q^\nu$ is the transferred energy and momentum, i.e., $p_{1}^\nu - p_{2}^\nu = p_p^\nu - p_0^\nu$, and the anomalous magnetic moment $\kappa = 1.79$ (-1.91) for the proton (neutron). 
For the proton, the form factors are related with the Sachs electric and magnetic form factors $G_E$ and $G_M$ such that
\begin{align}
F_1 &= \frac{G_E + \frac{Q^2}{4 m_p^2}}{1 + \frac{Q^2}{4 ma_p^2}}~, \\
\kappa F_2 &= \frac{G_M - G_E}{1 + \frac{Q^2}{4 m_p^2}}~,
\end{align}
for $Q^2 = - q^2 = 2(E_p - m_p) m_p$. 
Note that $q^2$ is negative for $E_p > m_p$ due to the metric $(1, -1, -1, -1)$. 
The Rosenbluth formula in conjunction with experimental measurements shows the following dipole approximation 
\begin{align}
G_E = \frac{G_M}{\mu_p} = \left(1 + \frac{Q^2}{0.71\,{\rm GeV}^2} \right)^{-2}~,
\end{align} 
up to around $Q^2 \sim 10\,{\rm GeV}^2$ with $\mu_p = (1 + \kappa_p) e / (2m_p) = 2.79$~\cite{Qattan:2004ht}.\footnote{Note, however, that the ratio of the electric and magnetic form factor $\mu_p G_E / G_M $ decreases from 1 for larger $Q^2 > 1\,{\rm GeV}^2$ from the high precision double polarization experiments~\cite{Perdrisat:2006hj,Galynsky:2012dp}. 
In our analysis, we assume that keeping the dipole approximation for $Q^2 \lesssim 4\,{\rm GeV}^2$ is reasonable in the region of elastic scattering of our proton target.} 
The form factors are also related as
\begin{align}
G_M = F_1 + \kappa_p F_2~, &\hspace{1cm} G_E = F_1 - \frac{Q^2}{4 m_p^2} \kappa_p F_2\,.
\end{align}

For the purpose of understanding several phenomena in a semi-analytic manner, we further provide the approximated formulas of the cross sections.
Recollect that for the proton channel, the differential cross section $d \sigma_{\chi_1 p} / d p_p$ is peaking toward $p_p \ll m_p$ so that the denominator of the matrix element stemming from the $t$-channel dark gauge boson exchange can be approximated to $\{2 m_p (E_2 - E_1) - m_X^2)\}^2 \simeq (p_p^2 + m_X^2)^2$. 
Also, the phase function $\lambda(s,m_p^2,m_1^2) = 4 m_p^2 (E_1^2 - m_1^2)$. 
Hence, the differential cross section is approximately written as
\begin{align}
\frac{d \sigma_{\chi_1 p}}{d p_p} &\simeq \frac{p_p}{\sqrt{m_p^2 + p_p^2}} \frac{m_p}{8 \pi \cdot 4 m_p^2 ( E_1^2 -  m_1^2 )} \frac{8 (\epsilon e g_{12})^2 m_p}{(p_p^2 + m_X^2)^2} \nonumber \\
& \hspace{0.5cm} \times 2m_p E_1^2 \left[ 1 - \frac{p_p^2}{2 m_p E_1} - \frac{p_p^2}{4 E_1} - \frac{m_1^2 p_p^2}{4 m_p^2 E_1^2} -\frac{\delta m}{2 E_1} \left\{ \frac{2 m_1}{m_p} + \frac{\delta m}{E_1} \left( 1 + \frac{E_1}{m_p} - \frac{p_p^2}{4 m_p^2} \right) \right\} \right]
\nonumber \\
&= \frac{p_p}{\sqrt{m_p^2 + p_p^2}} \frac{(\epsilon e g_{12})^2 m_p E_1^2}{2\pi ( E_1^2 -  m_1^2 ) (p_p^2 + m_X^2)^2} \left[ 1 - \frac{p_p^2}{2 m_p E_1} - \frac{p_p^2}{4 E_1^2} - \frac{m_1^2 p_p^2}{4 m_p^2 E_1^2}  \right. \nonumber \\
&\hspace{5cm} \left. -\frac{\delta m}{2 E_1} \left\{ \frac{2 m_1}{m_p} + \frac{\delta m}{E_1} \left( 1 + \frac{E_1}{m_p} - \frac{p_p^2}{4 m_p^2} \right) \right\} \right]\,,
\label{eq:papprox1}
\end{align}
up to $\mathcal O (p_p^4 / m_p^4)$, with the contributions by the form factors neglected.
Note that we used the chain rule $d \sigma_{\chi_1 p} / d p_p = (\partial E_p / \partial p_p) (d \sigma_{\chi_1 p} / d E_p)$ to obtain Eq.~\eqref{eq:papprox1} from Eq.~\eqref{eq:ETspec}.
The first factor compensates the increase of $d \sigma_{\chi_1 p} / d E_p$ in decreasing $p_p$ (for $p_p < m_p$) and thus the differential cross section culminates around a value of $p_p \ll m_p$ as in FIG.~2 of Ref.~\cite{Kim:2016zjx}.
Neglecting the extra terms suppressed by $p_p^2 / m_p E_1$, $p_p^2 / E_1^2$, and $p_p^2 / m_p^2$, we find that the differential cross section peaks around
\begin{align}
p_p &= m_p \sqrt{\frac{3 \left( -1 + \sqrt{1 + \frac{16 m_X^2}{9 m_p^2}}\right)}{8}} \\
&\simeq \frac{m_X}{\sqrt{3}}\,,~~{\rm if}~ m_X^2 \ll m_p^2\,.
\label{eq:maxpp}
\end{align}
In the parameter region holding the relations $\delta m \ll E_1$ and $m_1 \ll E_1$, Eq.~\eqref{eq:papprox1} can be further simplified to
\begin{align}
\left. \frac{d \sigma_{\chi_1 p}}{d p_p} \right|_{E_1 \gg \delta m, m_1} &\simeq \frac{p_p}{\sqrt{m_p^2 + p_p^2}} \frac{(\epsilon e g_{12})^2 m_p }{2\pi (p_p^2 + m_X^2)^2} \left[ 1 + \mathcal O \left( \frac{m_1^2}{E_1^2} , \frac{p_p^2}{E_1^2}, \frac{p_p^2}{m_p E_1}, \frac{\delta m m_1}{E_1 m_p}, \frac{(\delta m)^2}{ E_1^2} \right) \right]\,.
\label{eq:papprox2}
\end{align}
We warn that $E_1 \gg \delta m$ is, in principle, not always true in our parameter region, but the above approximation is still valid for {\it e}BDM and some of {\it i}BDM parameter space.

It is rather challenging to come up with a reasonably simple approximation for $e$-scattering since $m_e$ is too small for the differential cross section to drop rapidly as increasing $p_e$.
We nevertheless provide the following expressions for the differential cross section in $E_e$, namely,
\begin{align}
\frac{d \sigma_{\chi_1 e}}{d E_e} &= \frac{m_e}{8 \pi \cdot 4 m_e^2 (E_1^2 - m_1^2)} \frac{8 (\epsilon e g_{12})^2  m_e}{(2 m_e^2 - 2 m_e E_e - m_X^2)^2} \nonumber \\
&\hspace{1cm} \times \left[ m_e \left\{ 2 E_1^2 + E_e^2 - 2 E_1 E_e + m_1^2 \right\} + m_e^2 \left\{ 2 E_1 - 3 E_e \right\} + 2 m_e^3 - m_1^2 E_e \right. \nonumber \\
&\hspace{2.5cm} \left. + \, \delta m \left\{ 2 m_1 E_1 + \delta m \left( E_1 - \frac{E_e}{2} + m_e \right) \right\} \right] \nonumber \\
&= \frac{(\epsilon e g_{12})^2 m_e E_1^2}{2\pi (E_1^2 - m_1^2) (2 m_e^2 - 2 m_e E_e - m_X^2)^2} \nonumber \\
&\hspace{1.5cm} \times \left[ 1 + \frac{m_e - E_e}{E_1} + \frac{E_e^2 + m_1^2 - 3 m_e E_e + 2 m_e^2}{2 E_1^2} - \frac{m_1^2 E_e}{2 E_1^2 m_e} \right. \nonumber \\
&\hspace{2cm} \left. + \frac{\delta m}{E_1} \left\{ \frac{m_1}{m_e} + \frac{\delta m}{2 E_1} \left( \frac{E_1}{m_e} + 1 - \frac{E_e}{m_e} \right)\right\} \right]\,,
\label{eq:dsigdEe}
\end{align}
and the differential cross section in $p_e$, 
\begin{align}
\left. \frac{d \sigma_{\chi_1 e}}{d p_e} = \frac{p_e}{\sqrt{m_e^2 + p_e^2}} \, \frac{d \sigma_{\chi_1 e}}{d E_e} \right|_{E_e \to \sqrt{m_e^2 + p_e^2}}\,.
\end{align}
Note that the term $p_e / \sqrt{m_e^2 + p_e^2}$ is almost one as long as $p_e \gg m_e$.
Hence, the differential cross section peaks around $p_e \lesssim m_e$. 
Moreover, we numerically find that 
the distribution is rather smooth for $m_e \lesssim p_e \ll E_1$, in contrast to that of $d \sigma_{\chi_1 p} /  d p_p$.
Interestingly, for $E_1 \gg m_1, \delta m$ and $p_e \gg m_e, m_1$, the value of the differential cross section is almost flat up to 
\begin{align}
p_e &= \frac{\left( E_1 + \frac{m_1^2}{2 m_e} + \frac32 m_e \right)(m_X^2 - 2 m_e^2) + 4 m_e E_1^2 + 4 m_e^2 E_1 + 2 m_e m_1^2 + 8 m_e^3}{m_X^2 + 2 m_e E_1 + m_1^2 + m_e^2} \\
&\simeq \frac{E_1 (m_X^2 + 4 m_e E_1 + 4 m_e^2)}{m_X^2 + 2 m_e E_1}\,,~~{\rm if}~ m_X^2 \gg m_1^2 > m_e^2 \,,
\label{eq:pemax}
\end{align}
as long as it is smaller than the maximum $E_e^{\rm max}$ in Ref.~\cite{Kim:2016zjx}.
The flat region becomes wider in increasing $E_1$.

\subsection{Deep inelastic scattering}\label{sec:derivationDIS}

We can write the amplitude squared for deep inelastic $\chi_1 + N\to \chi_2 + X$, as
\begin{equation}
  \overline{|\mathcal{A}|}^2 = \left(\frac{g_{12}^2 e \epsilon}{q^2-m_{X}^2}\right)^2 \left|D^{\mu\nu}W_{\mu\nu}\right|^2\,,
\end{equation}
where $D^{\mu\nu}$ and  $W^{\mu\nu}$ are the tensors related to the $\chi_{1,2}$ and hadronic vertices, respectively. 
We call $p_{1,2}$ the four-momentum of $\chi_{1,2}$  and $q^\mu\equiv p_1^\mu-p_2^\mu$ is the momentum transfer. 
Taking into account non-vanishing masses for the dark sector fermions, we obtain
\begin{equation}
  D^{\mu\nu} = g^{\mu\nu}\left[q^2-(m_1-m_2)^2\right] - 2(p_1^\mu q^\nu + q^\mu p_1^\nu  )+4p_1^\mu p_1^\nu\,,
\end{equation}
while we parameterize ($k_\mu$ is the four-momentum of the target nucleon and $k^2=M^2$)
\begin{equation}
  W_{\mu\nu}=-W_1\left(g_{\mu\nu} -\frac{q_\mu q_\nu}{q^2}\right)+\frac{W_2}{M^2}\left[ k_\mu-\frac{(q\cdot k) q_\mu}{q^2} \right]
  	\left[ k_\nu-\frac{(q\cdot k) q_\nu}{q^2} \right].
\end{equation}
The structure functions $W_{1,2}$ are written in terms of the parton distribution functions (PDFs): 
\begin{equation}\label{eq:structure-functions}
  W_1 = \frac{f(x)}{2M}\,,\qquad  W_2 = \frac{x f(x)}{E_1-E_2}\,,
\end{equation}
where $x$ is the momentum fraction carried by the parton and $f(x)$ is the associated PDF. A sum over all partons is implicit in Eq.~(\ref{eq:structure-functions}). Proceeding with a 
further calculation we obtain
\begin{align}
  \frac{d^2\sigma_{\rm DIS}}{dx\,dy} &= g_{12}^2 e^2 \epsilon^2\frac{4 f(x)}{32 \pi E_1 (Q^2+m_{X}^2)} \times  \Big\{  2E_1^2 M x (2-2y+y^2)  - M x (m_1-m_2)^2 \\
  &\hspace{5.5cm}+\left[(m_1^2-m_2^2)(2-y)-2m_1 m_2 y -2M^2 x^2 y\right]E_1  \Big\}\,. \nonumber
\end{align}


\begin{table}[h]
\centering
\hspace*{-1.2cm}
\scalebox{0.6}{
\begin{tabular}{ c | c  c  c  c  c  c  c  c  c  c  c }
\hline \hline
{\bf Dark Matter} & Target & \multicolumn{2}{c}{Volume [t]} & Depth & $E_{\rm th}$ & \multicolumn{3}{c}{Resolution} & \multirow{2}{*}{PID} & Run & \multirow{2}{*}{Refs.} \\
{\bf Experiments} & Material & Active & Fiducial & [m] & [keV] & Position [cm] & Angular [$^\circ$] & Energy [\%] &  & Time &  \\
\hline
DarkSide & LAr & 46.4 & 36.9 & 3,800 & \multirow{2}{*}{$\mathcal{O}(1)$} & \multirow{2}{*}{$\sim0.1-1$} & \multirow{2}{*}{$-$} & \multirow{2}{*}{$\lesssim10$} & \multirow{2}{*}{$-$} & \multirow{2}{*}{2013-} & \multirow{2}{*}{\cite{Agnes:2014bvk}} \\
-50 & DP-TPC & kg & kg & m.w.e. &  &  &  &  &  &  & \\
\hline
DarkSide & LAr & \multirow{2}{*}{23} & \multirow{2}{*}{20} & 3,800 & \multirow{2}{*}{$\mathcal{O}(1)$} & \multirow{2}{*}{$\sim0.1-1$} & \multirow{2}{*}{$-$} & \multirow{2}{*}{$\lesssim10$} & \multirow{2}{*}{$-$} & goal: & \multirow{2}{*}{\cite{Aalseth:2017fik}} \\
-20k & DP-TPC &  &  & m.w.e. &  &  &  &  &  & 2021- & \\
\hline
\multirow{2}{*}{XENON1T} & LXe & \multirow{2}{*}{2.0} & \multirow{2}{*}{1.3} & 3,600 & \multirow{2}{*}{$\mathcal{O}(1)$} & \multirow{2}{*}{$\sim0.1-1$} & \multirow{2}{*}{$-$} & \multirow{2}{*}{$-$} & \multirow{2}{*}{$-$} & 2016 & \multirow{2}{*}{\cite{Aprile:2017aty, Aprile:2018dbl}} \\
 & DP-TPC &  &  & m.w.e. &  &  &  &  &  & -2018 & \\
\hline
\multirow{2}{*}{XENONnT} & LXe & \multirow{2}{*}{5.9} & \multirow{2}{*}{$\sim4$} & 3,600 & \multirow{2}{*}{$\mathcal{O}(1)$} & \multirow{2}{*}{$\sim0.1-1$} & \multirow{2}{*}{$-$} & \multirow{2}{*}{$-$} & \multirow{2}{*}{$-$} & goal: & \multirow{2}{*}{\cite{Aprile:2017aty}} \\
 & DP-TPC &  &  & m.w.e. &  &  &  &  &  & 2020- & \\
\hline
DEAP & SP LAr & \multirow{2}{*}{3.26} & \multirow{2}{*}{2.2} & \multirow{2}{*}{2,000} & \multirow{2}{*}{$\mathcal{O}(10)$} & \multirow{2}{*}{$<10$} & \multirow{2}{*}{$-$} & \multirow{2}{*}{$\sim10-20$} & \multirow{2}{*}{$-$} & \multirow{2}{*}{2016-} & \multirow{2}{*}{\cite{Amaudruz:2014nsa, Amaudruz:2017ekt, Amaudruz:2017ibl}} \\
-3600 & S1 only &  &  &  &  &  &  &  &  &  & \\
\hline
DEAP & SP LAr & \multirow{2}{*}{150} & \multirow{2}{*}{50} & \multirow{2}{*}{2,000} & \multirow{2}{*}{$\mathcal{O}(10)$} & \multirow{2}{*}{15} & \multirow{2}{*}{$-$} & \multirow{2}{*}{$-$} & \multirow{2}{*}{$-$} & \multirow{2}{*}{$-$} & \multirow{2}{*}{\cite{Amaudruz:2014nsa}} \\
-50T & S1 only &  &  &  &  &  &  &  &  &  & \\
\hline
LUX- & LXe & \multirow{2}{*}{7} & \multirow{2}{*}{5.6} & \multirow{2}{*}{1,500} & \multirow{2}{*}{$\mathcal{O}(1)$} & \multirow{2}{*}{$\sim0.1-1$} & \multirow{2}{*}{$-$} & \multirow{2}{*}{2.5 MeV: 2} & \multirow{2}{*}{$-$} & goal: & \multirow{2}{*}{\cite{Mount:2017qzi, Akerib:2019fml}} \\
ZEPLIN & DP-TPC &  &  &  &  &  &  &  &  & 2020- & \\
\hline
\hline
{\bf Neutrino} & Target & \multicolumn{2}{c}{Volume [kt]} & Depth & $E_{\rm th}$ & \multicolumn{3}{c}{Resolution} & \multirow{2}{*}{PID} & Run & \multirow{2}{*}{Refs.} \\
{\bf Experiments} & Material & Active & Fiducial & [m] & [MeV] &  Vertex [cm] & Angular [$^\circ$] & Energy [\%] &  & Time &  \\
\hline
\multirow{2}{*}{Borexino} & organic & \multirow{2}{*}{0.278} & \multirow{2}{*}{0.1} & 3,800 & \multirow{2}{*}{$\sim0.2$} & \multirow{2}{*}{$\sim$9-17} & \multirow{2}{*}{$-$} & \multirow{2}{*}{$\frac{5}{\sqrt{E\,({\rm MeV})}}$} & \multirow{2}{*}{$-$} & $>5.6$ & \multirow{2}{*}{\cite{Back:2012awa}} \\
 & LS &  &  & m.w.e. &  &  &  &  &  & year & \\
\hline
\multirow{2}{*}{KamLAND} & \multirow{2}{*}{LS} & \multirow{2}{*}{1} & \multirow{2}{*}{0.2686} & \multirow{2}{*}{1,000} & \multirow{2}{*}{0.2 - 1} & \multirow{2}{*}{$\frac{12 - 13}{\sqrt{E\,({\rm MeV})}}$} & \multirow{2}{*}{$-$} & \multirow{2}{*}{$\frac{6.4 - 6.9}{\sqrt{E\,({\rm MeV})}}$} & \multirow{2}{*}{$-$} & $\sim10$ & \multirow{2}{*}{\cite{Gando:2014wjd, Asakura:2015bga}} \\
 &  &  &  &  &  &  &  &  &  & year? & \\
\hline
\multirow{3}{*}{JUNO} & \multirow{3}{*}{LS} & \multirow{3}{*}{$-$} & \multirow{3}{*}{20} & \multirow{3}{*}{700} & $<1$, & \multirow{3}{*}{$\frac{12}{\sqrt{E\,({\rm MeV})}}$} & $\mu$: & \multirow{3}{*}{$\frac{3}{\sqrt{E\,({\rm MeV})}}$} & $\mu^\pm$ vs $\pi^\pm$, & goal: & \multirow{3}{*}{\cite{An:2015jdp, Djurcic:2015vqa, Guo:2019vuk}} \\
 &  &  &  &  & goal: 0.1 &  & $L>5$ m: $<1$, &  & $e^\pm$ vs $\pi^0$: & 2021- & \\
 &  &  &  &  &  &  & $L>1$ m: $<10$ &  & difficult & \\
\hline
\multirow{5}{*}{DUNE} & \multirow{5}{*}{LArTPC} &   &   & \multirow{5}{*}{1500} &   & \multirow{5}{*}{$\lsim1-2$} &  & $e: 20$ ($E<0.4$ GeV), &   &   & \multirow{5}{*}{\cite{Abi:2018alz, Abi:2018rgm, Abi:2020wmh, Abi:2020evt, Abi:2020loh, DeRoeck:2020ntj}} \\
  &   & Total: &   &   & $e: 30$, &   &   & 10  ($E<1.0$ GeV), & good & 10 kt: & \\
 &  & $17.5$ & $\gtrsim10$ &  & $p:$ &  & $e, \mu: 1$, & $2+\frac{8}{\sqrt{E/{\rm GeV}}}$ ($E>1.0$ GeV) & $e, \mu, \pi^\pm, p$ & 2026-, & \\
 &  & $\times4$ & $\times4$ &  & 21-50 &  & $\pi^\pm, p, n: 5$ & $p: 10$ ($E<1.0$ GeV), & separation & 20 kt: & \\
 &  &  &  &  &  &  &  & $5+\frac{5}{\sqrt{E/{\rm GeV}}}$ ($E>1.0$ GeV) &  & 2027- & \\
\hline
\multirow{3}{*}{SK} & Water & Total: & \multirow{3}{*}{22.5} & \multirow{3}{*}{1,000} & $e:\, 5$, & 5 MeV: 95, & 10 MeV: 25, & 10 MeV: 16, & $e, \mu$: &  $\gtrsim14$ & \multirow{3}{*}{\cite{Abe:2010hy, Abe:2014gda, Richard:2015aua}} \\
 & Cherenkov & 50 &  &  & $p: 485$ & 10 MeV: 55, & 0.1 GeV: 3, & 1 GeV: 2.5  & good & year & \\
 &  &  &  &  &  & 20 MeV: 40 & 1.33 GeV: 1.2 &  &  &  & \\
\hline
\multirow{4}{*}{HK} &  & Total: &  & Japan: &  & 5 MeV: 75, &  &  & $e, \mu$: &  & \multirow{4}{*}{\cite{Abe:2011ts, Abe:2016ero, Abe:2018uyc}} \\
 & Water & 258 & 187 & 650, & $e:\, <5$, & 10 MeV: 45, & similar & better & good, & goal: & \\
 & Cherenkov & $\times2$ & $\times2$ & Korea: & $p: 485$ & 15 MeV: 40, & to SK & than SK & $\pi^0, \pi^\pm$: & 2027- & \\
 &  &  &  & 1,000 &  & 0.5 GeV: 28 &  &  & mild &  & \\ 
\hline
\hline
{\bf Neutrino} & Target & \multicolumn{2}{c}{Effective} & Depth & $E_{\rm th}$ & \multicolumn{3}{c}{Resolution} & \multirow{2}{*}{PID} & Run & \multirow{2}{*}{Refs.} \\
{\bf Telescopes} & Material & \multicolumn{2}{c}{Volume [Mt]} & [m] & [GeV] &  Vertex [m] & Angular [$^\circ$] & Energy [\%] &  & Time & \\
\hline
\multirow{2}{*}{IceCube} & Ice & \multicolumn{2}{c}{100 GeV: $\sim30$,} & 1,450 & \multirow{2}{*}{$\sim100$} & vertical: 5, & $\mu$-track: $\sim1$, & 100 GeV: 28, & only & 2011- & \multirow{2}{*}{\cite{Aartsen:2016nxy, Aartsen:2013vja}} \\
 & Cherenkov & \multicolumn{2}{c}{200 GeV: $\sim200$} & Ice &  & horizontal: 15 & shower: $\sim30$ & 1 TeV: 16 & $\mu$ & (2008) & \\
\hline
\multirow{2}{*}{DeepCore} & Ice & \multicolumn{2}{c}{10 GeV: $\sim5$,} & 2,100 & \multirow{2}{*}{$\sim10$} & \multirow{2}{*}{better} & $\mu$-track: $\sim$1, & \multirow{2}{*}{$-$} & only & 2011- & \multirow{2}{*}{\cite{Collaboration:2011ym, Aartsen:2016nxy}} \\
 & Cherenkov & \multicolumn{2}{c}{100 GeV: $\sim30$} & Ice &  &  & shower: $\gtrsim10$ &  & $\mu$ & (2010) & \\
\hline
IceCube & Ice & \multicolumn{2}{c}{\multirow{2}{*}{$-$}} & 2,150 & \multirow{2}{*}{$\mathcal{O}(1)$} & much & 5 GeV: $\sim20$, & \multirow{2}{*}{$-$} & only & goal: & \multirow{2}{*}{\cite{Ishihara:2019aao}} \\
Upgrade & Cherenkov &  &  & Ice &  & better & 10 GeV: $\sim15$ &  & $\mu$ & 2023 & \\
\hline
\multirow{2}{*}{PINGU} & Ice & \multicolumn{2}{c}{1 GeV: $\gtrsim1$,} & 2,100 & \multirow{2}{*}{$\sim1$} & much & 1 GeV: 25, & 1 GeV: 55, & only & $>$ & \multirow{2}{*}{\cite{Aartsen:2014oha, TheIceCube-Gen2:2016cap}} \\
 & Cherenkov & \multicolumn{2}{c}{10 GeV: $\sim5$} & Ice &  & better & 10 GeV: 10 & 10 GeV: 25 & $\mu$ & 2023 & \\
\hline
\multirow{2}{*}{Gen2} & Ice & \multicolumn{2}{c}{\multirow{2}{*}{$\sim10$ Gt}} & 1,360 & $\sim50$ & \multirow{2}{*}{worse} & $\mu$-track: $<1$ & \multirow{2}{*}{$-$} & only & \multirow{2}{*}{$-$} & \multirow{2}{*}{\cite{Aartsen:2014njl}} \\
 & Cherenkov &  &  & Ice & TeV &  & shower: $\sim15$ &  & $\mu$ &  & \\
\hline \hline
\end{tabular}
}
\caption{A summary of detector specifications in various large-volume dark matter and neutrino experiments.
The numbers in the Active Volume entry for DUNE, SK, and HK are the values of their total volume.
Note that $p_p=1.07$ GeV corresponds to $E_{\rm th}\simeq485$ MeV for SK/HK.
[PID: particle identification, LAr/LXe: liquid argon/xenon, DP/SP: dual/single phase, TPC: time projection chamber, S1: prompt primary scintillation, LS: liquid scintillator, m.w.e.: meter water equivalent].
}
\label{table:exp}
\end{table}

\subsection{``Short-cut'' scheme \label{app:shortcut}}
We first prepare an acceptance table for the 10 kt DUNE-like detector module as a function of mean decay lengths of arbitrary long-lived particles. 
To this end, an arbitrary coordinate (regarded as a primary vertex) is generated in the detector fiducial volume, a random direction is generated,\footnote{Note that an assumption behind this scheme is that the flux of incoming boosted $\chi_1$ is cumulatively isotropic.} and then an actual flight distance is also generated according to the decay law with respect to a given mean decay length. 
From the coordinate, the direction, and the actual flight distance, the secondary vertex point is identified. 
If it is within the fiducial volume, it is accepted, otherwise rejected.
For each of 10,000 decay length values, 100 million pairs of the coordinate and the direction are produced to obtain the acceptance (denoted by $A_\ell$). 
Not all decay length values are probed, but we have found that the decay length vs. inverse acceptance is rather well described by a combination of a second order polynomial and a straight line jointed at some point which can be determined by fit:
\bea
\frac{1}{A_\ell}=\left\{
\begin{matrix*}[l]
c_1(\ell - c_2)+c_3 & \hbox{ for }\ell \geq c_2 \\[0.4em] 
c_4\ell^2+\left(\frac{c_3-1}{c_2}-c_2\cdot c_4 \right)\ell+1 & \hbox{ for } 0 \leq \ell <c_2
\end{matrix*}
\right.\,,
\eea
where $c_i$ ($i=1,2,3,4$) denote fit parameters.
Therefore, the acceptance for any non-investigated decay lengths can be calculated with this empirical fit function. 
Now for each mass point, we calculate the maximum laboratory-frame mean decay length of the associated long-lived particle (denoted by $\bar{\ell}^{\max}_{\rm lab}$), and find the corresponding acceptance by feeding $\bar{\ell}^{\max}_{\rm lab}$ to the fit function.
Finally, the eventual signal efficiency is given by the product of efficiencies attached to the above-bulleted items. 


\section{Information of Detectors}
\label{sec:detectors}

Here we collect specifications and useful information of detectors adopted at various large-volume ($\gsim$ 1 ton) dark matter and neutrino experiments, and tabulate in Table~\ref{table:exp}. 
One can take advantage of this table in combination with the main messages of our data analyses, when designing experimental strategies for searching for BDM-induced or similar signatures.


\bibliographystyle{JHEP}
\bibliography{biblio}

\end{document}